%% file: POWALKA_ApJPaper_GCcolors_vs_Models.tex
\documentclass{emulateapj}

\input{color.pac}

\shorttitle{NGVS/NGVS-IR: Globular cluster colors}
\shortauthors{M. Powalka et al.}

\begin{document}


\title{The Next Generation Virgo Cluster Survey (NGVS). XXV. Fiducial panchromatic colors of Virgo core globular clusters and their comparison to model predictions}


\author{Mathieu Powalka\altaffilmark{1}, Ariane Lan\c{c}on\altaffilmark{1}, Thomas H. Puzia\altaffilmark{2}, 
Eric W. Peng\altaffilmark{3,4}, Chengze Liu\altaffilmark{5,13}, Roberto P. Mu\~noz\altaffilmark{2}, John P. Blakeslee\altaffilmark{6}, Patrick C\^ot\'e\altaffilmark{6}, Laura Ferrarese\altaffilmark{6}, 
Joel Roediger\altaffilmark{6}, R\'uben S\'anchez-Janssen\altaffilmark{6}, Hongxin Zhang\altaffilmark{2,3}, Patrick R. Durrell\altaffilmark{8}, Jean-Charles Cuillandre\altaffilmark{7}, Pierre-Alain Duc\altaffilmark{7},
Puragra Guhathakurta\altaffilmark{9}, S. D. J. Gwyn\altaffilmark{6}, Patrick Hudelot\altaffilmark{12},
Simona Mei\altaffilmark{10,11,15}, Elisa Toloba\altaffilmark{9,14}}

\email{mathieu.powalka@astro.unistra.fr}

\affil{$^1$ Observatoire Astronomique de Strasbourg, Universit\'e de Strasbourg, CNRS, UMR 7550, 11 rue de l'Universit\'e, F-67000 Strasbourg, France}
\affil{$^2$ Departamento de Astronomía y Astrofísica, Pontificia Universidad Católica de Chile, 7820436 Macul, Santiago, Chile}
\affil{$^3$ Department of Astronomy, Peking University, Beijing 100871, China}
\affil{$^4$ Kavli Institute for Astronomy and Astrophysics, Peking University, Beijing 100871, China}
\affil{$^5$ Center for Astronomy and Astrophysics, Department of Physics and Astronomy, Shanghai Jiao Tong University, Shanghai 200240, China}
\affil{$^6$ National Research Council of Canada, Herzberg Astronomy and Astrophysics Program, Victoria, BC, V9E 2E7, Canada}
\affil{$^7$ AIM Paris Saclay, CNRS/INSU, CEA/Irfu, Université Paris Diderot, Orme des Merisiers, 91191 Gif sur Yvette cedex, France}
\affil{$^8$ Department of Physics and Astronomy, Youngstown State University, Youngstown, OH, USA}
\affil{$^9$ UCO/Lick Observatory, University of California, Santa Cruz, 1156 High Street, Santa Cruz, CA 95064, USA}
\affil{$^{10}$ GEPI, Observatoire de Paris, PSL Research University,  CNRS, University of Paris Diderot, 61, Avenue de l'Observatoire 75014, Paris France}
\affil{$^{11}$ University of Paris Denis Diderot, University of Paris Sorbonne Cit\'e (PSC), 75205 Paris Cedex 13, France}
\affil{$^{12}$ Institut d’Astrophysique de Paris, UMR 7095 CNRS \& UPMC, 98bis Bd Arago, F-75014 Paris, France}
\affil{$^{13}$ Shanghai Key Lab for Particle Physics and Cosmology, Shanghai Jiao Tong University, Shanghai 200240, China}
\affil{$^{14}$ Physics Department, Texas Tech University, Box 41051, Lubbock, TX 79409-1051, USA}
\affil{$^{15}$ California Institute of Technology, Pasadena, CA 91125, USA}


\begin{abstract}
The central region of the Virgo cluster of galaxies contains thousands
of globular clusters (GCs), an order of magnitude more than the numbers found
in the Local Group. Relics of early star formation epochs in the
universe, these GCs also provide ideal targets to test our understanding of
the Spectral Energy Distributions (SEDs) of old stellar populations.
Based on photometric data from the Next Generation Virgo cluster Survey
(NGVS) and its near-infrared counterpart NGVS-IR, we select a robust
sample of $\approx 2000 $ GCs with excellent photometry and spanning the
full range of colors present in the Virgo core. The
selection exploits the well defined locus of GCs in the $uiK$ diagram
and the fact that the globular clusters are marginally resolved in the
images. We show that the GCs define a narrow sequence in 5-dimensional
color space, with limited but real dispersion around the mean sequence.
The comparison of these SEDs with the predictions of eleven
widely used population synthesis models highlights differences between
models, and also shows that no single model adequately matches
the data in all colors. We discuss possible causes for some of these
discrepancies. Forthcoming papers of this series will examine how best to
estimate photometric metallicities in this context, and compare the
Virgo globular cluster colors with those in other environments.
\end{abstract}





\section{Introduction}

Globular clusters (GCs) are among the most thoroughly studied stellar populations 
in the sky. Their analysis has helped us understand stellar evolution,
and their ages have set essential constraints on cosmological models.
Since their formation, at times possibly as remote as the epoch of re-ionization, 
they have been affected by the numerous physical processes that shaped baryonic structure 
in the universe, and they have witnessed the dynamical and chemical
evolution of their host galaxies (e.g \citealt{pota2013}, \citealt{carretta2010}).

Historically, the stars bound in clusters have long been described 
as examples of coeval and chemically uniform stellar populations. 
This picture has naturally made GCs targets 
for the validation of population synthesis models (\citealt{renzini1988}).

Recent studies of color-magnitude diagrams and stellar
surface chemistries in nearby GCs have demonstrated that
the historical picture is only an approximation (\citealt{bedin2004},
\citealt{gratton2004}, \citealt{piotto2007}, \citealt{goudfrooij2009},
\citealt{piotto2012} or \citealt{renzini2015}). Possible processes
responsible for internal spreads in stellar properties
include self-enrichment, the merging of individual proto-clusters,
or the formation of nuclear clusters in galaxies that are
subsequently disrupted by tidal fields. Detailed studies of these
effects will be best carried out with resolved observations, but
the diversity of possible properties of GCs can also be tested,
out to much larger distances, with precise integrated multi-band photometry.

Over time, several studies have targeted nearby galaxies, producing 
catalogs of integrated photometry for GC samples.
To mention just a few:
the {\it McMaster catalog} of \citet[][and references therein]{harris1996} collects UBVRI colors of several dozen teams for 157 GCs of the Milky Way; the work of \citet{searle1980}
on the Magellanic Clouds has provided the {\it uvgr} colors of 61
star clusters, on which the popular (but now somewhat outdated)
SWB (Searle Wilkinson Bagnuolo) classification was based;
the Revised Bologna Catalog \citep{galleti2004} lists optical
and near-infrared photometry for several hundred
globular cluster candidates around M\,31, of which however not all have
been confirmed by more recent homogeneous surveys \citep{huxor2014}.
Finally, global studies of GC populations
as a function of host galaxy properties have yielded optical photometry
for samples of a few to a few $10^2$ globular cluster
candidates in various environments (e.g. \citealt{lotz2004};
\citealt{kundu2001}), but typically only for two
photometric passbands.

An important result of such surveys was the identification of color
subpopulations among GCs, as already suspected by \citet{kinman1959}.
In 1985, \citeauthor{zinn1985} showed that two distinct GCs subpopulations coexist in the Galaxy, and linked this to a metallicity bimodality.
In the following decades, it was shown that the distribution of blue (metal poor) GCs
is mostly associated with the stellar halo of galaxies,
while the red (more metal rich) GCs are mostly located in the central regions
(e.g \citealt{geisler1996}, \citealt{cote2001}, \citealt{forte2005},
\citealt{tamura2006} or \citealt{durrell2014}).
Although the shape of the color-metallicity relation remains a matter
of debate (\citealt{yoon2006}, \citealt{blakeslee2012}, \citealt{usher2015}),
it is now generally accepted that globular clusters
are found spread over three orders of magnitude in metallicity,
and that their mean metallicities are related
to their host galaxy stellar mass or luminosity (e.g \citealt{peng2006}).

Large, deep and well resolved surveys are critical for the definition of representative
GC samples with limited contamination, and they are progressively becoming available.
Targets like the Virgo, Coma or Fornax clusters are now being studied in detail.
The path was opened with surveys of the Hubble Space Telescope
Advanced Camera for Surveys (HST/ACS), which was used with two bandpasses (F475W and F850LP)
to scrutinize galaxies in the Virgo cluster (ACSVCS, \citealt{cote2004}),
the Fornax cluster \citep{jordan2007} or the Coma cluster \citep{carter2008}.
In Virgo alone, 12\,763 GCs were identified in pointed observations of 100 galaxies
\citep{jordan2009}. This catalog established the relationship between the shape of color
distributions and host mass for early type galaxies in Virgo \citep{peng2006} and 
served to characterize GC sizes as a function of environment \citep{jordan2005}.
Recently, \citet{forte2013} provided photometry in several additional optical passbands for about
800 GCs in an area of $\sim 30$ square arcminutes just South of Virgo's central galaxy, M87, 
and \citet{bellini2015} published deep HST photometry for GCs in $\sim 7$ square arcminutes
around the very core of M87.

In this paper, we exploit a recent ground-based wide field survey of the Virgo galaxy cluster,
the Next Generation Virgo cluster Survey, NGVS  \citep{ferrarese2012} and its
near-infrared follow-up NGVS-IR  \citep{munoz2014}.
The NGVS is currently the deepest photometric
survey of the Virgo cluster and it provides magnitudes
from the near UV to the near-IR in the $u^*$, $g$, $r$, $i$, $z$ and $K_s$ bands 
of the Canada-France-Hawaii-Telescope (CFHT) wide field imaging system.
Recent results based on the NGVS include a description of the population
of faint galaxies in Virgo \citep[][]{liu2015,zhang2015,sanchez2016},
the study of individual galaxies affected by the dense environment in Virgo \citep{paudel2013,liu2015b}, but also a tomography of the Milky Way halo towards Virgo \citep{lokhorst2016}
and a description of the high redshift background \citep[][ and Licitra et al. in prep.]{raichoor2014}.
Thanks to the exhaustive sky coverage of this survey (which
contrasts with the pointed HST/ACS observations), we have access to a
complete picture of the area including the
globular cluster population. Optical photometry allows a first selection
of thousands of GC candidates in this survey (\citealt{durrell2014},
\citealt{oldham2016}). As shown by \citet{munoz2014},
the combination of optical and near-IR photometry drastically
improves the rejection of contaminants, and this advantage is used here
extensively.
\smallskip

The two main purposes of this paper are (i) to present a catalog of robust,
well-calibrated colors for luminous globular clusters in the Virgo core
region, from the near-UV to the near-IR, and (ii) to compare their locus in
color-color space with the predictions of 11 commonly-used models of
synthetic stellar populations. We also use the data to provide fiducial
spectral energy distributions for Virgo core GCs, at any location along
the main color-sequence the sample defines. The comparison with models
remains qualitative in this article, as the color-color diagrams by themselves
contain much information that had not been highlighted in the past.
The new GC data form a tight locus in color-color space, with respect to
which discrepancies between models are highly significant.
No model is found to represent the observed trends adequately across all colors.
Consequences, in particular for photometric metallicity estimates,
will be quantified in a following paper.

This article is organized as follows.
Section \ref{sec_data} is devoted to an overall
summary of the NGVS data reduction, with the intent of
allowing the reader to assess the accuracy of the
photometric calibration. Two photometric calibration methods are
described in detail, one based on existing point source catalogs, the other on synthetic
photometry and several collections of theoretical or semi-empirical stellar spectra.
The first is given preference in this paper.
In Section \ref{GCsample.sec} we describe the selection of our
robust GC sample for the Virgo core region.  The average properties of the sample
and fiducial GC energy distributions are provided there,
together with a budget of possible systematic errors in the GC
photometry.  Section \ref{modelsec} presents the population
synthesis models we have considered. We compare these with the empirical data
in Section \ref{sec_results}.  Finally, in Section \ref{discussion.sec}, we discuss
causes of some of the discrepancies between models and some implication of our results.
We conclude the paper in Section \ref{conclusion.sec}.

An appendix provides additional figures and details in three areas:
(1) position-dependent terms in the photometric calibration of the NGVS data
for this paper; (2) color-color trends obtained for the observed GCs
when using the second of the photometric calibration methods described
in Section \ref{sec_data}; and (3) additional projections of the
GC color-color distribution, that are not discussed in the text to avoid redundancy,
but that early readers of this article suggested for the convenience of future
comparisons with other data sets.

\section[]{The data}
\label{sec_data}

\subsection{Optical and near-infrared images}

The Next Generation Virgo Cluster Survey (NGVS, \citealt{ferrarese2012}) is a deep imaging survey of 104~deg$^2$
of the sky towards the Virgo galaxy cluster (located at 16.5 Mpc distance, \citealt{mei2007}), 
carried out with the MegaCam wide field imager on CFHT \citep{boulade2003}. 
In this article, we focus on the core region of the Virgo cluster, 
an area of 3.62~$\rm{deg}^2$ roughly centered on M87 for which 
$K_s$-band data have been obtained with the CFHT/WIRCam instrument as part of the 
NGVS-IR project \citep{munoz2014}.

The processing of the MegaCam images is described in \cite{ferrarese2012}.
Four MegaCam pointings cover the core region of Virgo, and NGVS images for these 
are available in the $u^*, g, r, i$ and $z$ bands\footnote{The filter designation 
follows \cite{ferrarese2012}. The $i$ filter used is 
the one installed on the instrument in October 2007 (sometimes referred to as $i2$). 
As of 2015, the MegaCam filters have been replaced. In the new nomenclature, the filters
used in NGVS would be designated as $uS, gS, rS, iS, zS$, the $S$ referring
to the manufacturer, SAGEM.}.
Several methods of background subtraction and image combination 
were used by \cite{ferrarese2012} to produce image stacks for the individual pointings 
of the survey. Among these, we chose to work with the stacks built using the {\em MegaPipe} 
global background subtraction 
and combined with the artificial skepticism algorithm \citep{stetson1989}. 
These provide highest accuracy photometry for sources of small spatial extent, 
and therefore they also served as a basis for the analysis of ultra-compact dwarf galaxies of
\citet{liu2015}. The limiting magnitudes for point-sources are of
26.3 in the $u^*$ band, 26.8 in $g$, 26.7 in $r$, 26 in $i$, and 24.8 in $z$ 
(5$\sigma$; \citealt{ferrarese2012}). Over the core region, 
the average seeing in the stacked images is better than 0.6\arcsec\ in $i$, 
around 0.7\arcsec\ in $g$ and $r$ and around 0.8\arcsec\ in $u^*$ and $z$. 
All final images have the same astrometric reference frame, tied to the positions of stars 
in the Sloan Digital Sky Survey, and the same grid of pixels, 
with a scale of 0.186$\arcsec$/pixel. 

The processing of the NGVS-IR $K_s$ images is described
by \cite{munoz2014}. Nine WIRCam fields are required to
cover the area of each one of the four MegaCam pointing of the core region. 
Of the 36 WIRCam pointings hence requested, only 34 were actually observed, leaving out an
area of $40\arcsec \times 20\arcsec$ at the extreme South-West of
the core area (see \citealt{munoz2014} for an image of the footprint). 
Any raw images with a seeing worse than 0.7\arcsec\ were rejected before stacking, 
which typically resulted in 80 individual dithered images being combined for each
WIRCam field. This made it possible to produce stacked images with the same pixel scale 
as the MegaCam stacks, although the original WIRCam pixel scale is of 0.3$\arcsec$/pixel.
The stacking of sky-subtracted images was performed with
the Swarp software \citep{bertin2002}, using Lanczos-2
interpolation. Over the area of the Virgo core region, the
mean $K_s$ seeing is similar to that of the $i$-band MegaCam
images.

The diffuse light of the giant elliptical galaxy M87 extends 
over a significant fraction of the core region of Virgo,
and makes the automatic detection of star clusters difficult
in the central parts. Therefore, this light was modeled and
subtracted from the stacks of the M87 area in all passbands
before the object detection and the photometric measurements were performed. 
A simple galaxy model based on elliptical isophotes was found sufficient for this purpose.

\subsection{Overview of the photometric calibration procedures}

The photometric analysis of GC stellar populations relies on 
comparisons between observed and synthetic colors. 
Hence we endeavour to characterize our empirical and synthetic photometry in detail.
As in previous publications of the NGVS collaboration, we work with AB magnitudes
in the native passbands of the NGVS and NGVS-IR observations.

Before proceeding, it is worth recalling that empirical and synthetic photometry 
have different sources of systematic errors. While the former depends on the 
nightly choice of photometric standard stars and the previous absolute calibration of 
these in the passbands of interest, synthetic  photometry is a 
direct implementation of the AB magnitude definition. Synthetic photometry thus
provides the exact AB photometry associated with any given spectral energy distribution (SED),
as long as the adopted transmission curves are adequate. The latter condition, of course,
is never perfectly met. And when used for calibration purposes, synthetic photometry is 
limited by uncertainties on both the transmission curves and the assumed SEDs. 
Empirical AB magnitude systems are also imperfect. They depend on the adopted SEDs
of rare primary standards, on networks of secondary standards, on corrections for 
variable extinction, on aperture corrections and on transformation
equations to or from the systems in which the standards were initially measured. 
Even data sets as widely used as the Sloan Digital Sky Survey, 
to which the NGVS/MegaCam photometry is tied, are described as approximate AB systems 
in the literature (\citealt{schlafly2011}, \citealt{betoule2013}, 
SDSS calibration pages\footnote{{\tt https://www.sdss3.org/dr10/algorithms/fluxcal.php}}).
\smallskip


A brief outline of the steps followed to measure and calibrate the magnitudes 
of globular clusters is given here, to guide the reading of the details 
provided in the remainder of Section \ref{sec_data}.

(a) The first calibration step is part of the construction of image stacks.
Before combining individual MegaCam images \citep{ferrarese2012}, 
a comparison of the instrumental magnitudes of point 
sources with SDSS magnitudes is used to determine individual zero points for
each of the 36 detector chips of the camera. This corrects first order 
changes in transmission related to position within the field of view
(see \citealt{betoule2013} for a different approach),
as well as differences in the atmospheric extinction. 
For this procedure, point sources are selected via a cross-match 
with the SDSS point source catalog.

The WIRCam stacks of \cite{munoz2014} are calibrated using 2MASS point
sources as a reference. Again, differences in zero points between the
detector chips of the camera are accounted for.

(b) We then proceed to determine local aperture corrections
for point sources (Section~\ref{sec_ptsourcephot}). The sample
of point sources used for this step is cleaned of contaminants
using the near-UV to near-IR photometry and a measure of compactness.

(c) Using the stars selected in step (b), we compare 
the aperture corrected magnitudes respectively 
to PSF-magnitudes in SDSS and to aperture corrected magnitudes 
in the UKIRT Infrared Deep Sky Survey \citep{lawrence2007, casali2007}, 
to improve the calibration relative to these external surveys
(Section~\ref{photo_ext_cat}). Note that we transform the 
external photometry to the MegaCam and WIRCam systems before comparison, 
and not the reverse. The zero points of each image stack are reajusted 
at this step, based on all the stars of one field of view.
This provides our first set of final data. 
Systematic uncertainties on the AB magnitudes obtained this 
way come from departures of the SDSS and UKIDSS photometry
from a true AB system, as well as from
the transformations between these systems and the NGVS passbands.

(d) With the purpose of offering a color calibration independent 
of the SDSS and UKIDSS surveys, a second calibration method is implemented:
the observed stellar locus in color-color space is forced 
to match the stellar locus obtained from synthetic AB photometry of theoretical stellar SEDs.
This provides our second set of final colors. 
Systematic uncertainties here do not depend on SDSS or UKIDSS
but rather on the choice of adequate synthetic stellar spectra 
and filter transmission curves. 

Globular cluster photometry from steps (c) and (d) are made available
with this article (see Section \ref{Prop_GC}).  
A budget of systematic errors is given in Section \ref{error_budget}.
We use the first of the two calibrations by default in the main body of this paper, but
provide further comments on the second in the Appendix.

\subsection{Point source photometry}
\label{sec_ptsourcephot}

To measure aperture magnitudes, we used the SExtractor 
software \citep{bertin1996}. The local background subtraction of 
SExtractor was switched on for these measurements, using a sky annulus 
of $\sim 5\arcsec$ width around the sources. The sky is locally very flat, 
in particular after subtraction of M87, and the sky subtraction contributes 
negligible random errors except in areas contaminated by the halos of bright/saturated
stars, or near other galaxies (in total a few percent of the Virgo core area).
Work on the one-by-one subtraction of more galaxies is ongoing but not available 
as yet.

Aperture corrections for point sources were computed separately 
for four image stacks, each corresponding to the area of one MegaCam 
pointing. For this purpose, the star sample was cleaned 
on the basis of magnitude (bright but not saturated), compactness
in the NGVS images, and the relative location in a preliminary 
$uiK$ diagram \citep{munoz2014}. The latter criterion is very effective 
at rejecting contaminants, as illustrated in Section\,\ref{GCsample.sec} in the
context of the selection of GCs.  Point source fluxes were measured 
in a series of apertures, and aperture corrections were 
computed using the curves of growth \citep[as in][]{liu2015}. 
The average aperture corrections vary significantly between the four MegaCam 
pointings of the Virgo core region due to seeing differences. 
Typical aperture correction maps for one MegaCam pointing are shown 
in Figures \ref{iapcormap.fig} and \ref{rapcormap.fig} of the Appendix. 
The discrete maps were smoothed with a gaussian kernel ($\sigma = 1.6\arcsec$) to
provide corrections at any location.

In the WIRCam $K_s$ image stacks, the spatial variations of the aperture corrections 
mainly echo seeing differences between the individual WIRCam 
pointings that compose one MegaCam field of view (Figure  \ref{Kapcormap.fig} of the Appendix). 
The number of 2MASS stars per WIRCAM field with reasonable 
signal-to-noise is too small to measure aperture correction 
variations within a pointing reliably, 
and UKIDSS (which would provide a denser star grid) is not available systematically 
over the whole area. We note that the $K_s$ point-source size (FWHM) is more 
dispersed over the area of {\em one} MegaCam pointing than the $i$-band size.
But globally, over the whole area of the Virgo core region, 
the $K_s$ aperture corrections are more uniform than the optical ones 
because only images with a seeing better than 0.7\arcsec were used in WIRCam stacks.

In the remainder of the paper, we use apertures of 7 or 8 pixels in diameter 
(1.3\arcsec or 1.48\arcsec, i.e. about twice the seeing) as the basis for any
aperture-corrected photometry of stars. Globular cluster measurements
are discussed in Section \ref{subsecapphotgc}.

Our photometric error estimates are based on SExtractor errors,
with a correction for the correlation between neighbouring
pixels that results from the geometrical transformations applied to the original 
images before stacking.
For the MegaCam images, the stacks roughly preserve the initial pixel size 
and are computed with Lanczos-3 interpolation. In that
case, a correction factor of roughly 1.5 should be applied to
the error bars for point and point-like sources 
\citep{ilbert2006,coupon2009,raichoor2012}\footnote{ Note that \cite{bielby2012}
recommend a factor of 3
for the $r$ and $i$ bands in the CFHT Legacy Survey.}.

For the WIRCam images, the correction factor to be applied 
to SExtractor errors is larger because the final pixels are significantly smaller than the
original ones. The artificial star experiments we performed
to estimate completeness \citep{munoz2014} show that
SExtractor errors for point sources should be multiplied by
a factor of 2.5. This is consistent with the findings in \citet{bielby2012} (factor 2.49) or \citet{mccracken2010} (factor
2).

In the following, the term ``SExtractor errors'' refers to the error 
values before application of the recommended factors. But ``errors" refer
to the corrected values, and these are applied in any analysis.

\subsection{Photometric calibration against external catalogs}
\label{photo_ext_cat}
The first version of the photometry we provide is calibrated on external 
survey catalogs. 
For MegaCam, the Sloan Digital Sky Survey Data Release DR10 is used as a 
reference \citep{ahn2014}.
The SDSS PSF-magnitudes of stars common to both surveys (mostly main sequence
stars of spectral types later than F)  are converted to the MegaCam
system using the transformation in \citet{ferrarese2012}.
The NGVS aperture corrected point source magnitudes are then 
compared with these transformed SDSS magnitudes, to derive one
zero point offset per field of view. This zero point correction then applies
to all sources in that field of view, be they stars or other objects. 

\begin{figure*}
\begin{center}
\includegraphics[width=18cm]{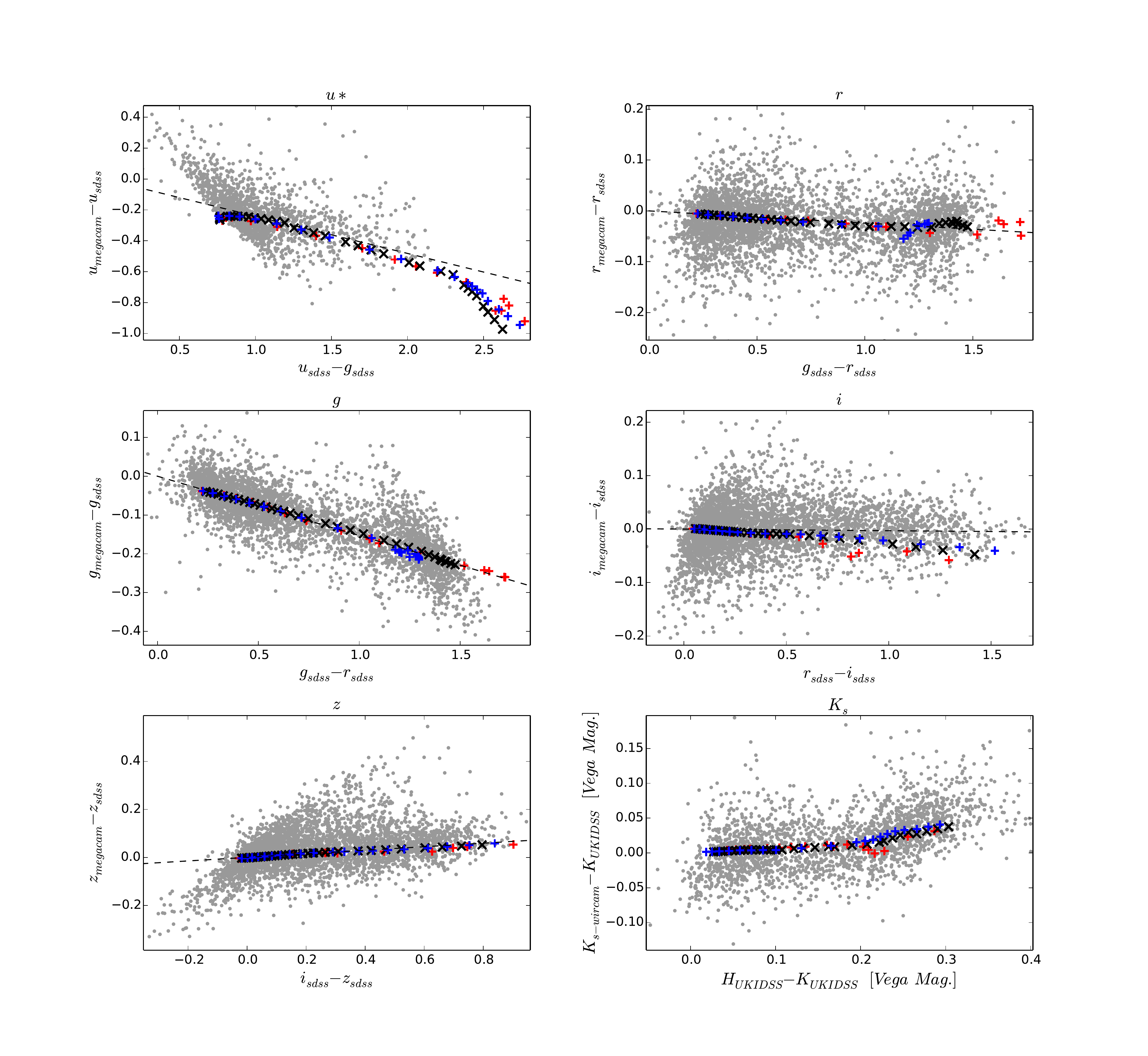}
\caption{\label{transfo_plot} Transformation between photometric systems. 
Star in common between NGVS and SDSS, or
NGVS and UKIDSS (for $K_s$) are shown as grey dots. For NGVS and UKIDSS, aperture
corrected magnitudes are used; for SDSS, psf magnitudes are adopted. 
The NGVS magnitudes are the final values obtained as described in
Section \ref{photo_ext_cat}.  In
each panel, a different subset of stars is plotted, restricted so the
SDSS or UKIDSS magnitudes used in the panel have errors as follows: $\sigma$($u^*$)\,$<$\,0.20, $\sigma$($g$)\,$<$\,0.10, $\sigma$($r$)\,$<$\,0.10, $\sigma$($i$)\,$<$\,0.10,
$\sigma$($z$)\,$<$\,0.15, $\sigma$($H_{\mathrm{UKIDSS}}$-$K_{\mathrm{UKIDSS}}$)\,$<$\,0.04 mag. Synthetic photometry based on model
dwarf stars is shown with crosses, based on energy distributions from
MARCS models (blue), PHOENIX models (black) and the BaSeL library
(red).  Along these sequences, stellar parameters are taken from the
Besan\c{c}on model of the Milky Way. The dashed lines in the five first
panels follow the equations quoted in \citet{ferrarese2012}.}
\end{center}
\end{figure*}

Because the transformations are an important element of the calibration 
of the magnitude zero points in this section,
we display them in the first five panels of Figure \ref{transfo_plot} 
together with stars common to NGVS and SDSS. 
The amplitude of the dispersion is primarily due to the
random photometric erros in SDSS.
Only one zero point per image is derived in the calibration 
against SDSS, hence the relevant errors  are the average differences between 
the various displayed loci (over the range of colors most populated with stars). 
The NGVS magnitudes used in the figure are taken after
calibration, hence by construction the stars are located, on average, 
on the calibration line, with (sample dependent) mean offsets smaller than 0.01~mag.

The transformations are also compared with those obtained from synthetic photometry
in Figure \ref{transfo_plot}.  
We used three libraries of synthetic stellar 
spectra: the MARCS library of \citet{gustafsson2008}, the  BaSeL 3.1 library (\citealt{lejeune1997}, \citealt{lejeune1998} and \citealt{westera2002}) and the PHOENIX library of \citet{husser2013}.
The assumed stellar temperatures, surface gravities and metallicities along
the NGVS stellar locus are obtained from the Besan\c{c}on model of the
Milky Way (\citealt{robin2003}, \citealt{robin2004}), to which adequate magnitude 
cuts were applied.  For reasons that will become apparent 
in Section \ref{sec_SLR}, our preferred library is the PHOENIX library. 

The transmission curves for the synthetic photometry were taken from 
\citet{betoule2013} for MegaCam\footnote{Betoule et al. provide transmissions
for various annuli around the center of the MegaCam field of view. 
We use the fourth radius (70\,mm from the center of the filter), which
within a few millimagnitudes is equivalent to using an area-weighted 
average of the local transmissions.}.
It includes all telescope and instrument components as well as 
typical telluric absorption features\footnote{The transmission curves are available 
with the online version of this paper.}.

Although the transformation equations are not actually fits to the
synthetic data, the similarity is quite impressive. Average residuals between
the synthetic data and the reference lines (over the range of colors most populated 
with NGVS+SDSS stars and hence most relevant to the calibration) are smaller than 0.01\,mag,
i.e. smaller than the dispersion expected from the photometric 
errors of SDSS. We note that there are essentially no stars of type F and hotter
in the calibration sample. Had there been many, a linear transformation equation would have
been inadequate for $u^*$.  Indeed, $u^*_{\mathrm{MegaCam}}-u_{\mathrm{SDSS}}$ rapidly deviates 
from a straight line when $(u-g)<0.7$, as a consequence of the strong Balmer jump in the 
spectra of hotter stars.
\medskip

In the near-infrared, we tied the WIRCam $K_s$ photometry to UKIDSS DR8 
(\citealt{hewett2006}, \citealt{dye2006}, 
\citealt{hodgkin2009})\footnote{The aperture-corrected magnitudes provided 
in UKIDSS catalogs as {\tt kAperMag3} are used for stars}.  
Although shallower by about 3 magnitudes than NGVS-IR,
the UKIDSS point source catalog is deeper and more precise than 2MASS. 

Both UKIDSS and 2MASS $K$ band transmissions have larger effective 
wavelengths than the WIRCam $K_s$ filter (for which an all inclusive
transmission curve is given in \citealt{munoz2014}). Over the range of colors
of stars in common with NGVS-IR, i.e. $0$\ $<$\ $(H-K)_{\mathrm{UKIDSS}}$\ $<$\ $0.35$ mag, 
the quantity $\Delta_K = K_{\mathrm{WIRCam}}-K_{\mathrm{UKIDSS}}$ varies
with a global dependence on color given by $0.27 \times (H-K)_{\mathrm{UKIDSS}}$ \citep{munoz2014}. 
Note that $(H-K)_{\mathrm{UKIDSS}}$ in this expression is the native UKIDSS
value, in Vega magnitudes, while we use AB magnitudes everywhere else
in this paper ($K_{\mathrm{WIRCam}}$ [AB]\ =\ $K_{\mathrm{WIRCam}}$ [Vega] + 1.827, 
\citealt{munoz2014}). The actual relation between $\Delta_K$ and color 
is not linear but shows curvature over the whole color range, and starts
off essentially flat for $(H-K)_{\mathrm{UKIDSS}}\ <$\ 0.2 mag 
(last panel of Fig.\,\ref{transfo_plot}).  We have used the synthetic
values of $\Delta_K$ in this restricted range of colors for the re-calibration 
of the NGVS-IR zero point, because all the collections of stellar spectra
agree there, while cool M dwarf models become progressively more 
uncertain at lower temperatures.

\medskip

{\em The NGVS photometry obtained here is used as a default in the remainder of 
this paper.} 
A budget of systematic errors is given in the context of 
globular cluster photometry in Section \ref{error_budget}
(subsections \ref{error_budget_sdss} to \ref{error_budget_ext},
and Tab.~\ref{tab_bes}). 
An alternative calibration based on the direct comparison of empirical 
and synthetic stellar loci in color-color planes is considered in 
Section \ref{sec_SLR}, but then only used as a second choice in the Appendix.

\subsection{Extinction correction}
\label{ext_part}

The foreground extinction towards the Virgo core region is low. 
\citet{schlegel1998} report 0.06\ $<$\ A(V)\ $<$\ 0.16,
while \citet{schlafly2011} produce values that are typically 15\,\% lower.
Over 90\,\% of the field, including the M87 region,
 A(V)\ $<$\ 0.10.

Extinction coefficients for the MegaCam and WIRCam
filters were provided in an appendix of \citet{munoz2014},
using the extinction law of \citet{cardelli1989} with R(V)\ =\ 3.1 and
stellar spectra of a variety of spectral types.  We have used the values
they derived for a solar type star. Changes between 
{\em extreme} stellar types lead to changes in $A(\lambda)/A(V)$ 
smaller than 0.02 in $r$, $i$, $z$ and $K_s$, than 0.03 in $u^*$ 
and than 0.07 $g$. Towards Virgo, errors on $A(\lambda)$ 
due to the  color-dependence of extinction coefficients
are therefore smaller than 0.01\,mag.

Based on the above, typical reddening corrections amount 0.06\,mag in $(u-i)$,
and 0.04\,mag in $(g-i)$ and $(i-K_s)$.  Rescaling A(V) from the value of \citet{schlegel1998} 
to that of \citet{schlafly2011} reduces $(u-i)$ towards M87 by 0.011\,mag and
$(i-K_s)$ by 0.007\,mag. 
In the following, when correcting for extinction on individual 
lines of sight, we have used the values of E(B-V) of \citet{schlegel1998} 
for consistency with previous publications of the NGVS collaboration.

\subsection{Alternative calibration via Stellar Locus Regression}
\label{sec_SLR}

As mentioned earlier, we have explored a second calibration method, that
relies on synthetic colors of stars instead of the stellar fluxes of external surveys.
Although this new method looks promising, the choice of an external spectral library 
as a reference remains a limiting factor. Hence, we restrict this section to
a description of the method and its key ingredients, 
and to an assessment of the differences with the the previous photometry. 
We then use the calibration in Section \ref{photo_ext_cat} for the analysis of GCs. 
Further details relevant to the alternative calibration method, 
and a repetition of some of the GC analysis with that calibration, 
are made available in the Appendix.

\subsubsection{The SLR Method}
\label{SLR2_m}

Stellar Locus Regression (SLR) was introduced under this name by \citet{high2009}, who
used it to calibrate colors of new photometric surveys against 
colors in pre-existing, supposedly well-calibrated ones. In brief, the method forces 
the loci of point sources in color-color space to agree in the two surveys,
assuming this locus is (at least roughly) universal. It does not provide an 
absolute flux calibration, but explicitely focuses on colors. Here, we have adapted 
the method to attach the NGVS/NGVS-IR stellar locus to the locus predicted 
by theoretical stellar spectra.

In principle, it makes sense to require a good match between empirical
and synthetic stellar colors whenever the final purpose is to compare empirical
colors of stellar {\em populations} with synthetic ones.
However, in practice this test is not as relevant as it may seem: 
the stars we see in surveys such as the NGVS are essentially all on the lower main sequence, 
while the red and near-IR light of globular clusters or galaxies comes mostly from red giants. 
Here, we explore this second calibration simply as
an alternative to the calibration against SDSS and UKIDSS. As a side product, this 
allows us to assess model spectra of cool dwarf stars.

In the SLR of \citet{high2009}, the color transformation equation is written as
\begin{equation}
 \label{SLR_eq}
 c\ =\ \kappa\,+\,(1\,+\,B)\,c_0
\end{equation}
where $c$ is a vector of new (possibly uncalibrated) colors, 
$c_0$ is the vector of assumed true colors
(the reference color locus), $\kappa$ accounts for zero point shifts due for instance
to atmospheric extinction and differences between the effective wavelengths of the 
used and reference filters, and $(1+B)$ is the color transformation matrix. The method assumes that
the color transformations between the reference and adopted passbands are known, 
i.e. $(1+B)$ is known (from standard star observations). The problem is then essentially reduced to 
searching for the optimal offsets $\kappa$.

In our case, we use synthetic photometry as a reference and we assume the 
NGVS and NGVS-IR transmission curves are well known, 
so Eq.\,\ref{SLR_eq} reduces to $c\ =\ \kappa\,+\,c_0$.

The stellar locus regression has been implemented as in \citet{high2009}: 
we minimize the weighted sum of the color-distances between the dereddened 
empirical stellar colors, after shifting with $\kappa$, 
and the respectively closest point on the synthetic locus. The photometric
errors are used for the inverse-variance weighting.

\subsubsection{Choice of a reference library and of fitted colors}

The main difficulty in the application of the SLR is the choice of the reference stellar locus.
The results also depend on the choice of colors used in the fit.

\citet{high2009} advise against using SLR for the $u$ filter 
due to the large dependence of $u$ band fluxes 
on stellar metallicity and galactic dust extinction. 
Thus, we have decided to determine SLR shifts only for 
$(g-r)$, $(r-i)$, $(i-z)$, and $(i-K_s)$ at first,
which sample the energy distribution from $g$ to $K_s$.
The effects of including the $u$ band in the SLR calibration
procedure are briefly assessed in the Appendix (Fig.\,\ref{u_prob}
and corresponding text).
We confirm that the near-ultraviolet raises stronger issues than other bands.

The first three of the colors listed above 
are the ones also used by \citet{high2009}. 
As these authors highlight, the colors must be chosen so the   
stellar locus displays a kink in at least one color-color plane, otherwise
the fit is not well constrained (the offsets {\em along} the stellar locus would be arbitrary). 
Our choice satisfies this requirement as subsequent figures will show.

As a source of stellar spectra for synthetic photometry, we have used the 
collections already mentioned in Section \ref{photo_ext_cat}:
the PHOENIX theoretical spectral library of \citet{husser2013}, the MARCS model collection
of \citet{gustafsson2008}, and the semi-empirical library BaSeL 3.1.
We also considered the empirical library of \citet{pickles1998}, which
has robust colors for near-solar metallicity, but we ended up not using it because
its sampling of metallicity is too scarce.
While all these libraries agree rather well for the colors of main sequence stars
of types F to K, their colors fan out in very different ways at cool temperatures,
where the molecular bands of M dwarfs become increasingly important.

The typical stellar properties of the NGVS stars vary along the stellar locus from
Milky Way halo-like at the blue end, to thin and thick disk-like
at the red end. The stellar parameters we used are derived from the Besan\c{c}on model 
of the Milky Way \citep{robin2003,robin2004}\footnote{Version available on line
in early 2015.} in the NGVS footprint, 
taking into account the saturation and detection limits  
of the survey in all passbands. Besan\c{c}on model
stars were sorted into bins of 500\,K width, from $\approx$3000 K to $\approx$6500 K. 
The statistical properties of log(g), [Fe/H], [$\alpha$/Fe] that we have used to choose 
spectra for each bin are listed in Table \ref{prop_lib}. We note that the BaSeL
library has only solar abundance ratios, so in that case 
changes of [$\alpha$/Fe] were not accounted for.

\begin{table}
  \begin{center}
    \begin{tabular}{cccc}
\hline
Teff (K) & log(g) & [Fe/H] & [$\alpha$/Fe] \\
\hline
3100 & 5 & 0.0 & 0.0 \\
3600 & 5 & -0.5  & 0.2   \\
4000 & 5 & -1 & 0.2 \\
4500 & 5 & -1.5 & 0.4 \\
5000 & 4.5 & -1.5  & 0.4 \\
5500 & 4.5 & -1.5 & 0.4 \\
6000 & 4.5 & -1.5 & 0.4 \\
6400 & 4.5 & -2 & 0.4 \\
\hline
    \end{tabular}
   \caption{\label{prop_lib} Rounded average statistical properties of stars along NGVS 
stellar locus, based on the Besan\c{c}on model.}
      \end{center}
\end{table}

\begin{figure*}
\begin{center}
\includegraphics[width=16cm]{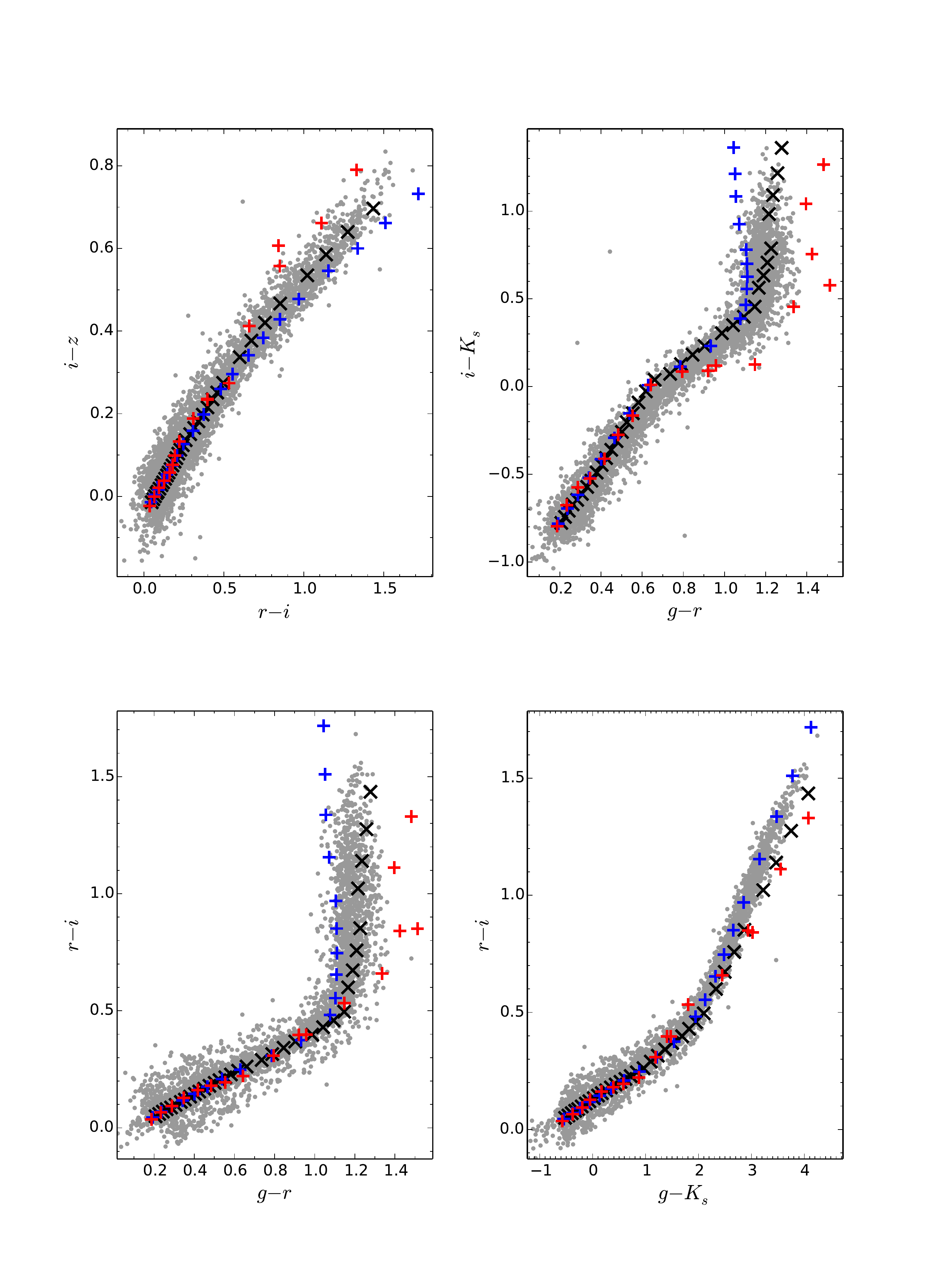} 
\caption{\label{SLR_difflib} NGVS stellar locus and predictions for the 
PHOENIX (black), MARCS (blue), and BaSeL (red) libraries. The observations are calibrated as in Section \ref{photo_ext_cat} and dereddened. The stellar templates 
are chosen in accordance with the Besan\c{c}on Milky Way model predictions (Tab.\,\ref{prop_lib}), except for the BaSeL library which has only solar [$\alpha$/Fe] ratios.}
\end{center}
\end{figure*}

In Figure \ref{SLR_difflib}, the BaSeL (red), PHOENIX (black) and MARCS (blue) libraries 
are shown superimposed to our NGVS stellar locus. At the red end, 
the discrepancies between those libraries are large.  
The PHOENIX library fits the {\em shape} of our 
empirical distributions well in all color-color diagrams, with only a small tilt of 
the M-dwarf sequence with respect to observations in the plot of 
$(r-i)$ vs. $(g-K_s)$.
 As only shifts and not change of shape are allowed in the SLR calibration, we conclude that only the
PHOENIX library is appropriate for our purpose, and we discard other libraries
in the remainder of this section.

Important features in these color-color diagrams are
the kinks seen in all but the $riz$ diagrams. The locus of these kinks controls shifts
along the color-color sequences of stars. As these shifts are also applied to globular clusters, 
they directly affect the metallicity estimates of the latter.

\subsubsection{SLR results}
\label{SLR2_r}

\begin{figure*}
\begin{center}
\includegraphics[width=16cm]{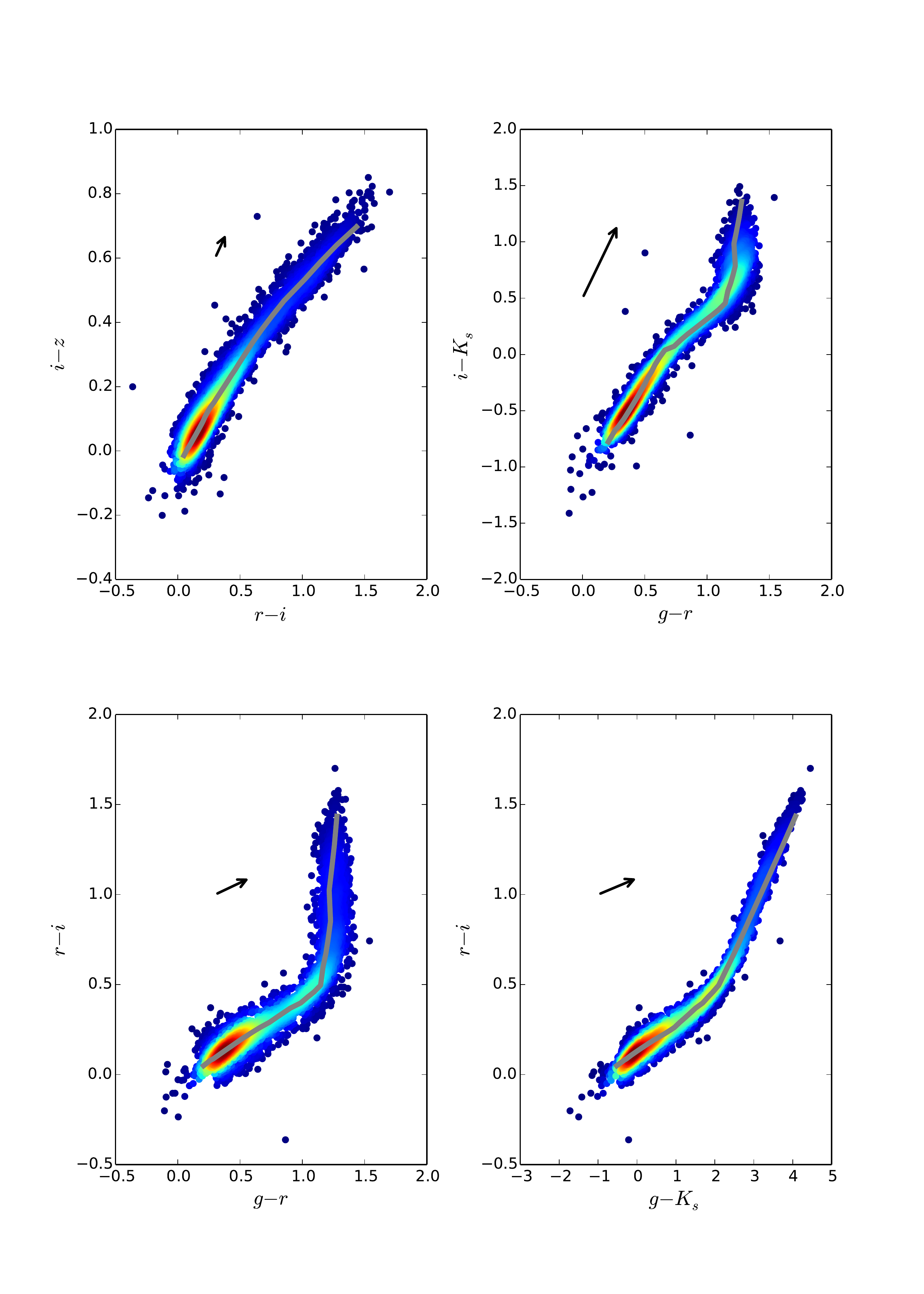}
\caption{\label{SLR_fig} NGVS stellar locus shifted by the SLR vector $\kappa$, with the stellar predictions of the PHOENIX library (gray line) superimposed. The black
arrow shows the SLR vector $\kappa$ multiplied by a factor of 5 for better visibility. 
The model stellar parameters are based on the Besan\c{c}on model of the Milky Way.
The color coding of the NGVS stellar locus maps the density of stars.}
\end{center}
\end{figure*}

To account for spatial variations of extinction over the area of the survey,
we deredden NGVS stars before estimating the best vector of corrections, $\kappa$.
We then apply these corrections to all objects in the NGVS data set.
Figure \ref{SLR_fig} shows the stellar locus obtained after the SLR calibration and the arrow illustrates the displacement applied.

The SLR offsets found with the method above are \\
$\kappa=[(g-r):\,0.058,(r-i):\,0.019,(i-z):\,0.016,(i-K_s):\,0.133]$.
The offsets in $(r-i)$ and $(i-z)$ are marginally consistent with our estimated bounds on errors in Section \ref{error_budget}.
 The shifts in $(g-r)$ and $(i-K_s)$ are larger than expected.
Figure\ \ref{SLR_fig}  shows this may be related to the slight tilt of the slope of the PHOENIX sequence in the $gri$ and $gKri$ planes.
The slope on the red side of the kink in the stellar locus differs between models and the observations. $\kappa$ partly 
compensates for the difference this generates at low temperatures.

\subsubsection{Summary of the photometric calibration}

Our default photometric calibration rests on three steps: the construction of
image stacks that account for differences in photometric zero points between detector chips, 
the computation of local aperture corrections for point sources, and the 
comparison with SDSS and UKIDSS (after transformation to the NGVS passbands).
We use the extinction map of \citet{schlegel1998} but have provided the
comparison with \citet{schlafly2011} in Section \ref{ext_part}.

We have also implemented an alternative calibration of the
colors, based on stellar spectral libraries, a model for the stellar population
of the Milky Way, and synthetic photometry.
Because some of the color shifts suggested by that calibration 
are large, we suspect biases exist even in the best models 
for the colors of lower main sequence stars on the line of sight towards Virgo. 
Our preferred calibration to date is the first one.

\section{The globular cluster sample}
\label{GCsample.sec}

\subsection{Selection}
\label{GCselect.sec}
The selection of the globular clusters is a crucial point in our study. 
Our purpose is to provide
typical globular cluster colors and SEDs as a benchmark for comparisons with model predictions, not to discuss the number distribution of
globular clusters over the range of possible colors.{\textit{Therefore, our main concern is to limit contamination by
foreground stars or background galaxies and to work with objects that have good photometry. Completeness is
not a target, except that we wish to sample the whole range of colors along the main direction of the GC
color sequence.} }

\begin{figure}
\begin{center}
\includegraphics[width=8.5cm,angle=0]{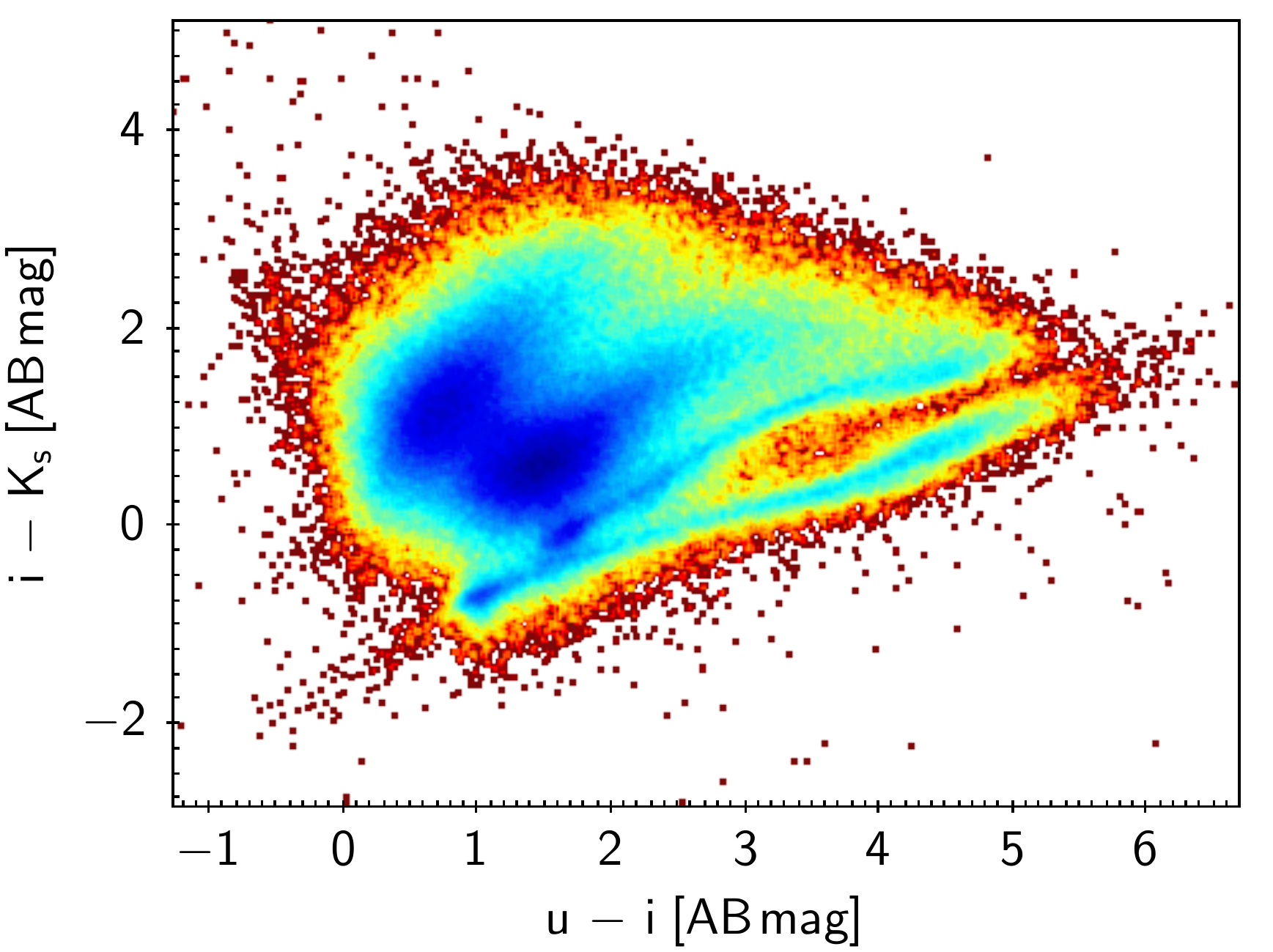}  
\caption{\label{col_col_dia} NGVS pilot-field detections in the $uiK$ diagram.
The colors shown are from measurements in 
8\,pixel diameter apertures (1.48\arcsec) to which point source aperture corrections are applied (i.e. they are
representative only of central colors for extended sources). Most sources are background galaxies.
The foreground stars are mainly along the bottom sequence, and the globular clusters are located just above it 
and below the two large regions of galaxies. At the red end of the GC sequence ($u-i$\ $>$\ 3), redshifted passively evolving galaxies are found.
 The color coding maps the density of objects.}
\end{center}
\end{figure}

Our starting point is a merged NGVS + NGVS-IR catalog of over a million sources in the Virgo core region.
Preliminary processing includes the rejection of objects that lack data in one or more filters 
(catalog magnitude\ $>$\ 60), the rejection of sources with magnitude error larger than 0.5\,mag, and the removal of duplicate or erroneous objects 
in regions of overlap between pointings.
Figure \ref{col_col_dia} shows this catalog in the $uiK$ diagram \citep{munoz2014}. 
From red to blue $(i-K_s)$ colors, the
most conspicuous sequences in this diagram correspond to background galaxies with various star forming
histories at redshifts up to $\sim\!1.5$, globular clusters (which merge into the redshift sequence 
of passive galaxies at the red end), and foreground main sequence stars. Although the $uiK$ diagram
provides a better separation between sequences than any other color-color diagram, there is a significant overlap
between populations in this deep and exhaustive catalog.

\begin{figure}
\begin{center}
\includegraphics[width=8.5cm,angle=0]{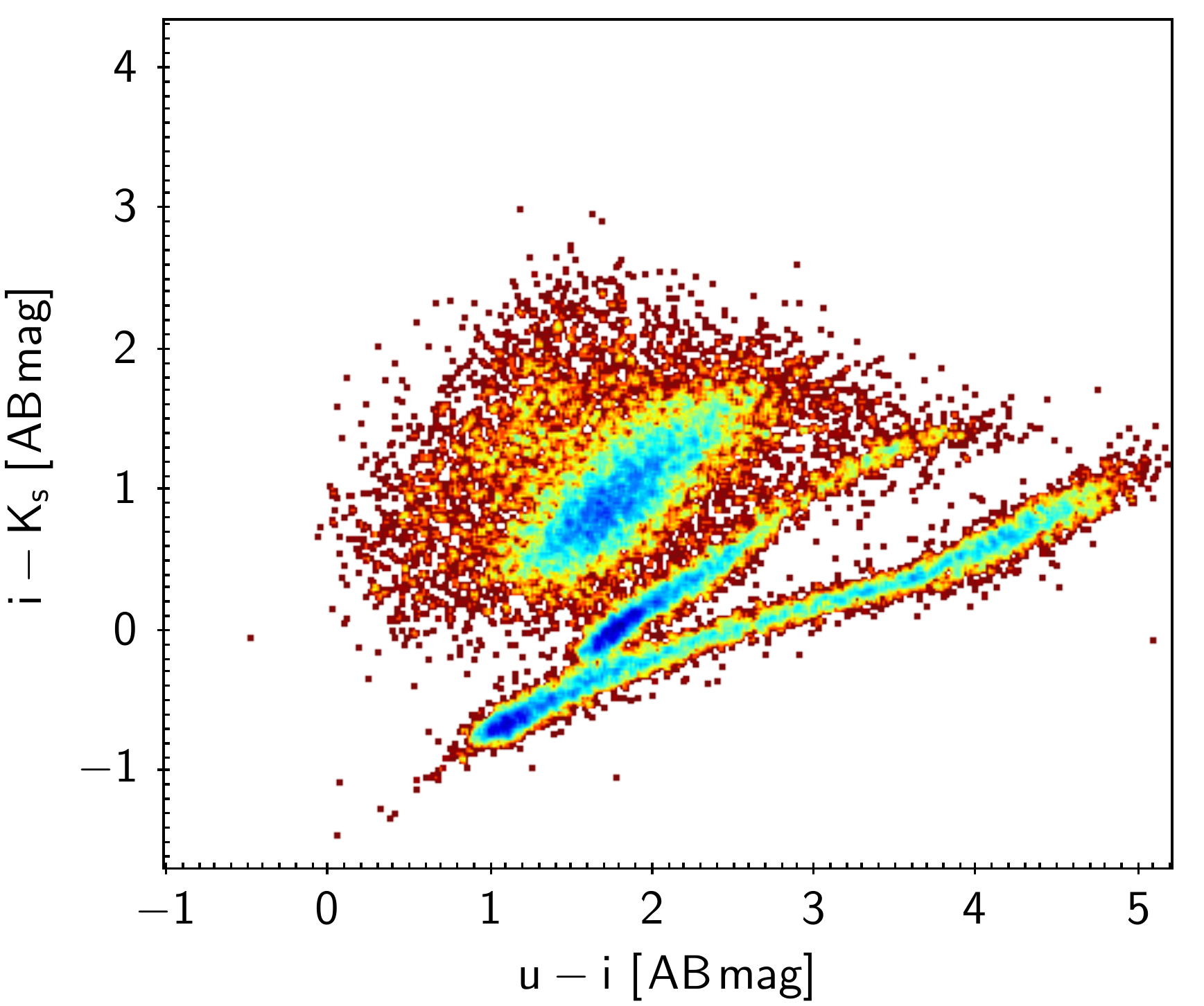}  
\includegraphics[width=8.5cm,angle=0]{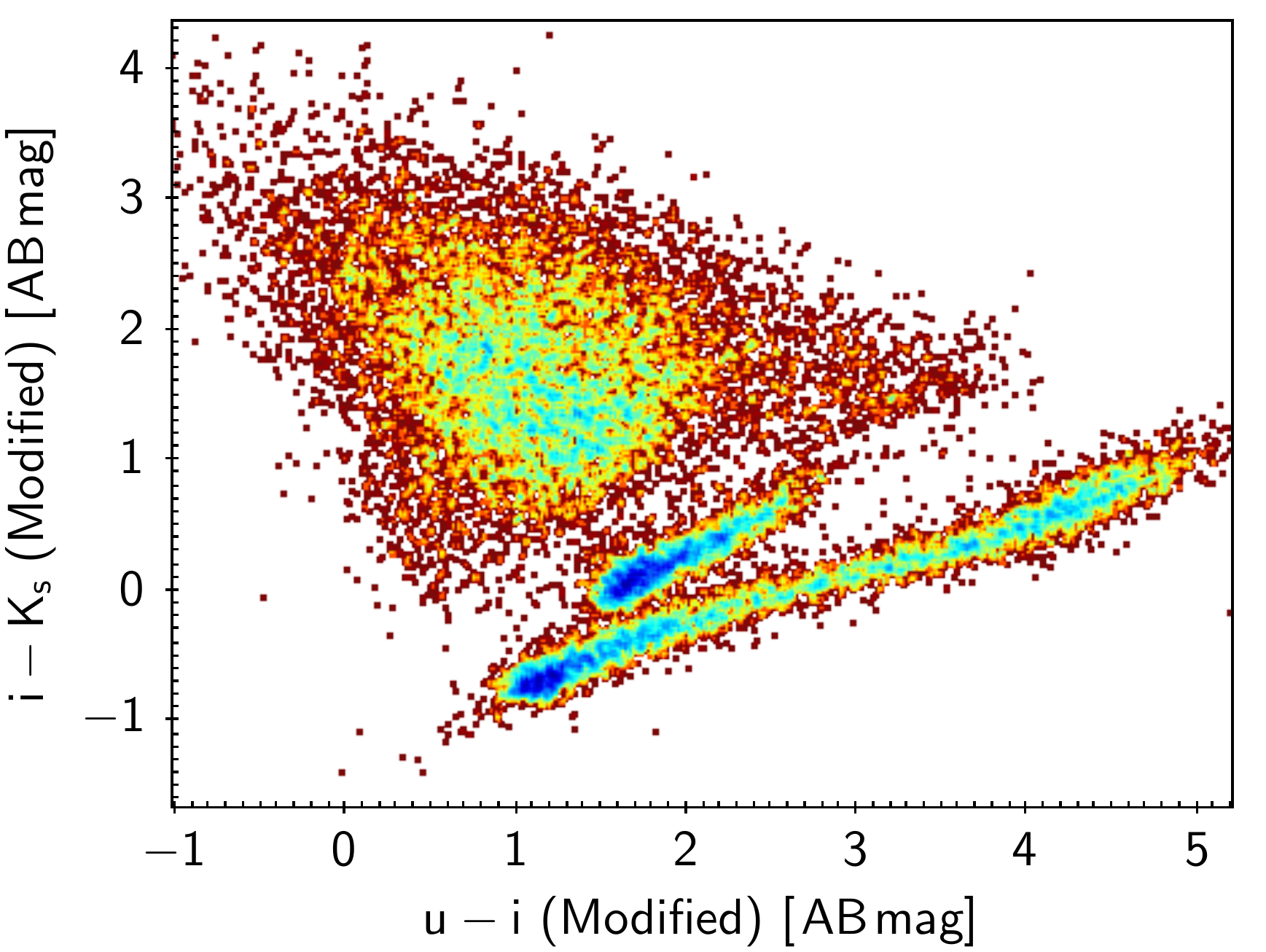} 
\caption{\label{uik_modi} Top: the $uik$ diagram after removing from the sample shown in 
Figure \ref{col_col_dia} saturated stars, large objects and sources with large errors.
Bottom: the $uiK$ diagram modified using a compactness criterion 
as explained in Section \ref{GCselect.sec}.
The color coding in both panels maps the density of objects.}
\end{center}
\end{figure}

At this point, we applied stricter selection criteria on our sample to remove saturated sources (the limit depends on the filter
and on the seeing but it is typically around 18 mag [AB] whatever the filter), large objects (half-flux radius\ $>$\ 4\,pixels) and sources with large errors (SExtractor errors\ $>$\ 0.06\,mag in 
any filter).
The sources surviving these cuts are shown in the top panel of Figure \ref{uik_modi}.

Our final cleaned selection then exploits both the $uiK$ diagram and size information. Massive globular clusters and DGTOs (Dwarf Galaxy Transition Objects) in Virgo are 
marginally resolved in images with 0.6\arcsec\ seeing (48\,pc), such as the NGVS $i$ and NGVS-IR $K_s$ images.
Absolute sizes vary accross the pilot region because the
various individual fields were observed in different seeing
conditions. A good way to quantify whether an object is more spatially extended 
than a star is to compute the difference between two aperture-corrected magnitudes in the same filter. We will write such differences APCORn-APCORm, with n and m standing for the aperture diameters in pixels. 
These differences are on average zero for stars (the local aperture correction absorbs
any spatial variations of the PSF), but are positive for extended sources. We have used both APCOR4-APCOR8 and
APCOR4-APCOR16, finding that both behave similarly.
In the standard $uiK$ diagram [$(i-K_s)$ on the y-axis,
$(u-i)$ on the x-axis], extended objects tend to lie to the upper left
of the stellar sequence. By adding (APCOR4-APCOR8)(i)
to $(i-K_s)$ and subtracting that quantity from $(u-i)$, extended
sources are efficiently moved away from the stellar sequence.
Moreover, this translation effect can be improved by adding a non-linear function of (APCOR4-APCOR8)($i$).
Our implementation depends more strongly on compactness outside the supposed range of GC colors as indicated in Eq.\,\ref{compacteq} and Eq.\,\ref{compacteq2}. This may bias slightly against possible unresolved blue clusters, but improves the 
rejection of stellar contaminants.

Between $C_{i,\mathrm{min}}$ and $C_{i,\mathrm{max}}$:

\begin{equation}
\label{compacteq}
 C_i\ =\ (\mathrm{APCOR4}-\mathrm{APCOR8})(i)
\end{equation}

Outside this range:

\begin{equation}
 \label{compacteq2}
 C_i'\ =\ \exp(1+C_i)+ K
\end{equation}

\noindent where the constant is set by the requirement of continuity at $C_{i,\mathrm{min}}$ 
and $C_{i,\mathrm{max}}$. In our case, $C_{i,\mathrm{min}}\ =\ 0.02$ and 
$C_{i,\mathrm{max}}\ =\ 0.2$. 

This `modified' $uiK$ diagram is shown in the bottom panel of 
Figure \ref{uik_modi}.

In the standard $uiK$ diagram, the GC sequence suffers from contamination by halo main sequence stars, in particular at the blue end. It is fortunate that blue clusters
tend to be the most extended \citep{jordan2005}: taking size into account
therefore effectively separates halo stars from blue globular clusters.
At the red end, many extended passive galaxies are also
efficiently moved away from the globular cluster sequence.

\begin{figure}
\begin{center}
\includegraphics[width=8.5cm,angle=0]{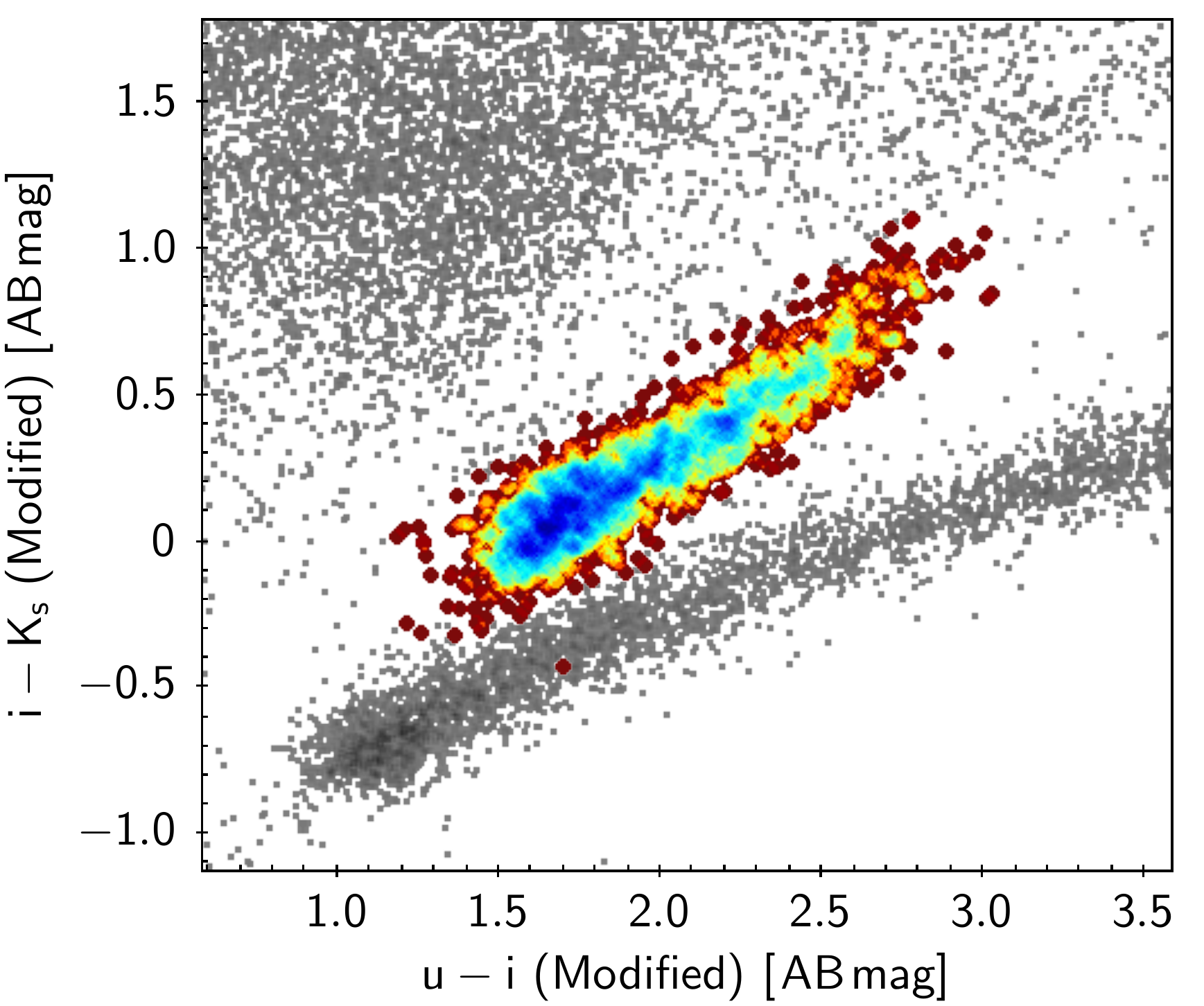}  
\caption{\label{uik_closer} Final selection of the NGVS GCs in the modified $uiK$ diagram. The color coding for the GC maps the density of objects.}
\end{center}
\end{figure}

The final selection of GCs, shown in Figure\,\ref{uik_closer}, is obtained 
by applying a conservative sigma clipping algorithm in the modified $uiK$ diagram. 
We use a polynomial fit of the current GC locus as a reference and broaden it by 0.1\,mag 
in both colors.  GCs distant from this broad locus by more than 3 times the uncertainty 
on their colors are rejected. We are left with 2321 globular clusters 
with median errors in $u,g,r,i,z,K_s$ of, respectively, 
0.02, 0.008, 0.008, 0.01, 0.02 and 0.08\,mag.
In the following subsection, we compare our selected globular clusters 
to several spectroscopic datasets from the literature. 

\subsection{Comparison with spectroscopic samples}

\begin{figure*}
\begin{center}
\includegraphics[width=8.5cm,angle=0]{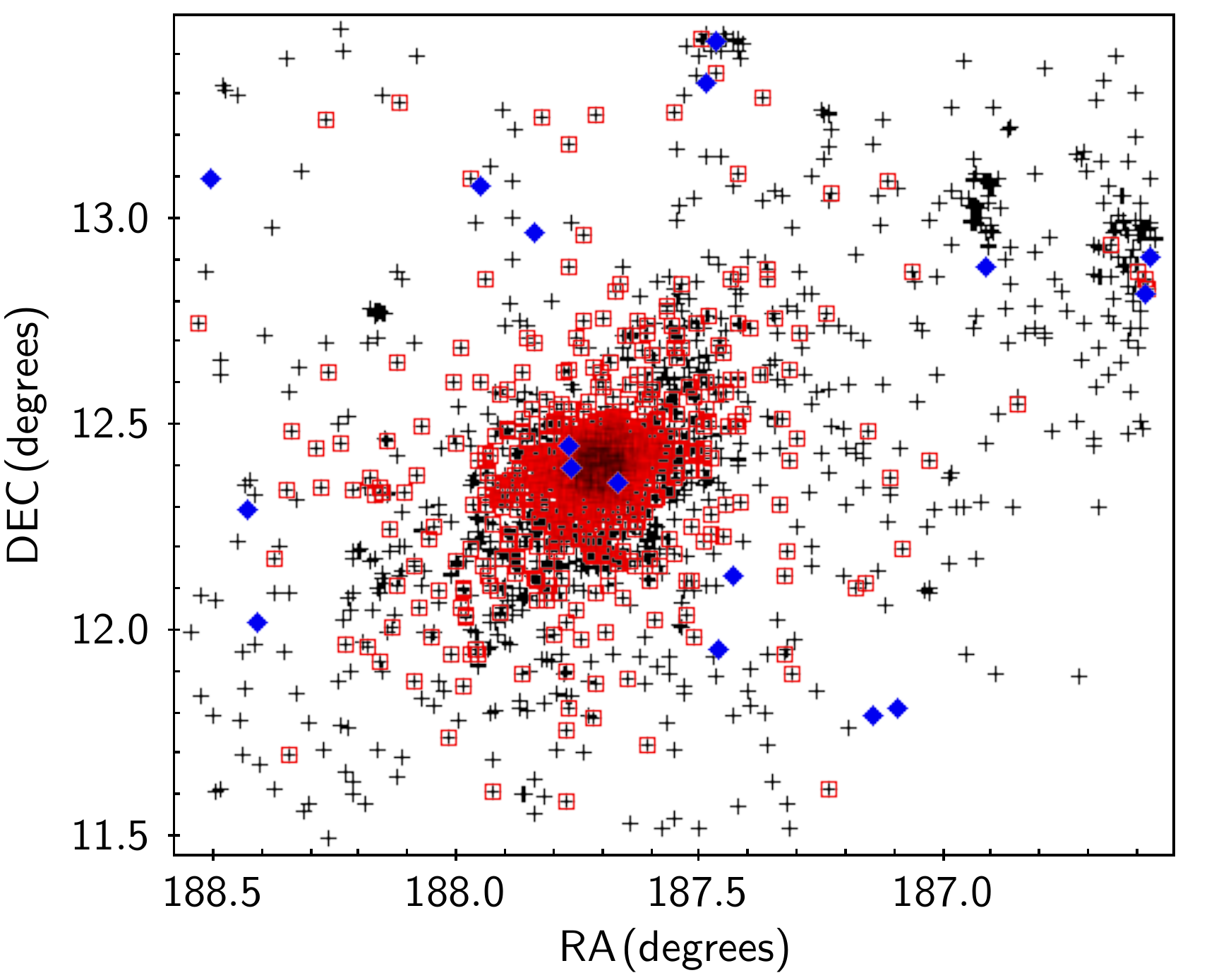} 
\includegraphics[width=8.5cm,angle=0]{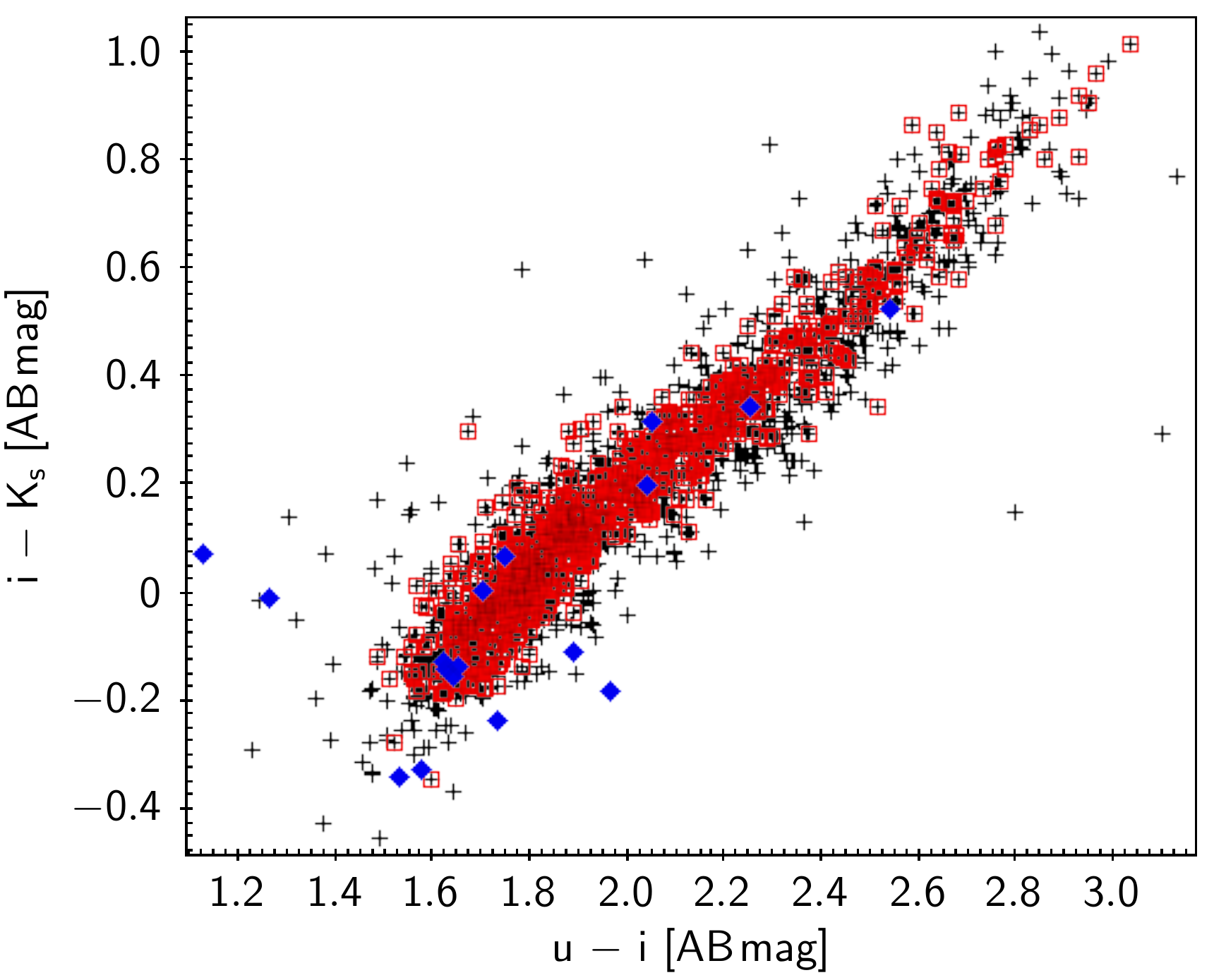} 
\caption{\label{GC_comparison}  Left: Selected NGVS GC candidates from our modified $uiK$ diagram (black), spectroscopically confirmed GCs based on matching to literature data (red), and the false positive matches (blue).
Right: the NGVS GC candidates in the $uiK$ diagram (black), spectroscopically confirmed GCs matched to the literature (red), and the false positive matches (blue).}
\end{center}
\end{figure*}

The NGVS collaboration maintains a `master spectroscopic catalogue' that includes all objects within the NGVS footprint with measured redshifts,
collected from the literature, or part of the NGVS collaboration efforts to target objects in the field. In particular,
data from the literature includes the SDSS DR10 release, the NASA Extragalactic Database for extended objects \citep{binggeli1985}, and catalogues of \citet{hanes2001}, and \citet{strader2011}.
Spectroscopic campaign were carried out by the NGVS team using Anglo-Australian Telescope 2dF observations and
Multiple Mirror Telescope Hectospec observations by E. Peng and Keck DEIMOS observations by R. Guhathakurta.

Among our selection of 2321 globular clusters, 783 have a measured redshift. All but 17 
are bona fide Virgo globular clusters according to the spectroscopic data. Among those 17, 5 are considered galaxies and 12 stars. Figure \ref{GC_comparison} shows 
our sample together with the matched spectroscopic targets in RA-DEC (left) and in color-color space (right). 
Globular cluster overdensities are visible near M87 but also around NGC 4473, NGC 4438 or M86 
(in the North-West corner of the field,  from the East to the West). We note that the spectroscopic catalog has no objects associated with NGC 4438 and NGC 4435.

Extrapolating from this test, we estimate the contamination of our full GC sample to be limited to about 50 objects
out of 2321 (i.e. about 2\,\%). Eyeball estimates based on the distribution of sources in the modified $uiK$ diagram 
(Fig.\,\ref{uik_modi}) would allow contaminations of up to 100 objects, i.e. about 5\,\%.

We note that the colors of the matched spectroscopic sample span the whole range of colors of our photometric GC catalog (Fig.\,\ref{GC_comparison}).
This provides confidence that our reddest objects are not background ellipticals and our bluest ones not foreground stars.

\subsection{Aperture photometry of globular clusters}
\label{subsecapphotgc}
As globular clusters are marginally resolved sources in the NGVS survey,
the point source aperture corrections do not
strictly apply to them. However, these aperture corrections
efficiently absorb the spatial variations of the PSF (mostly due
to seeing variations with time), and we can limit any bias in
{\em color} measurements by applying
aperture corrections to relatively large measurement apertures.

To test this assertion, we have compared aperture-corrected magnitudes
(APCOR-magnitudes hereafter) and the corresponding colors (APCOR-colors)
as a function of the compactness parameter already used earlier
(APCOR4-APCOR8 in the $i$ band).
As expected, the comparison between globular cluster APCOR-colors
measured in 2 apertures of which one is small, shows a difference
that depends strongly on compactness.  For example for apertures of
4 and 8 pixels the amplitude of this trend along the GC compactness-sequence
exceeds 0.1\,mag for $(g-i)$, $(u-i)$ and $(i-z)$.  However,
the amplitude of the trend drops to 0.01\,mag or less in all colors $(X-i)$
when APCOR-colors in 7 and 8 pixels are compared
(we note that the difference between APCOR-magnitudes from 7 and 8 pixels 
still changes by $\sim$0.03\,mag along the compactness-sequence).
For apertures larger than 8 pixels, for instance APCOR-colors measured
in 8 and 16 pixels, no systematic trends with compactness are
detected. The discussion in this paper is based on APCOR-colors measured in 8 pixel apertures, APCOR16 colors being noisier for faint objects.

\subsection{r band seeing issues}

\begin{figure}
\begin{center}
\includegraphics[width=8.5cm,angle=0]{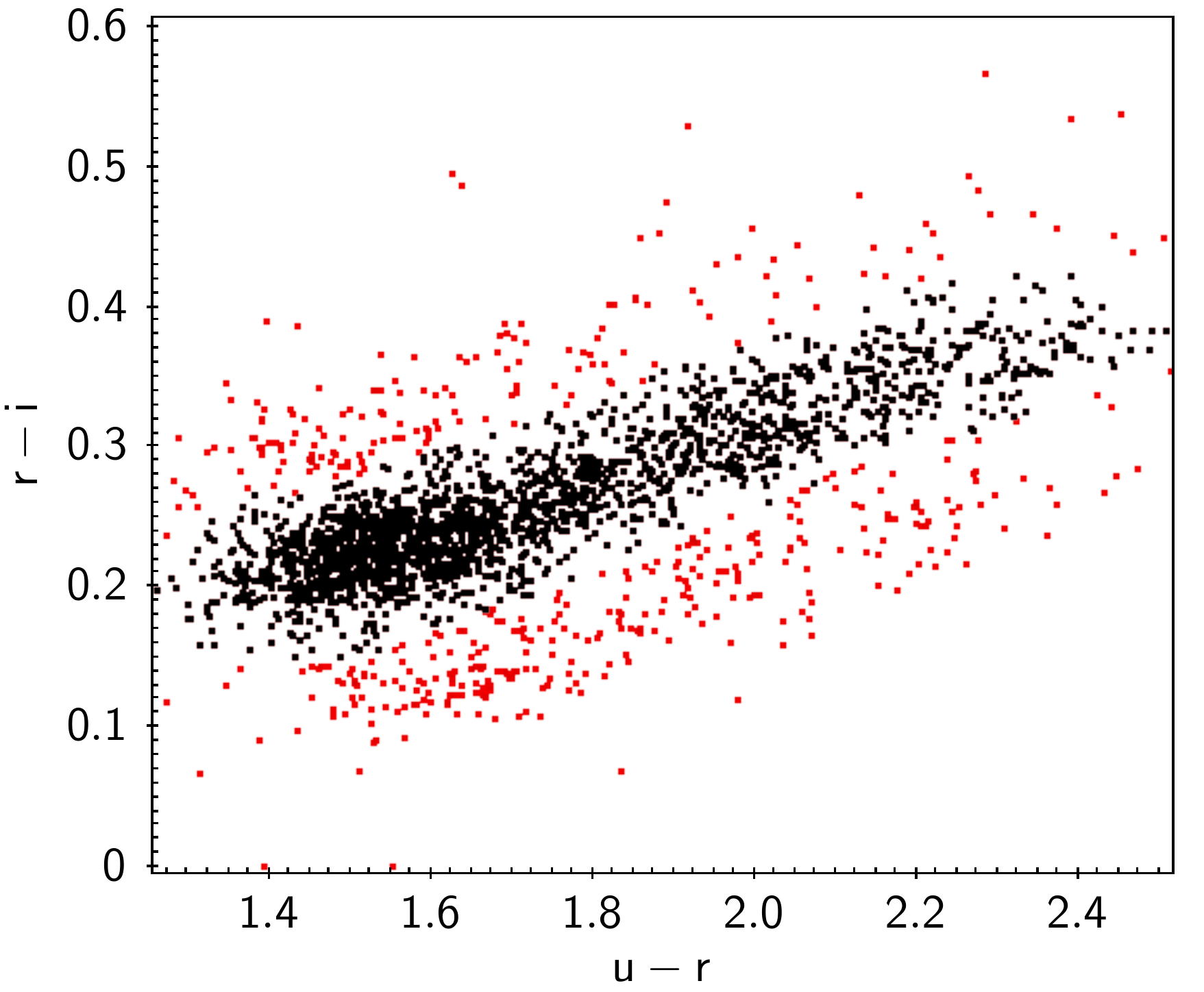} 
\caption{\label{r_band_issues} The $r$ band issues caused by different seeing in the 0-0 tile plotted in the uri diagram. The 2 external and mistaken branchs are highlighted in red.}
\end{center}
\end{figure}

During the data acquisition for NGVS pointing +0+0 in the $r$ band 
(the pointing containing M87), the seeing has varied significantly more than 
for all other pointings and filters. As a consequence, the point sources located
along the gaps between the individual rows of detectors have sizes that
differ from other locations (the number of exposures combined in these pixels
is smaller than elsewhere). The local aperture corrections cannot be determined
with a spatial sampling as small as these gaps. The consequences are 
outliers in color-color diagrams that involve the $r$ band.  
Figure \ref{r_band_issues} shows the effect on the globular cluster sequence:
two abnormal branches are seen on either side of the main locus.

Our goal is to have a clean sample of GC colors. Thus for the purpose of 
this paper we have removed all the objects with abnormal $r$-band photometry from
our reference sample. This last modification reduced our sample from
2321 to 1846 globular clusters.

\subsection{Properties of the GC sample}
\label{Prop_GC}

At this point, we have a clean sample of 1846 globular clusters.
As announced previously, this catalog is available in numerical
form and an extract is given in Tab.\,\ref{catalog_paper}.

\input{ext_table1.tex}

This GC sample is designed to provide a robust reference locus in color
space, as opposed to being complete in volume or magnitude. Each of the 1846
clusters was selected to have good photometry across the whole spectrum. The 
population of the red end of the GC sequence (metal rich clusters) is limited
by the requirement of good quality $u^*$ photometry, and the number of objects at the blue end by   
requirements in $z$ and $K_s$. The typical magnitudes of the GCs in the sample
are provided in Tab.\,\ref{tab_GCmags}. At the blue end of the sequence, this 
corresponds to typical masses of $1.6 \times 10^6\,M_{\odot}$ , and at the red end to masses of $3.3 \times 10^6\,M_{\odot}$.
These masses are typically a factor of 10 above the turn-over of the GC mass function 
\citep{jordan2007}.

\begin{table}
\caption{GC sample magnitudes}
\begin{center}
\begin{tabular}{ccccc} \hline
 & Mean & 10\,\% & 90\,\% & Mean errors\\
 &   (1) & (2) & (3) & (4)  \\ \hline
$u^*$ & 23.05  & 22.14 & 23.95 & 30\\
$g$ & 21.88  & 21.00 & 22.63 & 9 \\
$r$ & 21.32 & 20.46 & 22.06 & 8 \\
$i$ & 21.05 & 20.20 & 21.77 & 12 \\
$z$ & 20.87 & 20.02 & 21.58 & 21\\
$K_s$ & 20.90 & 20.02 & 21.66 & 83\\
\hline
\end{tabular}
\end{center}
\label{tab_GCmags}
Columns (2) and (3) provide the 10\,\% and 90\,\% percentiles of the 
distributions. The mean photometric errors are given in mmag.
\end{table}

\begin{figure}
\begin{center}
\includegraphics[width=9.2cm]{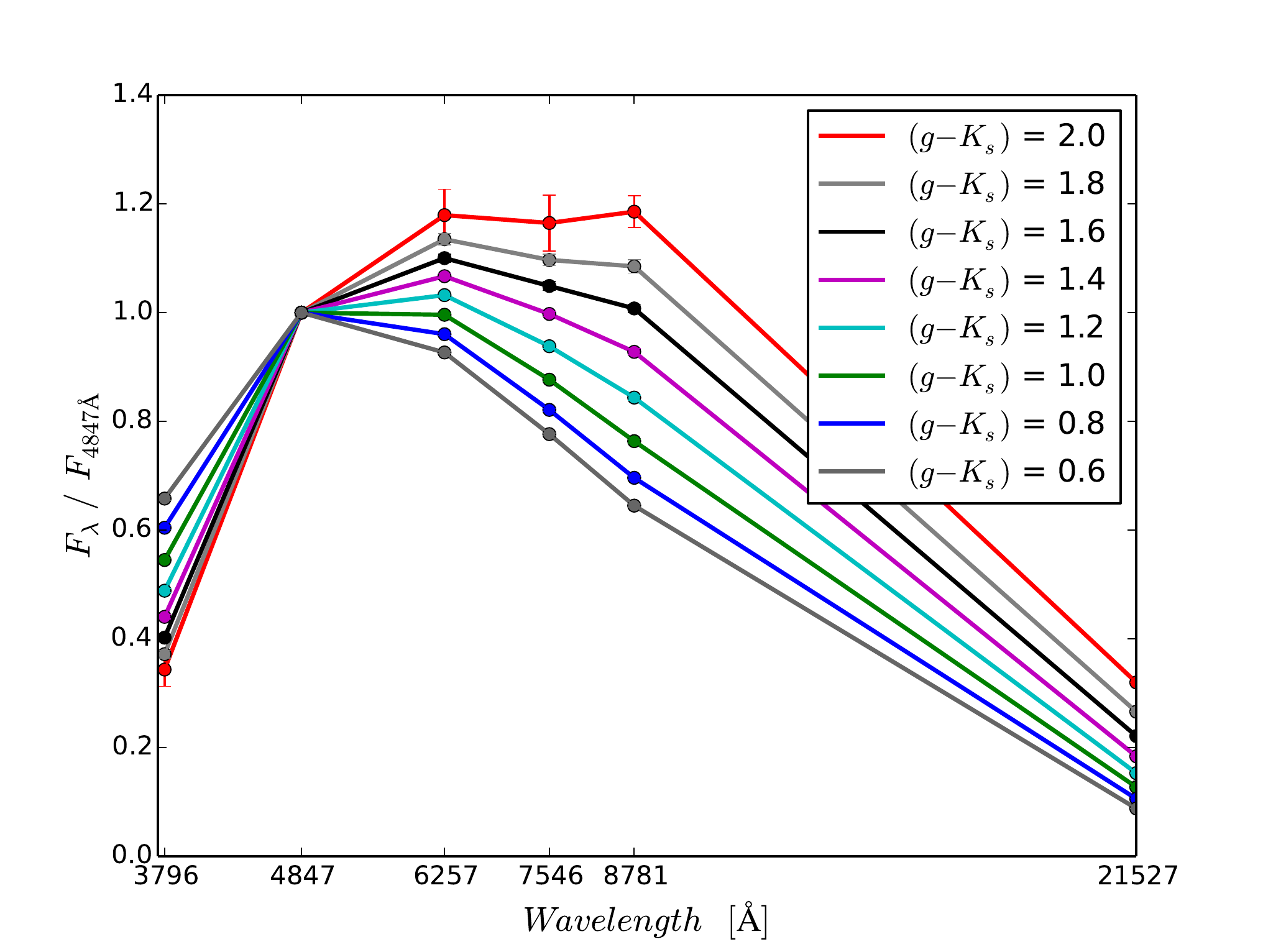} 
\caption{\label{SED_NGVS_s} Fiducial SEDs for the NGVS globular cluster sample 
described in Section \ref{GCsample.sec}. The corresponding values 
of $(g-K_s)$ [AB] are indicated. They are representative of the range of GC metallicities. 
These SEDs are available in Tab.\,\ref{sed_ngvs_full}.}
\end{center}
\end{figure}

Color distributions of the GCs will be discussed in detail 
in Section \ref{sec_results}. Using these, we have determined fiducial
loci in color-color diagrams, and fiducial SEDs for various locations 
along the GC sequence. The purpose of these is to provide an easy 
graphical reference when comparing color distributions with models 
(Section \ref{sec_results}). As the Virgo core region contains GCs with a 
variety of histories and environements, 
one may expect different GCs to contain different stellar
populations (age, metallicity, chemical abundances, etc), and we warn that
the fiducial SEDs would not capture this diversity. 

The fiducial SEDs are based on maximum likelihood polynomial fits to GC 
color-color distributions. The likelihood $L$ of a polynomial is the probability 
of obtaining the observed color-color distribution when drawing from this polynomial
parent distribution, taking into account the errors 
on the colors and their covariance 
(we treat the errors as gaussian in this process). Numerically, the polynomial 
$f$ is segmented into a large number of small segments $f_j$, which are 
here assigned equal prior probability (flat prior).
$$ L = \prod_{i=1}^{N_{GC}}\ p(c_i | f)\ = \prod_i \left[ \sum_j\ p(c_i | f_j) . p(f_j) \right]$$
Here $p(c_i | f)$ is the probability, for cluster $i$ in the sample, 
to be observed with colors $c_i$ if it originally was located on the polynomial,
and $p(c_i|f_j)$ is the same for location $f_j$ on the polynomial. 
$p(f_j)$ is a constant. 

Figure \ref{SED_NGVS_s} shows a set of fiducial SEDs obtained as a
function of $(g-K_s)$, a color with a large range of values 
compared to error bars. To first order, the sequence may be seen as an empirical
illustration of the effects of metallicity, combined with a possible effect of age.
To define these SEDs, polynomials were fitted respectively to 
the loci in diagrams of $(u-g)$, $(g-r)$, $(r-i)$, $(i-z)$ and $(z-K_s)$ versus $(g-K_s)$.
The values of the polynomials at a set of 
$(g-K_s)$ colors define the fiducial SEDs.

Thanks to the large
number of clusters and to the small individual photometric errors for each GC, 
the fiducial sequence is extremely well defined.
Bootstrap resampling provides
the 3\,$\sigma$ error bars shown in Fig.\,\ref{SED_NGVS_s} (most of these
are too small to see, and all are smaller than the systematic errors on the
GC photometry). The fiducial SEDs can also be modified by changing
the order of the adopted polynomial, removing more or fewer outliers, and other
fitting process differences. However the modifications obtained with various 
reasonable variants of the fitting details are of small amplitude compared 
to the systematic effects we intend to discuss in the comparison with models 
later on. 

To conclude the description of the sample, we have tested whether or not the empirical GC color distribution is statistically compatible with an infinitely tight theoretical color-locus, given by the fiducial SEDs just described.
To quantify the goodness-of-fit, we used the following reduced $\chi^2$:
\begin{equation}
 \chi^2\ =\ \frac{1}{N_{GC}}\ \sum_{i=1}^{N_{GC}}\ \min_j \left[ (c_i - c_{f_j})^t\,\Sigma^{-1}\,(c_i - c_{f_j})\right]
\end{equation}
where $c_i$ contains the colors considered,
$\Sigma$ is the covariance matrix, $c_{f_j}$ holds the colors of a point of 
the fiducial locus, and $\min_j$ takes the minimum along the polynomial. $N_{GC}$ is 
a proxy for the number of degrees of freedom of the fit. 
 
We fitted two-color distributions with polynomials of orders 2 to 5, 
and explored the effects of removing up ten GCs with strongest individual impacts 
on the fit and the $\chi^2$. All in all, using several combinaisons of colors, we did not find a best $\chi^2$ below 1.23 for a single color-color diagram.
Conversely, we found a number of
color-color planes in which the best $\chi^2$ in these tests remained above 3,
for instance $(i-z)$ vs. $(g-i)$, or $(u-i)$ vs $(g-i)$.
For good representations
of the data, the $\chi^2$ would not exceed 1 by more than a few $(N_{GC})^{-1/2}$
(i.e. a few times 0.023). Hence there is real dispersion across the main locus
of the data.


\input{ext_table2.tex}
\subsection{Budget of systematic errors}
\label{error_budget}
The online catalog of GCs provides individual uncertainties on the
magnitude measurements, as described in Section \ref{photo_ext_cat}. In addition to these
random errors, we have mentioned a variety of sources of possible
systematic errors on the photometry. We provide a summary of these here,
with estimated bounds in Tab.\,\ref{tab_bes}.

\subsubsection{Systematic errors in the external reference catalogs}
\label{error_budget_sdss}
The NGVS MegaCam photometry is calibrated against SDSS stars, thus any
systematic errors in the SDSS photometry has a direct effect on NGVS.
Currently, the relative calibration within SDSS DR10 seems to be well
known, with studies by \citet{padmanabhan2008}, \citet{bramich2012}, \citet{schlafly2011}. The precision of the internal
calibration is estimated to be around 2\,\% in the $u^*$ band and 1\,\% in the
$g$, $r$, $i$ and $z$ filters. Regarding the absolute calibration of SDSS (which
is known not to be on an exact AB magnitude system), limits are more
difficult to set. The SDSS DR10
documentation\footnote{https://www.sdss3.org/dr10/algorithms/fluxcal.php} indicates
a likely offset of 0.04\,mag in $u^*$, in the sense that $u_{SDSS,AB}\ =\ 
u_{SDSS}-0.04$. An offset of 0.02\,mag in $z$ in the opposite direction is
also advocated there. These offsets are considered known to no better
than 0.01\,mag, and possibly slightly less precisely for $u^*$. 
We have not implemented these zero point shifts in our 
data but discuss their effect whenever 
necessary. In summary,
we adopt limits on the systematic errors of 0.04\,mag 
in the $u^*$ filter, 0.01\,mag in $g$, $r$, $i$, and 0.02\,mag in $z$, and note that there is a preferred direction
for the offsets in $u^*$ and $z$\footnote{More specifically, users who wish to
apply the conversions from SDSS to AB magnitudes suggested by the SDSS
web pages should remove 0.04\,mag from the NGVS $u^*$ values 
published in this paper and add 0.02\,
mag to the NGVS $z$ values. After these corrections, one may consider
reducing the SDSS calibration errors to 0.015 in $u^*$ and 0.01 in $z$}. \\
Similarly, the $K_s$ band is affected by any systematic errors in the
UKIDSS photometry (DR8). Based on the various tests presented by \citet{hodgkin2009}, we assign a bound of 0.02 magnitudes on systematic errors
to this photometry.

\subsubsection{Systematic errors in the calibration of NGVS with respect to the
external catalogs}
\label{error_budget_tie}

The calibration of NGVS relative to SDSS or UKIDSS is affected slightly
by the dispersion in the photometry of stars in common between the
surveys. The dispersion seen around the mean trend in the calibration
figures (Fig.\,\ref{transfo_plot}) is due mainly to dispersion in the SDSS and
2MASS/UKIDSS catalogs, NGVS being deeper. The number of stars available
for the calibration reduces errors on the mean to a few millimagnitudes
in all filters.

Small differences are seen in Fig.\,\ref{transfo_plot} between the reference
transformation curve used in the $u^*$, $g$, $r$, $i$, $z$ data processing and modern
synthetic photometry. We take this as an indication of possible systematics
in the transformation between systems. As an estimate of their amplitude, we adopt
the mean difference between the empirical and the synthetic loci, over
the range of colors of stars actually observed in the survey. The
offsets are smaller than 50\,mmag in $u^*$, 5\,mmag in $g$, 5\,mmag in $r$, 8\,mmag in $i$, and 2\,
mmag in $z$.

The transmission curves adopted for NGVS have impact on the locus of
synthetic colors in the calibration figures (Fig.\,\ref{transfo_plot}). A few
versions of the Megacam filters have been available on CFHT/CADC web
pages over the years, prior to the work of \citet{betoule2013}. Our
test for two extreme sets of filters results in discrepancies inferior
to 12\,mmag in $u^*$, 2\,mmag in $g$,$i$, 3\,mmag in $r$ and 8\,mmag in $z$. If we
estimate the main source of uncertainty in the WIRCam $K_s$ transmission is
telluric absorption, we find that reasonable changes in airmass/humidity
change the $K_s$ magnitudes by less than 5\,mmag.

Our calibration of the WIRCam $K_s$ photometry against UKIDSS involves a
conversion from AB magnitudes to Vega magnitudes. We have converted our
WIRCam $K_s$ AB magnitudes to Vega magnitudes for this purpose. The offset
used, determined in \citet{munoz2014}, is based to the best of our
knowledge, on the same reference Vega spectrum as used for UKIDSS
\citep{hewett2006}, i.e. a spectrum originally provided by \citet{bohlin2004}\footnote{That Vega spectrum was made available at the
time on the Hubble Space Telescope Science Institute web pages as {\tt
alpha.lyr.stis.003} or {\tt alpha.lyr.stis.005}, these two files leading
to identical results for the $K_s$ band.}. Therefore we take it that this
source of error adds little to those already included in the absolute
UKIDSS errors and the errors due to transmission changes, described above.

\subsubsection{Systematic errors in the reddening corrections}
\label{error_budget_ext}

Systematic errors can also occur in the dereddening process, with the
choice of a particular local value of A(V) or E(B-V), and of wavelength
dependent extinction coefficients. The different total extinction
estimates of \citet{schlegel1998} and \citet{schlafly2011}
translate into differences of 19\,mmag for $u^*$, 14\,mmag for $g$, 11\,mmag for
$r$, 8\,mmag for $i$, 6\,mmag for $z$ and 1\,mmag for $K_s$ when using the extinction
law of \citet{cardelli1989}, in the sense that the \citet{schlafly2011} reddening corrections are smaller (cf. Section \ref{ext_part}). The
use of extreme stellar types to derive extinction coefficients for a
given extinction law produces a span of extinction magnitudes in the
Virgo region inferior to 5\,mmag in $u^*$, 10\,mmag in $g$, and 2\,mmag for
the $r$, $i$, $z$ and $K_s$ filters. Using the extinction law 
of \citet{fitzpatrick1999} instead of \citet{cardelli1989} changes 
Virgo magnitudes by 10\,mmag maximum.

\begin{table*}
\caption{Budget of systematic errors}
\begin{center}
\begin{tabular}{ccccccc} \hline
  & \multicolumn{6}{c}{Maximum estimated errors in mmag.} \\
  &  $u^*$ & $g$ & $r$ & $i$ & $z$ & $K_s$ \\ \hline
SDSS calibration & 40 & 10 & 10 & 10 & 20 &  \\
UKIDSS calibration  &    & & & & & 20 \\
Transformation between systems & 50 & 5 & 5 & 8 & 2 &  \\
Color dependance of extinction coefficients & 5 & 10 & 2 & 2 & 2 & 2 \\
A(V): Schlafly vs Schlegel & 19 & 14 & 11 & 8 & 6 & 1 \\
Filter transmissions & 12 & 2 & 3 & 2 & 8 & 5\\
\hline
\end{tabular}
\end{center}
\label{tab_bes}
Note that uncertainties on A(V) creates systematics that are not independent between passbands. 
See Section \ref{error_budget_sdss} for the preferred direction of the systematic 
errors on the SDSS calibration in $u^*$ and $z$.
\end{table*}

\subsubsection{Systematic errors in the SLR method}

The SLR method relies on spectral libraries and transmission curves. The
use of a different library or a different set of stellar parameters
along the stellar locus can induce very large changes of the vector of
color-shifts, $\kappa$. For example, the differences between our preferred set
of parameters (black line in Fig.\,\ref{SLR_fig}) and a set composed of solar
metallicity stars all with [$\alpha$/Fe]\ =\ 0 and log(g)\ =\ 5.0 produces a color difference
of $\Delta\,\kappa$(mmag.)\ =\ [$\Delta\,$($g-r$)\ =\ 37, $\Delta\,$($r-i$)\ =\ 19,$\Delta\,$($i-z$)\ =\ 15, $\Delta\,$($i-Ks$)\ =\ 2]. The varying parameters given by the
Besan\c{c}on model of the Milky Way are more reasonable than a set with
uniform composition and gravity, reducing this source of systematics
somewhat.
\medskip

 Having described our empirical GC sample and its photometric accuracy, we
turn towards population synthesis models and predicted colors. The comparison  
between models and data in color-color diagrams (Section \ref{sec_results}) serves to
characterize the empirical color locus further, and is a fundamental step 
towards estimating the evolutionary parameters of the clusters. 

\section[]{The models}
\label{modelsec}

Numerous population synthesis models are available in the literature and can be used to estimate ages and 
metallicities of stellar populations from empirical SEDs. In this section, we describe the codes 
we have used, as well as the generic assumptions made to construct synthetic SEDs for globular clusters
with each of them. Comparisons between the resulting SEDs, and with the NGVS globular cluster colors, are
made in Section \ref{sec_results}.

In this paper, we consider only models for single stellar populations, containing stars of a single age
and chemical composition. This assumption is questionable, especially for a sample of massive clusters,
since photometric and spectroscopic studies of resolved massive clusters nearby 
revealed the existence of multiple subpopulations. 
Our analysis is meant to provide a reference point for future studies, in which these assumptions could be relaxed.

We have considered six commonly used stellar population synthesis codes
(SPS codes hereafter),
for which predictions can be obtained via dedicated webpages.

From each SPS code, we obtained a set of synthetic spectral energy distributions for single stellar populations, 
i.e. synthetic SSP models (sSSP hereafter).
[Fe/H] was varied from -2 to 0.17 (with three exceptions among the 11 sSSPs mentioned below), 
and ages between 6 and 13 Gyr. The majority of the GCs in the Virgo sample are assumed to be old.
 Nevertheless, these restrictions on age and metallicity must be kept in mind in the
comparisons below, and possibly be relaxed in future studies of individual objects. 

We adopted the initial mass function (IMF) of \citet{kroupa1998} 
or \citet{kroupa2001} as available with the codes. 
The discrepancies due to changes in the IMF are smaller than other discrepancies between model families, so
we will not show any assessments of these here.

Whenever possible, we used the SPS codes to compute synthetic spectra, and derived synthetic photometry from them
ourselves with the filters described in Section \ref{photo_ext_cat}. 
For codes that allow the input of customized transmission curves,
we compared our synthetic photometry with
the one produced by those codes, finding that differences were
negligible (less than 0.05\,\%).

To account for the redshift effect, we have computed all the model colors at the typical redshift of the Virgo cluster.
This correction (which reaches 15\,mmag in the $g$ band and 5\,mmag in the $i$ band) has been obtained directly by a computation of the colors on a redshifted spectrum,
or otherwise by the use of a redshift-correction based on the \citet{maraston2005} model and the {\sc P\'egase} models.

\smallskip

The sSSP models differ from each other by the stellar evolution tracks they rest upon and by the stellar library used to
predict spectrophotometric properties. We briefly describe these choices
below.

Two first sets of sSSPs, labelled {\bf BC03} and {\bf BC03B} hereafter, are taken from
\citet{bruzual2003}\footnote{We use the 2012 update made available
by the authors upon request, but it differs from the 2003 version only in
additional outputs that we have not used.}.
We selected the 1994 version of the Padova  isochrones as input \citep{alongi1993,bressan1993,fagotto1994a,fagotto1994b,girardi1996}.
The default synthetic spectra (BC03) combine optical stellar spectra from 
STELIB \citep{leborgne2003} 
between 3200 and 9500\,\AA\ with the BaSeL 3.1 spectral library outside this wavelength range (\citealt{lejeune1997}, \citealt{lejeune1998} and \citealt{westera2002}).
For the BC03B set, the BaSeL stellar library was used instead at all wavelengths.

Three sets of sSSPs, labelled {\bf C09PB}, 
{\bf C09BB} and {\bf C09PM}, 
are based on the Flexible Stellar Population Synthesis (v2.4) 
model of \cite{conroy2009}. The first one (C09PB) is computed with the 
Padova 2007 set of isochrones 
\citep{girardi2000,marigo2007,marigo2008}
and the BaSeL 3.1
spectral library. The second one (C09BB) is modeled with the BaSTI isochrones 
\citep{pietrinferni2004,cordier2007} and the BaSeL 3.1 library.
For the final one (C09PM), we used the Padova 2007 isochrones and the 
MILES spectral library \citep{sanchezblazquez2006}.
MILES spectra extend from ~3500 to ~7500\,\AA\
and can only provide fluxes in the $g$ and $r$ filters, so they
are extended with the BaSeL spectral library beyond this range.
The C09PM and C09BB sets do not reach down to [Fe/H]\ $\sim$\ -2\,dex, but instead respectively start at [Fe/H]\ =\ -1.39 and -1.82.

Two sets of sSSPs, labelled {\bf M05} and {\bf MS11}, were constructed using the models of \citet{maraston2005} and \citet{maraston2011}.
The former uses the Cassisi isochrones 
\citep{cassisi1997a, cassisi1997b, cassisi2000} and the BaSeL 3.1 library.
This model offers two options for the morphology of the horizontal branch (HB): a red HB and a bluer one.
The red HB produced a better representation of our observations, so the blue HB will not be shown in this paper.
The latter set (MS11) also uses the Cassisi isochrones and a combination of the MILES library and the BaSeL 3.1 library.
Both these models are computed using algorithms based on fuel-consumption instead of the more common isochrone synthesis.

We also considered one model from the web interface 
CMD 2.7\footnote{\tt http://stev.oapd.inaf.it/cgi-bin/cmd} 
provided by the Padova group
(labelled {\bf PAD} hereafter).
This model uses the PARSEC 1.2S isochrones 
\citep{bressan2012,tang2014,chen2014,chen2015}
and it is based on the PHOENIX BT-Settl library \citep{allard2003}
for effective temperatures lower than 4000\,K,
and on ATLAS9 ODFNEW \citep{castelli2004} otherwise.
This version of PARSEC isochrones does not take into account
thermally pulsing asymptotic giant branch (TP-AGB) stars.
Although the tendency is for the estimated TP-AGB contributions
to be revised downwards at old ages \citep{gullieuszik2008,
girardi2010,melbourne2012,rosenfield2014},
the complete lack thereof is expected to produce a lack of near-infrared flux.
Another issue with the PAD models is that no spectra are available
from the web site. The synthetic photometry is based on
filter transmissions older than the ones we now prefer.
Discrepancies such as these may produce offsets
of a few percent, in particular in $i$.

The sSSP labelled {\bf PEG} is produced with
{\sc P\'egase} \citep{fioc1997}\footnote{We use the code made
available as {\sc P\'egase.2} or {\sc P\'egase-HR} by \citet{leborgne2004}.}
The isochrones are the Padova 1994 set and the BaSeL 2.2 spectral library
provides the photometry.

Finally, we have considered two sSSPs based on the model of \cite{vazdekis2012}. 
We have chosen to compute these models with the Padova 2000 isochrones.
The first one, labelled {\bf VAZ\_MIUSCAT}, uses the MIUSCAT library 
with a wavelength range from 3464 to 9468\,\AA. It only allows us to compute the $g$, $r$ and $i$ magnitudes.
The second one, labelled {\bf VAZ\_MIUSCAT\_IR}, is based on the MIUSCAT-IR library 
which extends from 3464\,\AA\ to 49999\,\AA. However, 
the values of [Fe/H] currently available for the spectra
are restricted to -0.40, 0 or 0.22.  Due to the wavelength and [Fe/H] ranges, 
the display of these sSSPs is done separately.  

All these sSSPs are produced with the simplest assumptions
possible: there is no dust (so no extinction), we chose default
mass-loss parameters, zero binary fractions, etc.
Overall, the aim was to compute for each SPS code the same distribution of
GCs parameters.

In the following section, this set of models is compared
to our GC sample using color-color diagrams (Section \ref{ccd} and \ref{UVcolcol})
and SEDs (Section \ref{SED}).

\section[]{Models versus data}
\label{sec_results}
To provide GC ages and metallicities, one needs to connect the various model predictions with the empirical GC color distributions.
Because the age and metallicity information is hardwired in each model depending on the particular set of assumptions (see Section \ref{modelsec}),
we will pay particular attention to the differences in the derived GC age and metallicity distributions.

\subsection{Optical-NIR color-color diagrams}
\label{ccd}
Color-color diagrams provide a powerful global overview 
of all the assessed objects, as well as direct insight into the physical properties of GCs.
In the case of old globular clusters, 
even a single color carries information on metallicity 
(e.g. \citealt{cantiello2007}; \citealt{puzia2002}), 
although as we shall emphasize the relation between color and 
[Fe/H] remains model-dependent. Color-color diagrams of 
GC samples in principle allow access to a second parameter, typically
age. The distribution in the 5-dimensional color-space available for 
the NGVS clusters should improve the age and metallicity assessments. 
In practice however, they also highlight differences between models.

\begin{figure}
\begin{center}
\includegraphics[width=9.2cm]{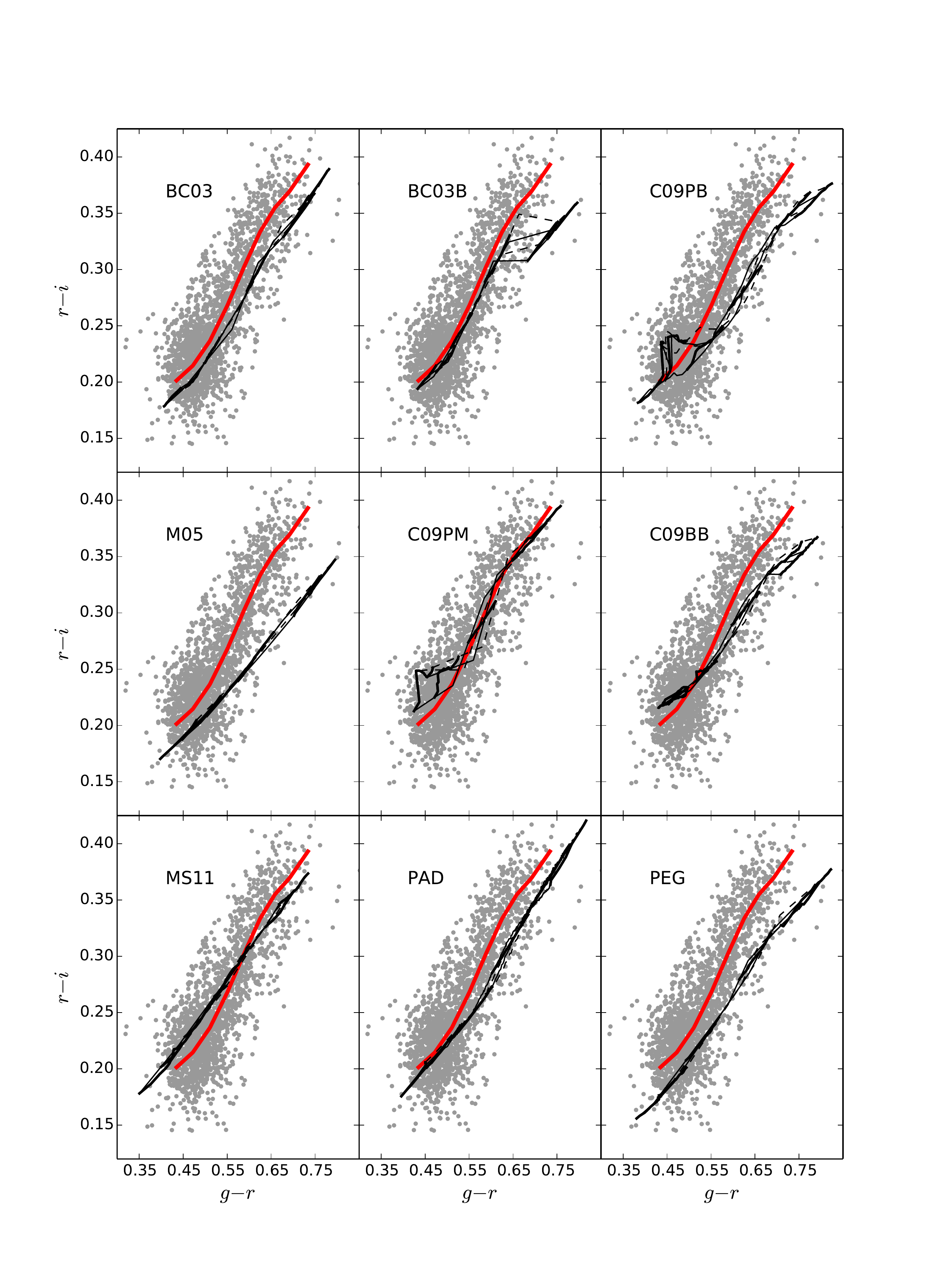} 
\caption{\label{ccd1} $gri$ diagram. The gray dots are the GCs of 
our robust Virgo sample, after correction for foreground extinction. The calibration is that of Section \ref{photo_ext_cat} (see appendix Figure \ref{ccd4s} for a version of these plots using the SLR calibration). 
For each model set, the thick solid and dashed 
lines represent the metallicity tracks at given ages, 
with metallicity increasing 
from the bottom left to the top right.
The metallicities of the grid nodes are [Fe/H]\ =\ [-2, -1.5, -1, -0.5, 0, 0.17] (except for C09BB and C09PM, see 
Section \ref{modelsec}).
Alternating thin solid and dotted lines 
connect models of constant age, at [6, 8, 10, 13] Gyr.
The red line represent the polynomial fit defined in Section \ref{Prop_GC}. 
The model grids are
degenerate in age and metallicity in this particular color-color
diagram.}
\end{center}
\end{figure}

\begin{figure}
\begin{center}
\includegraphics[width=9.2cm]{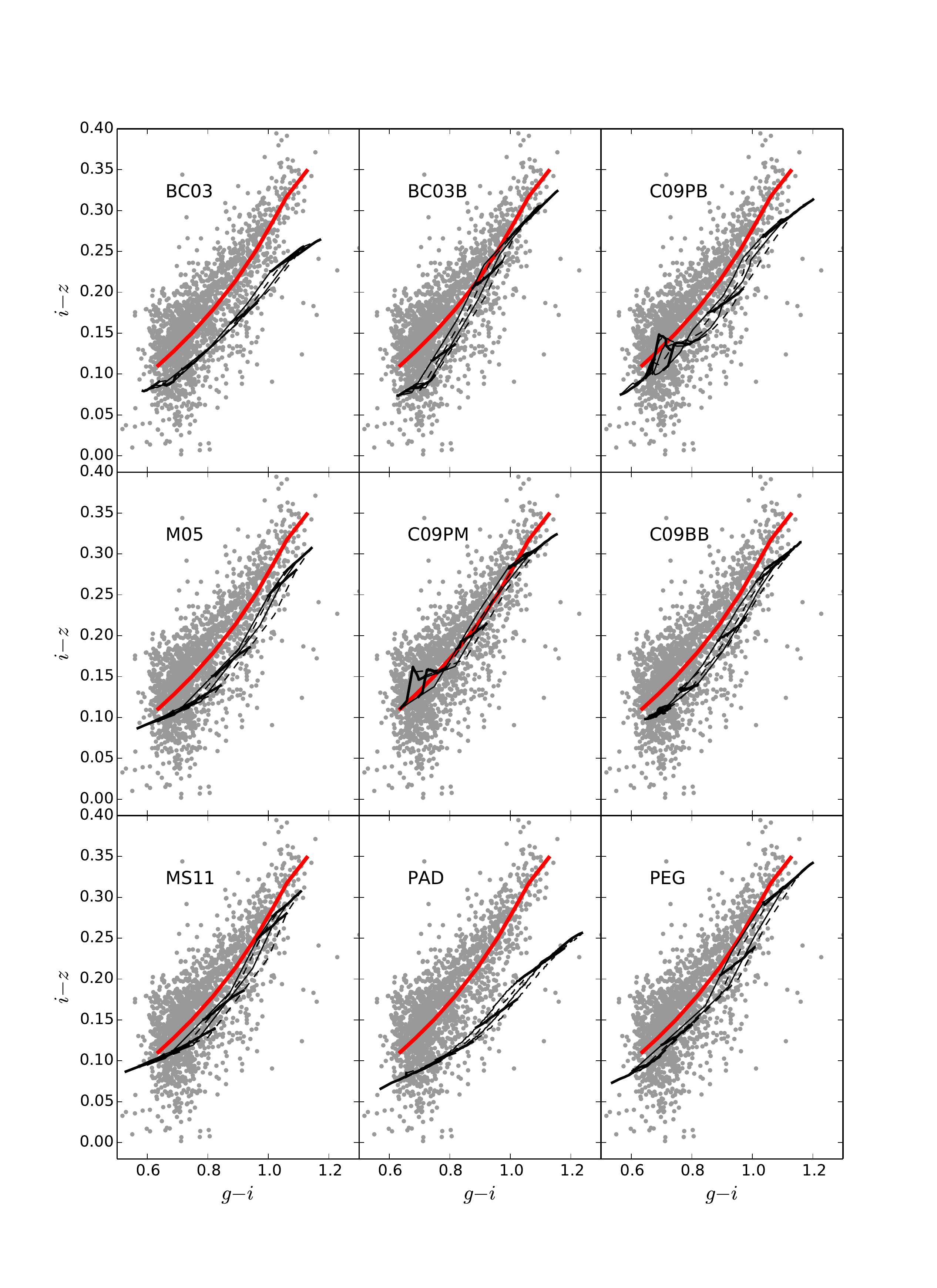} 
\caption{\label{ccd2} $giz$ diagram. 
Symbols and lines are as in Fig.\,\ref{ccd1}.
}
\end{center}
\end{figure}

The locus of sSSP models with respect to the robust NGVS globular cluster 
sample in various color-color diagrams is shown in
Figs.\,\ref{ccd1}, \ref{ccd2}, \ref{ccd3}, and \ref{ccd4} (additional color-color diagrams are shown in 
Figure \ref{ccd_appgik} of the Appendix).
The first two are restricted to the MegaCam colors $g$, $r$, $i$, and $z$, 
the last two include $K_s$.

In the $gri$ diagram ($r-i$ vs. $g-r$, Fig.\,\ref{ccd1}), 
age and metallicity are degenerate in the models. This is essentially
also the case in the $giz$ diagram ($i-z$ vs. $g-i$, Fig.\,\ref{ccd2})
and the $gKiz$ diagram ($i-z$ vs. $g-K_s$, Fig.\,\ref{ccd3}). 
The age-metallicity degeneracy is best broken in planes such as the
$riK$ diagram of Fig.\,\ref{ccd4} ($i-K_s$ vs. an optical color, here $r-i$).
This property has already been highlighted in the literature
(e.g. \citealt{puzia2002}).

Large discrepancies are seen between models in all
color-color planes, despite the fact that all model grids
cover the same range in age and [Fe/H] 
(with the exception of C09PM and C09BB, that lack the lowest
metallicities).  A model set that seems best in 
one color-color plane is not usually best in all the others.
\smallskip

The {\em range in color}\ spanned by the models 
in $(g-r)$ and $(g-i)$ corresponds quite well to the range observed  
(Figs.\,\ref{ccd1} and \ref{ccd2}). On the contrary, 
several models fall short of reproducing the observed range
of colors in $(i-z)$ (Figs.\,\ref{ccd2} and \ref{ccd3}),  $(r-i)$ 
(Figs.\,\ref{ccd1} and \ref{ccd4}) or $(i-K_s)$. 
The models M05 and M11, which are known to have among the strongest
AGB contributions in the literature at intermediate ages, struggle
at older ages to reach the optical-near-IR colors 
of the reddest observed clusters. 
The PAD  grid stops at an even smaller $(i-z)$ color index, about 0.1\,mag
bluer than the red end of the observed cluster distribution. 
This could seem a natural consequence of the 
lack of any TP-AGB stars in the PAD models, if this same model grid
did not extend right to the end of the observed distributions
in $(r-i)$, $(g-K_s)$ or $(i-K_s)$. 
In this case, this may argue for a systematic difference in the 
molecular absorption in the $z$ band between observed and synthetic GCs. 
The BC03 grid produces color ranges very similar to the PAD grid, although it is based
on different isochrones
and a different stellar library. BC03 and 
PAD however have in common that they do not use BaSeL, the
library that provides $z$ and $K_s$ band fluxes in all 
other cases.
\smallskip

Now looking at the {\em loci of the grids} instead of their range
in color, we find a variety of behaviors again.  It is important
to keep in mind that zero point offset errors in the NGVS photometry
could shift the distributions but not modify their shape.
Errors in individual extinction corrections would increase the 
dispersion.  The shapes of the model grids could, on the other hand, 
be affected by errors in the assumed filter transmission curves 
as well as the input stellar physics.

\begin{figure}
\begin{center}
\includegraphics[width=9.2cm]{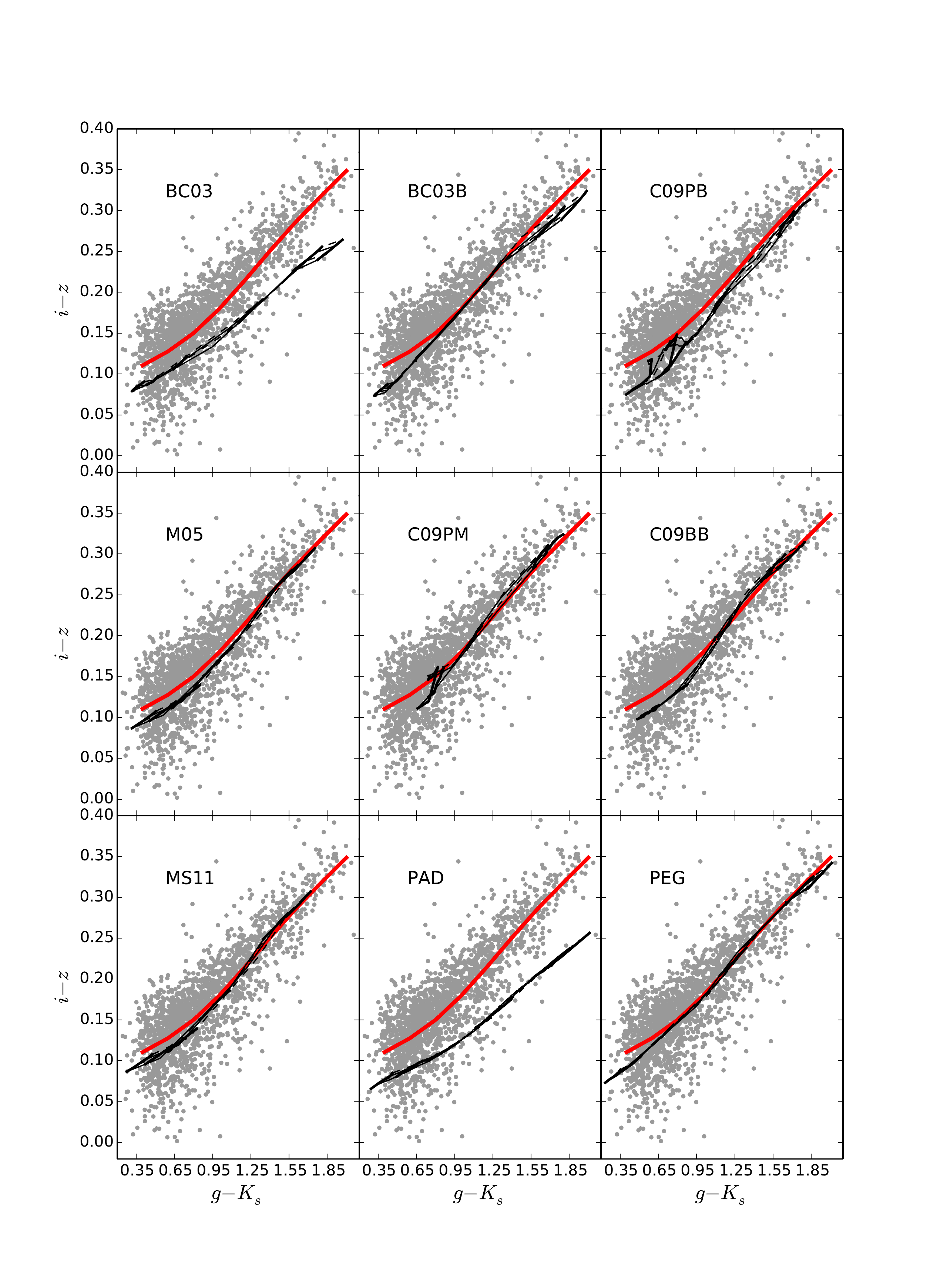} 
\caption{\label{ccd3} $gKiz$ diagram. 
Symbols and lines are as in Fig.\,\ref{ccd1}. }
\end{center}
\end{figure}

\begin{figure}
\begin{center}
\includegraphics[width=9.2cm]{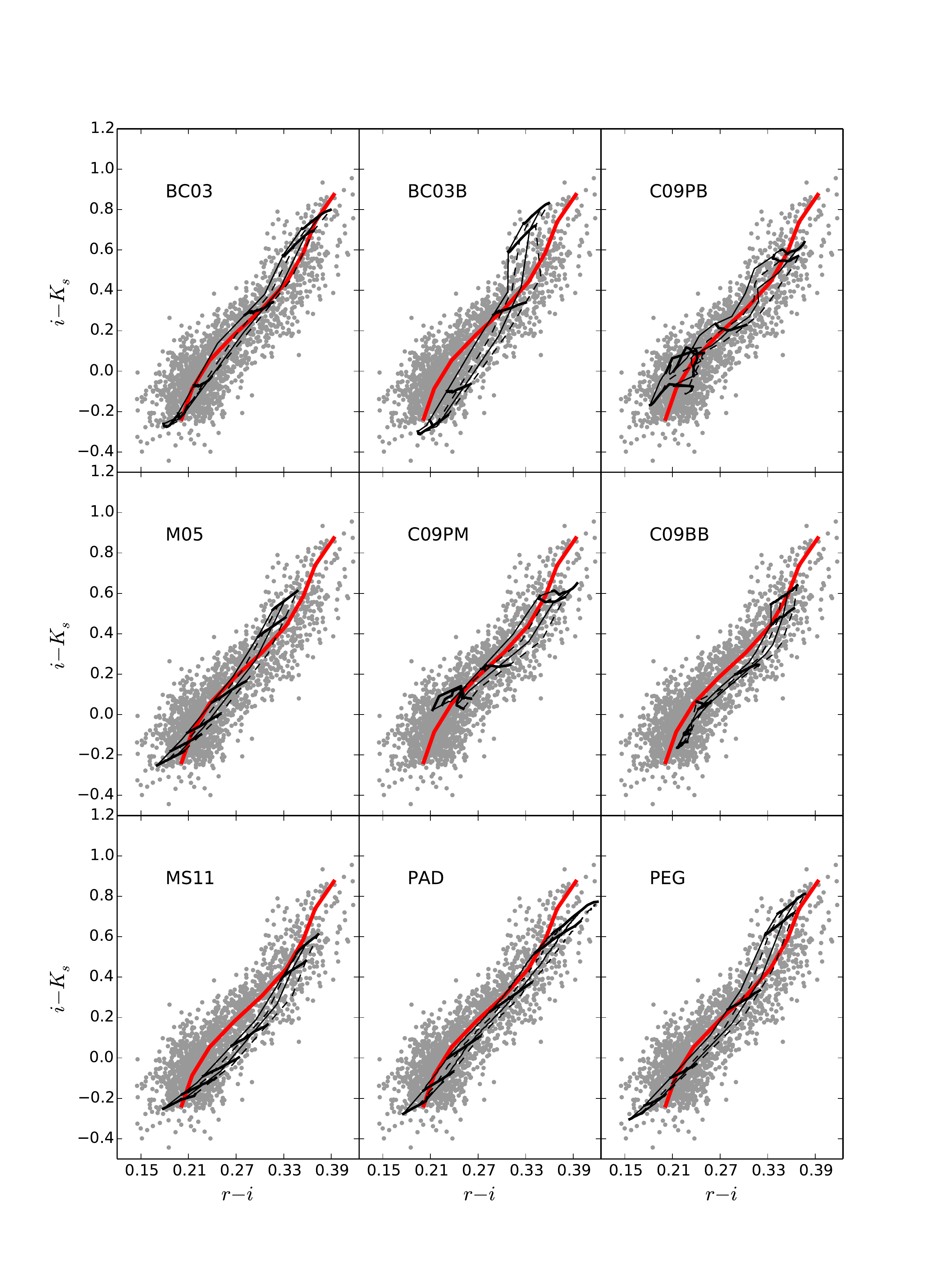} 
\caption{\label{ccd4} $riK$ diagram.
Symbols and lines are as in Fig.\,\ref{ccd1}. }
\end{center}
\end{figure}

\begin{figure}
\begin{center}
\includegraphics[width=9.2cm]{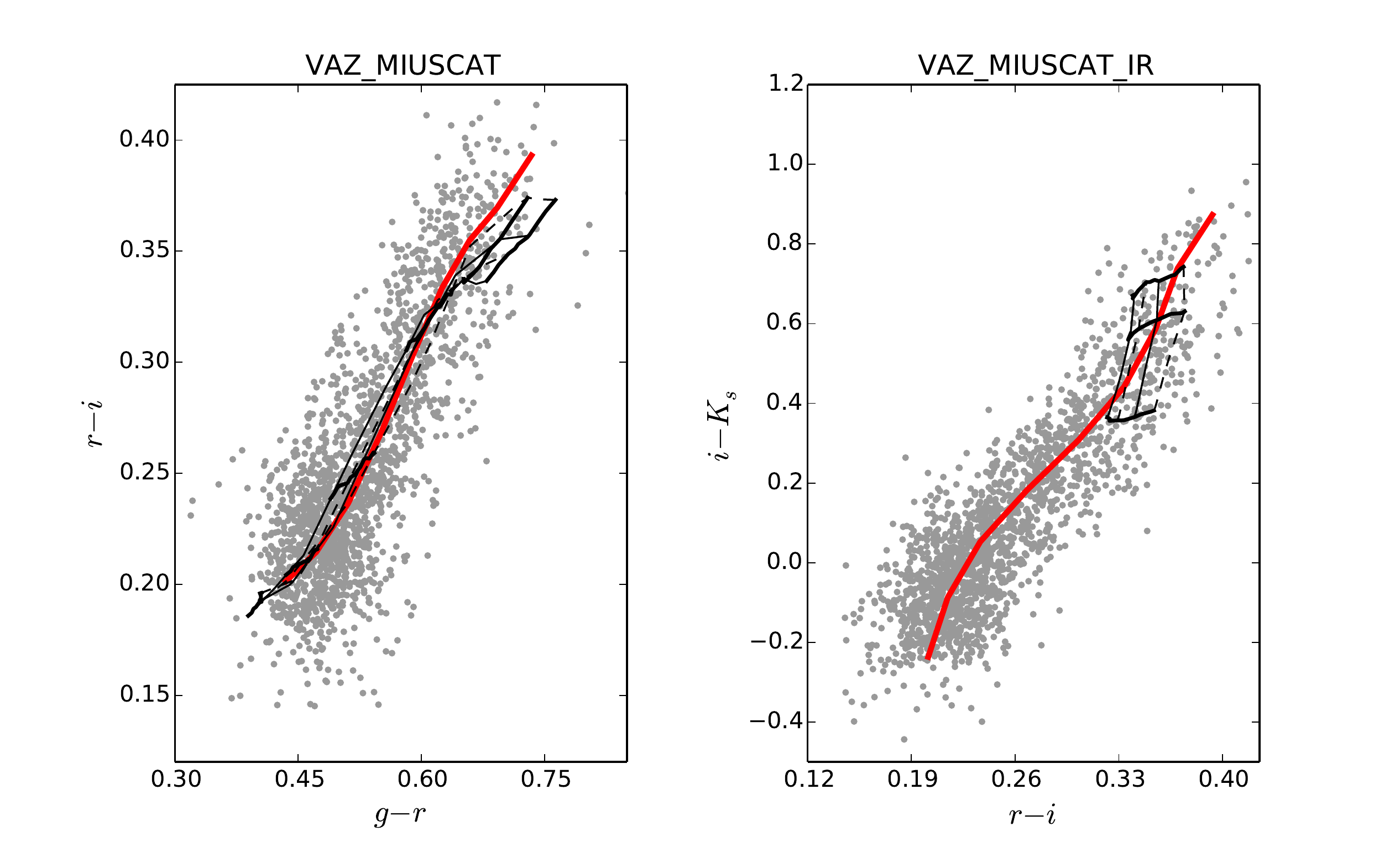} 
\caption{\label{vaz_col} Two color-color diagrams 
for the \citet{vazdekis2012} model. In the $gri$ diagram (left panel), 
the symbols and lines are as in Fig.\,\ref{ccd1}.
In the $riK$ diagram, the model is used with the MIUSCAT IR spectral
library, for which only metallicities near solar are available.
The grid nodes shown are [Fe/H]\ =\ [-0.39, 0, 0.17].}
\end{center}
\end{figure}

Surprisingly, it is in the $riK$ diagram (Fig.\,\ref{ccd4})
that the behaviour of the models is most uniformly satisfactory: the
model grids are located within the bounds
of the empirical color distribution,
though sometimes with significant deviations from the
fitted line of typical colors.
As the color spreads of the various model grids differ, any given cluster
could however be assigned rather different absolute metallicities and ages
depending on the model adopted. In the $gKiz$ diagram (Fig.\,\ref{ccd3}), the
model loci are satisfactory except for BC03 and PAD which, as
already mentioned, do not produce red-enough $(i-z)$ colors at high metallicity.
 In the $giz$ diagram (Fig.\,\ref{ccd2}), the shapes of the model grids
are mostly adequate, but a uniform offset in $(g-i)$ or in $(i-z)$ would seem
required to match the data. Applying the offset of 0.02\,mag in $z$ suggested
by the SDSS DR10 calibration pages would act in the right direction for most models
(see Section \ref{error_budget}).

The purely optical $gri$ diagram (Fig.\,\ref{ccd1})
is not uniformly well matched. In several model grids, the slope of $(r-i)$
as a function of $(g-r)$ is too shallow compared to the data. In general,
the models match the blue end of the GC distribution better than the red
end. This could be because the Milky Way globular clusters
frequently used to calibrate population synthesis models are mostly
metal poor.
However some models, such as C09PM or MS11, behave rather well at the
red end in the $gri$ plane. These two have in common that they exploit the
MILES spectral library at optical wavelengths, which has an effect on
the $g$ band fluxes (\citealt{maraston2011}
and also Section \ref{UVcolcol}). We confirmed this trend
with the MIUSCAT/MILES-based models of \citet{vazdekis2012} in Fig.\,\ref{vaz_col}. Finally, we note that the C09PM and C09PB models
display a complex dependence with age and metallicity at the blue end,
which is not seen in other model collections that also use Padova
isochrones.

\subsection{UV-optical-NIR color color diagrams}
\label{UVcolcol}

\begin{figure}
\begin{center}
\includegraphics[width=9.2cm]{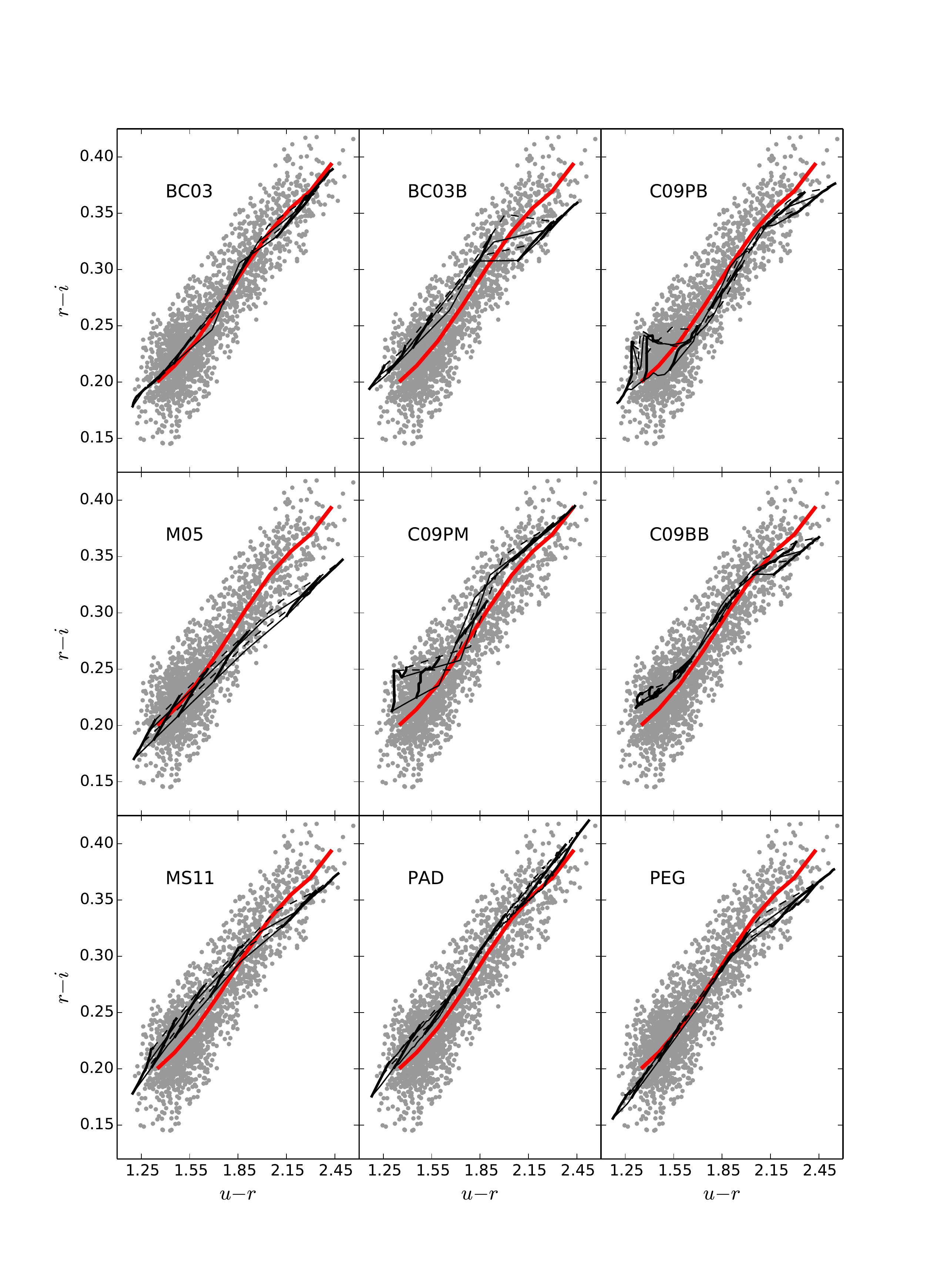} 
\caption{\label{ccd5} $uri$ diagram. 
Symbols and lines are as in Fig.\,\ref{ccd1}.
}
\end{center}
\end{figure}

\begin{figure}
\begin{center}
\includegraphics[width=9.2cm]{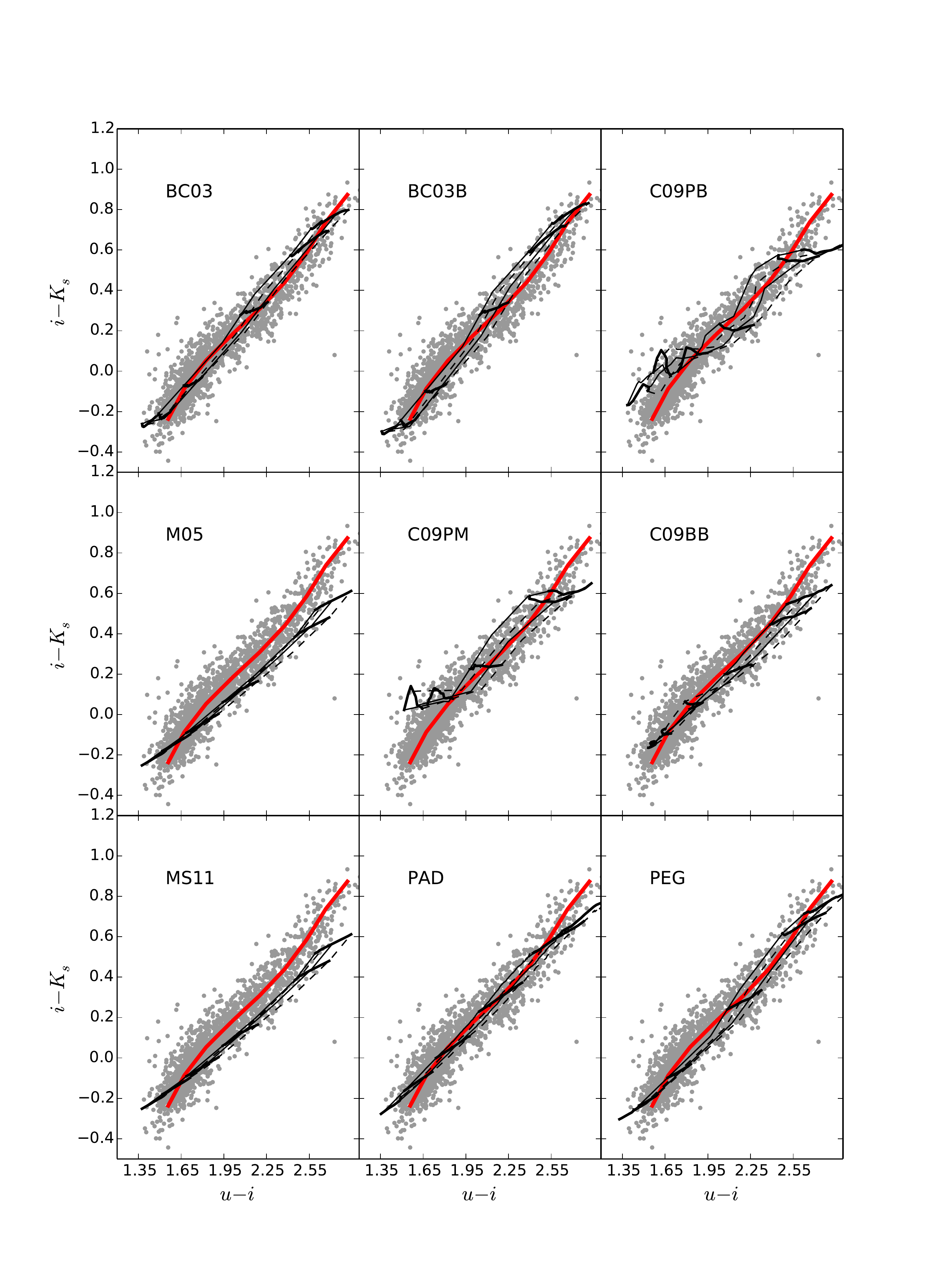} 
\caption{\label{ccd6} $uiK$ diagram. 
Symbols and lines are as in Fig.\,\ref{ccd1}. }
\end{center}
\end{figure}

\begin{figure}
\begin{center}
\includegraphics[width=9.2cm]{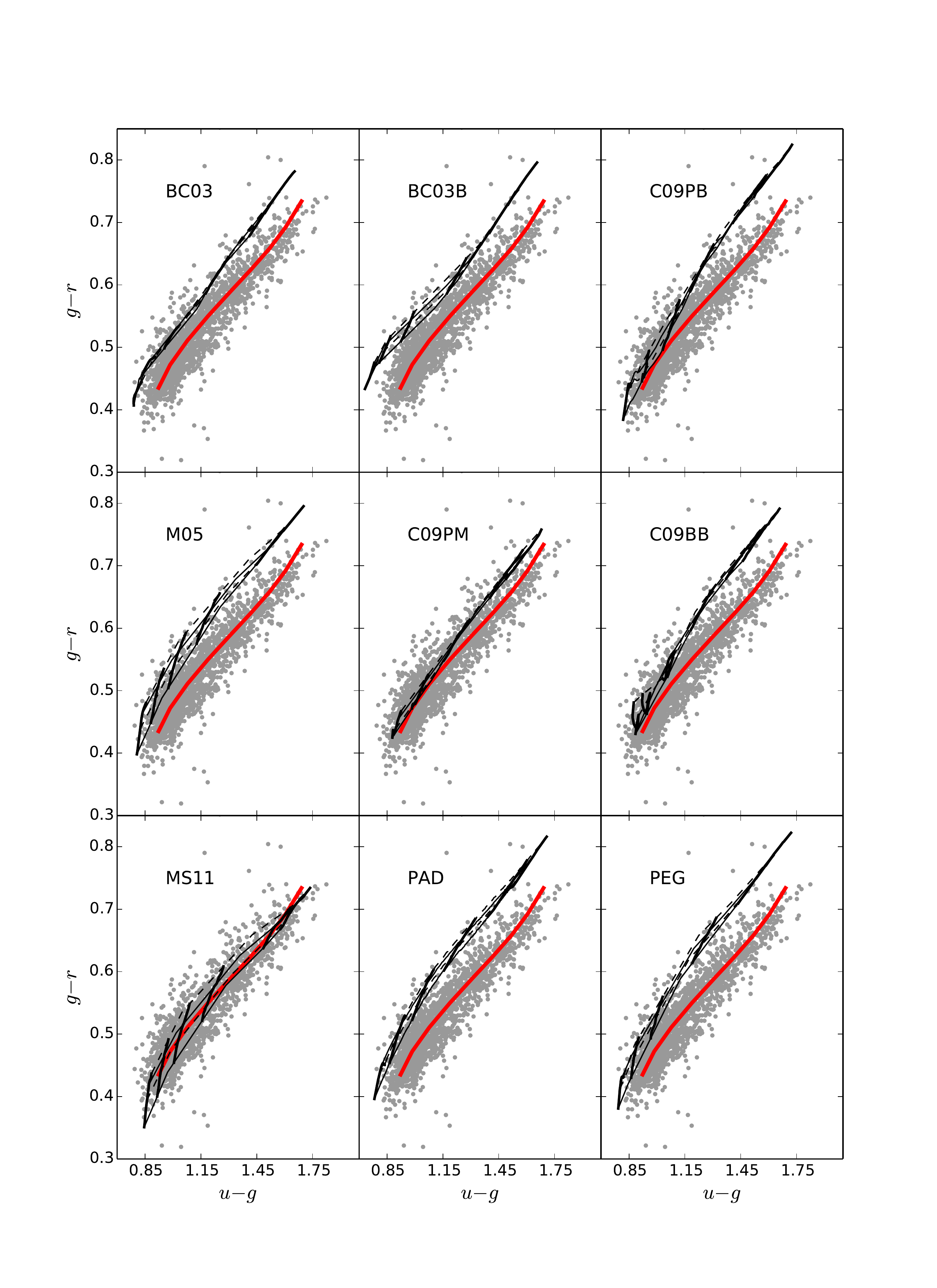} 
\caption{\label{ccd7} $ugr$ diagram.
Symbols and lines are as in Fig.\,\ref{ccd1}. }
\end{center}
\end{figure}

To study the effect of $u^*$ photometry on the relative locations of the empirical and
theoretical color distributions, we use the three color-color diagrams 
in Figs.\,\ref{ccd5}, \ref{ccd6}, and \ref{ccd7}.

The dynamic ranges of the synthetic $(u-i)$, $(u-g)$ and $(u-r)$ colors
agree well with the observed range. Any zero point errors compatible
with our error budget (including the possible 0.04 offset in $u^*$ between
SDSS and true AB magnitudes) would be small on the scale of the figures,
and would not affect any conclusion in this section.

The $ugr$ diagram (Fig.\,\ref{ccd7}) confirms that the spectral region around the
$g$ band is matched best by models built with the MILES spectral library (C09PM and MS11).
The $g$ magnitude is used in the two colors that define this diagram,
exacerbating the discrepancies already seen in $gri$ (Fig.\,\ref{ccd1}).
The majority of the models lack flux in $g$ at a given $u^*$ and $r$.

The locus of the empirical color distribution is very tight both in the $uiK$ and in
the $ugr$ diagrams, and this is reflected in the model grids.
As in previous diagrams, there are some irregularities in the
predictions of the C09 models at low metallicities, that can be traced
back to their internal interpolation procedures. Only a subset of
the models predicts that the addition of the $u^*$ band helps break
the age-metallicity degeneracy. According to MS11, this would be best done
in the $ugr$ diagram, while other models predict that the degeneracy is
best broken in $uiK$.

In summary, while many model colors are roughly satisfactory,
none of the theoretical sets we have examined, over the range
of ages and compositions we have explored, satisfactorily matches
the well-defined locus of the Virgo clusters in all the color-color diagrams. 
Each model grid has its strengths and weaknesses in the above comparison, 
and we have not found strong arguments to favour one over the others overall.

\subsection{Spectral Energy Distributions}
\label{SED}

\begin{figure*}
\begin{center}
\includegraphics[angle=90,height=23cm,width=15cm]{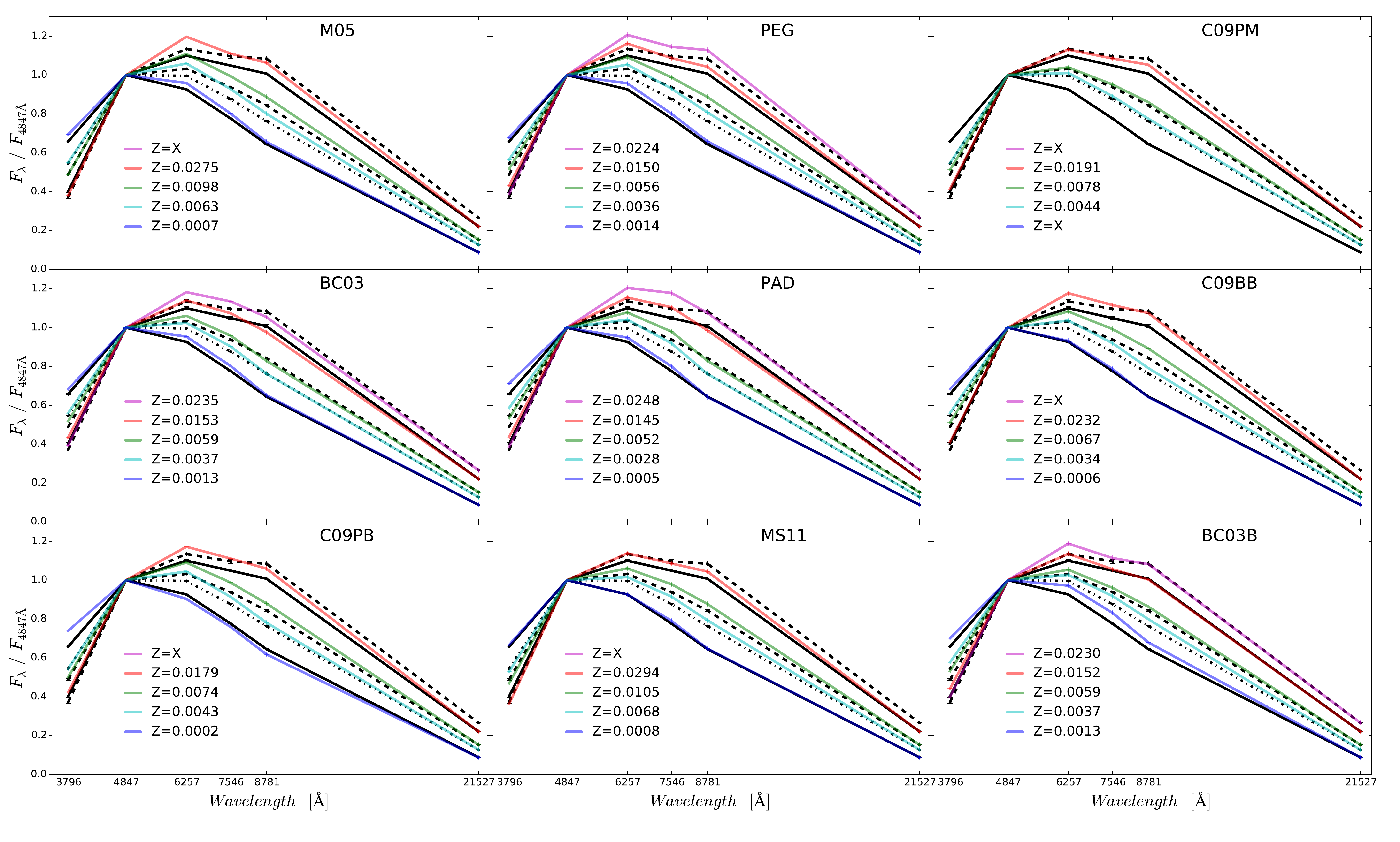}

\caption{\label{SED_plot} 
Comparison between empirical and synthetic SEDs. The fiducial 
SEDs for GCs in the Virgo core are shown in black, for 5 values of $(g-K_s)$ 
[0.6\ =\ Solid, 1.0\ =\ Dotted-Dashed, 1.2\ =\ Dashed, 1.6\ =\ Solid, 1.8\ =\ Dashed]. These SEDs are 
identical in all panels.
Model SEDs
of 10\,Gyr-old sSSPs are shown in color, for the same set of $(g-K_s)$. The associated 
(model-dependent) metallicities are listed. The special label Z\ =\ X is used 
when the required $(g-K_s)$ is not reached with a particular model set,
and no line is drawn.
}
\end{center}
\end{figure*}

While a color-color diagram provides
only two colors but for all GC ages and metallicities, one SED allows a view of a set 
of possible colors but for only one GC (of given metallicity and age). 

In Figure \ref{SED_plot} we compare the fiducial SEDs 
of Virgo core globular clusters defined in Section \ref{Prop_GC}, with nine
sets of synthetic energy distributions for 10\,Gyr-old stellar populations. 
The SEDs shown correspond to a given set of $(g-K_s)$
colors : [0.6\ =\ Blue, 1.0\ =\ Cyan, 1.2\ =\ Green, 1.6\ =\ Red, 1.8\ =\ Magenta]. 
The metallicity associated with each of the plotted models is also given, 
to facilitate the comparison between models.

A quick overview of these theoretical SEDs confirms the wide range of 
fluxes that different models can predict. These discrepancies between models and observations, which easily 
reach 10\,\%, were expected based on the inspection of the 
color-color diagrams.
They are larger at high metallicities than in the low metallicity regime. 
At low metallicities, models that match the bluest $(g-K_s)$ colors 
 tend to match also the rest of the SED. 
But this does not mean that the matching models all have the same metallicity.
As an extreme example, $(g-K_s)\ =\ 0.6$ is obtained 
with C09PB at $Z\ =\ 2\,\times\,10^{-3}$ and 
with PEG or BC03 at $Z\ \simeq\ 1.3\,\times\,10^{-2}$. The MILES-based model MS11 is intermediate.

Some sets of models do not reach $(g-K_s)$ values 
as high as 1.8\,mag for the metallicities we have computed
(in those cases less than five model SEDs are shown in Fig.\,\ref{SED_plot}). 
Our nine model grids extend to [Fe/H]\ =\ 0.17. For five of these, this is not sufficient
to reach $(g-K_s)\ =\ 1.8$ at an age of 10\,Gyr.  The reason why
$(g-K_s)\ =\ 0.6$ is not reached with C09PM is only that these models are not
available below [Fe/H]\ =\ -1.39.

In the color-color diagrams of Section \ref{ccd}, we had highlighted two main patterns: 
the relatively blue $(i-z)$ indices for the PAD and BC03 models at 
high metallicity, and the larger $g$ band flux relative to $u^*$ and $r$ 
in the C09PM and MS11 models.
Both patterns can be seen in the SEDs, by inspecting the slope of the colored lines
between $i$ and $z$, or the $u-g-r$ energy distributions.

\section[]{Discussion}
\label{discussion.sec}

\subsection{SLR calibration}
\label{SLR_res}

The comparison of the observed GC colors with models in Section \ref{sec_results} is 
based on the data calibration against SDSS and UKIDSS (Section \ref{photo_ext_cat}). 
Here we briefly discuss the effect of adopting, instead, the Stellar 
Locus Regression against synthetic stellar AB photometry (Section \ref{sec_SLR}).

The amplitude of the SLR color-corrections based on the 
PHOENIX library of stellar spectra and on stellar parameters from the 
Besan\c{c}on model of the Milky Way, is consistent with our budget of 
systematic errors for $(r-i)$ and $(i-z)$, but not for
$(g-r)$ and $(i-K_s)$. Although the PHOENIX spectral library, combined 
with the Besan\c{c}on model, performs better than the other libraries we 
have tested, we must keep in mind that it is only an approximate 
representation of the true stellar colors. Following \cite{high2009}, 
we have avoided using the $u^*$ band in the calculation of SLR 
color-corrections. We have however included the $K_s$ band. The large SLR 
corrections to $(g-r)$ and $(i-K_s)$ are driven by the stellar locus in 
the three last panels of Fig.\,\ref{SLR_difflib}. At a given $(r-i)$ (bottom panels), the 
data must be shifted to redder $(g-r)$ and $(g-K_s)$ to match the 
synthetic stellar locus. Then the plot of $(g-r)$ versus $(i-K_s)$ 
requires a significant shift in $(i-K_s)$.

The SLR color-correction vector, $\kappa$,
points towards redder colors for all color indices. Taken at face value this suggests the dereddening corrections 
we have applied to the data might be excessive. The amplitude of $\kappa$, 
however, is much larger than allowed by our estimated 
maximal uncertainties on A(V).  
In $(g-r)$ and $(i-K_s)$ the amplitude is even larger than the 
total reddening towards the Virgo core region. We doubt this correction 
would be correct.

Based on the above, we prefer the initial calibration of the data, and 
restrict the discussion of the effects of the SLR to a few main trends.
The re-calibrated data are compared with model 
grids in various color-color diagrams in Figure \ref{ccd4s}
of the Appendix.

The SLR corrections bring the Virgo core GC data closer 
to the models in the $gri$ diagram and in the $giz$ diagram 
(in the latter, one has to keep in mind 
that SDSS advocates 0.02\,mag be subtracted from $z$ magnitudes). 
The SLR-corrected colors seem to be too red compared to the models in 
the $riK$ and $gKiz$ diagrams.

In any analysis of cluster properties, the main effect of using 
SLR-corrected colors would be a higher metallicity estimate. After the 
SLR-correction, more of the red GCs find no match in the model grids, 
and the models at [Fe/H]\ =\ -2 lie outside the distribution of observed 
colors. For a further discussion of GC metallicities, we refer the reader to a following paper (Powalka et al., in preparation).

\subsection{Isochrones and libraries}
\label{isolib}

The synthetic colors of globular cluster models depend on the assumed stellar evolution prescriptions, the resulting isochrones, and
the adopted stellar spectral libraries. Our choice of models allows us to 
re-evaluate the validity of these ingredients in defining the $ugrizK$ colors of GCs in the Virgo core region.

The C09PB and the C09BB models of \citet{conroy2009} respectively use Padova 2007 and BaSTI 2007 
isochrones. This modification mainly affects the metal poor GCs in the 
$gri$, $giz$ and $riK$ diagrams (Figs.\,\ref{ccd1}, \ref{ccd2} and \ref{ccd3}), 
with resulting differences in color that reach 0.05 magnitudes. 
With that population synthesis code, the Padova 2007 isochrones produce 
a wider spread in colors than the BaSTI 2007 isochrones.
The BC03 model can also be computed with two different sets of isochrones,
which are Padova isochrones from 1993-1994 (shown in our figures) 
and Padova isochrones of \citet{girardi2000}, not shown). 
The resulting differences are below 0.05 magnitudes in all the colors in our diagrams.  

Now turning to the spectral libraries, we can quantify the effects of replacing
the BaSeL library with the MILES library at optical wavelengths, 
by comparing C09PB and C09PM
in the $gri$ diagram (Fig.\,\ref{ccd1}). The offset in $(g-r)$ is about 0.05 magnitudes. 
Between BC03 and BC03B in the same diagram, the effect is of similar
amplitude for red clusters, but smaller than in the Conroy implementation for blue clusters.
In the $gKiz$ diagram however (Fig.\,\ref{ccd3}) the replacement of one library with the
other has larger impact in the implementation of Bruzual \& Charlot than with the 
code of Conroy et al. Here, the critical issue is the algorithm used to connect 
the purely optical MILES spectra with those of the panchromatic BaSeL one.

The PEG and BC03B predictions are based on very similar ingredients, and in general 
they agree rather well (color-color diagrams, and SEDs of Fig.\,\ref{SED_plot}). Differences in the
details show the impact of particular implementation choices. Examples are the
interpolation scheme between the few metallicities for which Padova isochrones are
available, the interpolations between the spectra available in the original BaSeL 
library, or the detailed treatment of the TP-AGB. These numerous differences are
known to produce significant effects on spectra, and make it very difficult to 
trace back differences in colors to a single physical origin (see \citealt{koleva2008} for
a discussion of these subtle effects on the analysis of 
Milky Way globular cluster spectra).

Our inspection of the various models in color-color space does not allow us to point 
to a single ``best model" that would, within the range of ages and metallicities we have
considered, match the shapes and locations of the empirical distributions in all projections in a 
statistically acceptable way. However, we confirm that the careful flux calibration 
of the MILES library improves the modelling of optical spectra over pre-existing libraries. 
The BaSeL library on the other hand has the interesting property that it helps reaching
the reddest $(i-z)$ colors seen at high metallicity in the Virgo core region,
i.e. around M87. Direct consequences of the adopted stellar evolution prescriptions
are smaller, or at least more difficult to isolate, than these effects
of the spectral libraries.

\subsection{Width of the NGVS GC distribution}
\label{width_dis}

The width of the locus of the GCs in the multi-dimensional color space is larger than can be explained by the random
photometric errors. Several physical causes probably all contribute to 
this dispersion to some degree.

First, the observed GCs may have a range of ages. As we have shown in various 
color-color diagrams, the age-metallicity degeneracy is strong
and the dispersion expected from ages spread between 6 and 13\,Gyr is insufficient 
to explain the dispersion in the observed GC colors. One way to broaden the 
predicted distribution is to include ages younger than 6\,Gyr. Indeed, a fraction of the
clusters in the core of Virgo might have been born in relatively recent gas-rich
interaction events, as a side-effect of the hierarchical merging that progressively 
built up the stellar mass in the center of the galaxy cluster. 

The stochastic sampling of the stellar mass function in each cluster is 
another source of natural dispersion at a given age and metallicity.
The clusters selected in this paper are massive ($>10^6$\,M$_{\odot}$), 
hence these stochastic effects are small. Based on models by \citet{fouesneau2010}, we find
that they would be strongest in colors involving the $u^*$ and the $K_s$ filters,
and that they remain of the order of 0.01 magnitude or less for the ages and masses of interest here.
This does not suffice to explain the observed color spread.

To obtain even more variety, we may invoke more complex star formation histories
than a single chemically homogeneous star formation episode, 
or reconsider other simplifying assumptions of the models.
In the Milky Way, the assumption of a single age and a chemically 
homogeneous stellar population breaks down for essentially all the 
massive globular clusters that have been studied in
detail. Hence it is likely that massive clusters in Virgo would also host 
more complex populations than we have assumed here. 

Internal spreads in age and 
metallicity by themselves lead to modified integrated colors. Changes in
abundance ratios significantly add to this diversity, for instance via the effects
of CNO abundances on molecular bands in red giant spectra, or via the effects of
helium abundances on horizontal branch temperatures. 
The models of \citet{maraston2011} with a blue horizontal branch do not 
explain the observed colors as a whole better than the models with a red horizontal branch.
Using toy models constructed with {\sc P\'egase}, for which we artificially varied the 
temperature of horizontal branch stars, we reach the same conclusion.
On the other hand recent HST-based UV studies of M87 established the presence of hot stellar populations 
in its globular clusters (\citealt{sohn2006} and \citealt{bellini2015}), which is an 
indication for the presence of hot horizontal branch stars.
Consequently, a mix of horizontal branch morphologies is likely to contribute to 
the dispersion in colors we have seen. 
Similarly, varying fractions of blue stragglers, or
stellar rotation statistics, or strong changes in the stellar mass function will spread out the colors.

Finally, we note our GC sample in the Virgo core region combines objects located around M87, 
around smaller Virgo galaxies such as M86 or NGC 4438, and in the intracluster region of the Virgo core. 
The detailed properties of each of these subpopulations differ, indicating a link with environment
that we will describe in a subsequent article (Powalka et al., in preparation). 
Previous studies hinted at systematic differences between GC subsets. 
For instance, \citet{harris2009} showed  
that the typical color of the blue subsample of clusters around elliptical galaxies depends both on the 
distance to the galaxy center and on the GC luminosity, 
and \citet{forte2013} suggest the blue and red GC subpopulations of M87 
follow distinct color-color relations. The respective
role of age, metallicity and additional parameters in explaining these
remains to be clarified.

\subsection{Stellar population properties derived from colors}
\label{error_source}

The well-defined color-locus of the brightest Virgo core GCs has
allowed us to illustrate the amplitude of the variation in the predictions from different sSSP models.
Nevertheless, colors
will remain easier to measure than spectra for remote objects in the universe, 
and it therefore remains desirable to provide color-based estimates 
of stellar population properties such ages and metallicities.

In a subsequent paper, we will present SP properties of the Virgo core 
clusters based on a Bayesian analysis (Powalka et al., in preparation).
Using five colors that cover wavelengths from the $u^*$ to $K_s$, we find that the 
metallicities derived using each of the SSP models listed in this paper individually,
can differ systematically from one another by up to 0.5\,dex. 
An analysis that combines several models
allows to narrow the range of likely values significantly, although
the accuracy remains dependent on the absolute adequacy of the selected models.
The population synthesis models themselves clearly require more work
in the future.

\section{Conclusions}
\label{conclusion.sec}

In this paper, we have used the Next Generation Virgo Survey to provide 
near-ultraviolet to near-infrared colors for a representative sample of 
luminous GCs in the Virgo core. The sample was selected in a plane that combines the color
information of the $uiK$ diagram and a compactness index, measured on the
NGVS images with the best seeing ($i$ band). 
This hybrid plane separates globular clusters from foreground stars 
and background galaxies. Careful aperture corrections 
were applied to remove any systematic differences in colors due to 
seeing-variations between camera pointings.

The selection was designed to provide precise colors 
across the spectrum rather than ensure volume completeness. 
Hence the 1846 objects in the catalog are relatively bright objects, with typical masses just above $10^6$\,M$_{\odot}$.
The catalog defines a characteristic 
locus in color-color space, from which we derive fiducial SEDs that will be
useful for future quick comparisons with other samples\footnote{Detailed 
studies should use the full catalog rather than the fit.}.

As an accurate photometric zero point is essential 
for comparisons with other data sets or with model predictions, 
we have decided to provide and test two calibrations. 
The first is based on a set of stars common between SDSS and NGVS,
or between UKIDSS and NGVS-IR for the near-infrared.  
The second calibration is computed using 
a PHOENIX synthetic stellar library as a reference locus (Stellar Locus Regression). 
The PHOENIX library nicely matches the shape of the NGVS stellar locus
(better than other libraries we have tested). However, residual differences
between empirical and theoretical colors in the regime of low mass M dwarfs
drive suspiciously large zero point corrections in $(g-r)$ and $(i-K_s)$. 
Hence we prefer the first calibration.

To quantify the accuracy of our photometry, 
we have also assessed the systematic errors in the GC data. 
They primarily reflect uncertainties in the zero points of SDSS and UKIDSS, 
as well as uncertainties in the extinction towards the Virgo core.
Our random errors on the mean colors along the GC color distributions
are negligible compared to the systematic errors.

Accounting for our random errors, we have shown that the GC locus in 
color-color space cannot be explained by a single line. 
The width of the observed color-distribution is real.
Part of this spread could be due to age differences.
Other potential causes for the spread include varying binary fractions between clusters, 
varying stellar rotation statistics, the co-existence of sub-populations of stars
with different surface chemistries within each cluster, or the presence of 
subsets of clusters with different chemical peculiarities in our sample.
We will link color differences to the cluster environment in future work.
\medskip

In the second part of this paper, we have compared the
colors in our globular cluster catalog with 11 commonly used models
for single age, single composition stellar populations,
with ages between 6 and 13\,Gyr and [Fe/H] between $-2$ and 0.17. 
Despite a rough global agreement, 
we have shown that none of the 11 models provides a statistically satisfactory
match to the NGVS data. Moreover, this comparison has highlighted patterns 
in the predictions around the $z$ and $g$ bands which are likely related to the choice of 
stellar spectral library made in the models. 

Finally, we have not provided in this paper any ranking of the models
based on their ability to match the Virgo core GC data. This ranking would depend on the exact set of 
colors used in the assessment, and also on the method used to measure goodness-of-fit.
More importantly, it would convey the misleading impression that we recommend
using one particular model and discarding the others. The ability of any of the
models to represent the data is currently too imperfect to justify such a radical 
approach. Also, we will show in subsequent work that GCs in other environments than 
the Virgo core may have different color distributions, thus requiring a new assessment of the models. 

We hope that these data will help for future studies and to do so, 
we make them available in the catalog service of the Centre de Donn\'ees astronomiques de Strasbourg (CDS).



\acknowledgments

The authors gratefully acknowledge fruitful discussions with 
M. Betoule, N. Regnault, H.J. McCracken, R. Peletier, R. Ibata, N. Martin,
A. Nebot, E. Schlafly, P. Prugniel, S.C. Trager. 
Direct information on some model details was kindly provided by C. Maraston,
S. Charlot, G. Bruzual, A. Vazdekis, L. Girardi.

This work was supported in part by the French Agence Nationale de la Recherche (ANR) 
grant Programme Blanc VIRAGE (ANR10-BLANC-0506-01), and
research (CANFAR) which has been made possible by funding from CANARIE under the Network-Enabled Platforms program.
This research also used the facilities of the Canadian Astronomy Data Centre operated by the National Research Council of Canada with the support of the Canadian Space Agency. 
E. W. P. acknowledges support from the National Natural Science Foundation of China under Grant No. 11573002, and from the Strategic Priority Research Program,
``The Emergence of Cosmological Structures,” of the Chinese Academy of Sciences, Grant No. XDB09000105.
C.L. acknowledges the National Key Basic Research Program of China (2015CB857002), NSFC grants 11203017, 11125313.
E.T. acknowledges the NSF grants AST-1412504 and AST-1010039

Funding for SDSS-III has been provided by the Alfred P. Sloan Foundation, the Participating Institutions, the National Science Foundation, and the U.S. Department of Energy Office of Science.
The SDSS-III web site is {\tt{http://www.sdss3.org/}}.
SDSS-III is managed by the Astrophysical Research Consortium for the Participating Institutions
of the SDSS-III Collaboration including the University of Arizona, the Brazilian Participation Group,
Brookhaven National Laboratory, Carnegie Mellon University, University of Florida, the French Participation Group,
the German Participation Group, Harvard University, the Instituto de Astrofisica de Canarias, the Michigan State/Notre Dame/JINA Participation Group,
Johns Hopkins University, Lawrence Berkeley National Laboratory, Max Planck Institute for Astrophysics,
Max Planck Institute for Extraterrestrial Physics, New Mexico State University, New York University, Ohio State University,
Pennsylvania State University, University of Portsmouth, Princeton University, the Spanish Participation Group, University of Tokyo,
University of Utah, Vanderbilt University, University of Virginia, University of Washington, and Yale University.

This work is based in part on data obtained as part of the UKIRT Infrared Deep Sky Survey.

This research has made use of the VizieR catalogue access tool and the Aladin plot tool at CDS, Strasbourg, France, as well as the TOPCAT software available at {\tt{http://www.starlink.ac.uk/topcat/}}.






\appendix

\section{Aperture corrections}
\label{apcormap.app}

\begin{figure}
\begin{center}
\includegraphics[clip=,width=14cm]{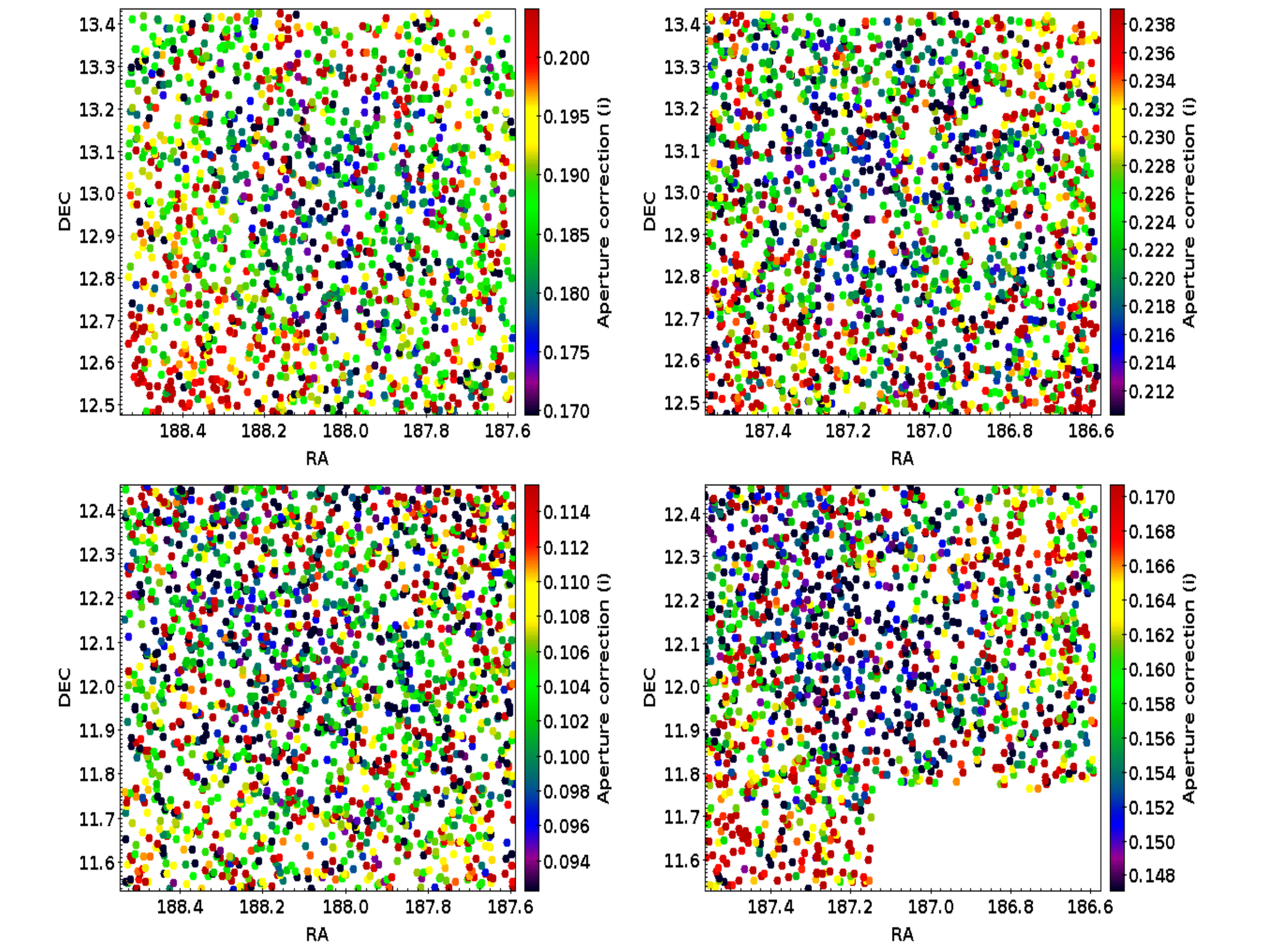}
\caption[]{Illustration of the spatial variations of the point source 
aperture corrections in $i$ within individual pointings. From left to right, then
top to bottom, the fields correpond to NGVS MegaCam pointings +0+1, -1+1, +0+0, and 
-1+0 \citep{ferrarese2012}. }
\label{iapcormap.fig}
\end{center}
\end{figure}

Figure \ref{iapcormap.fig} illustrates the spatial distribution 
of the aperture corrections applied in the $i$ band. Note that this figure differs 
from the one in \citet{liu2015}. Here, each star is colored using the difference between
magnitudes in two SExtractor apertures, without attaching the ``infinite" aperture
estimate to SDSS PSF-magnitudes (as was done in the figure of \citealt{liu2015}).
The varying ranges of the color scales of the four panels shows that the average aperture 
correction differs between pointings, due to differences in average seeing. 
Similar amplitudes of spatial patterns, and a similar spread between pointings,
are seen in the other photometric passbands, with one exception in the $r$ band
described below.

\begin{figure}
\centering
\begin{minipage}{0.45\linewidth}
\begin{center}
\includegraphics[clip=,width=8cm]{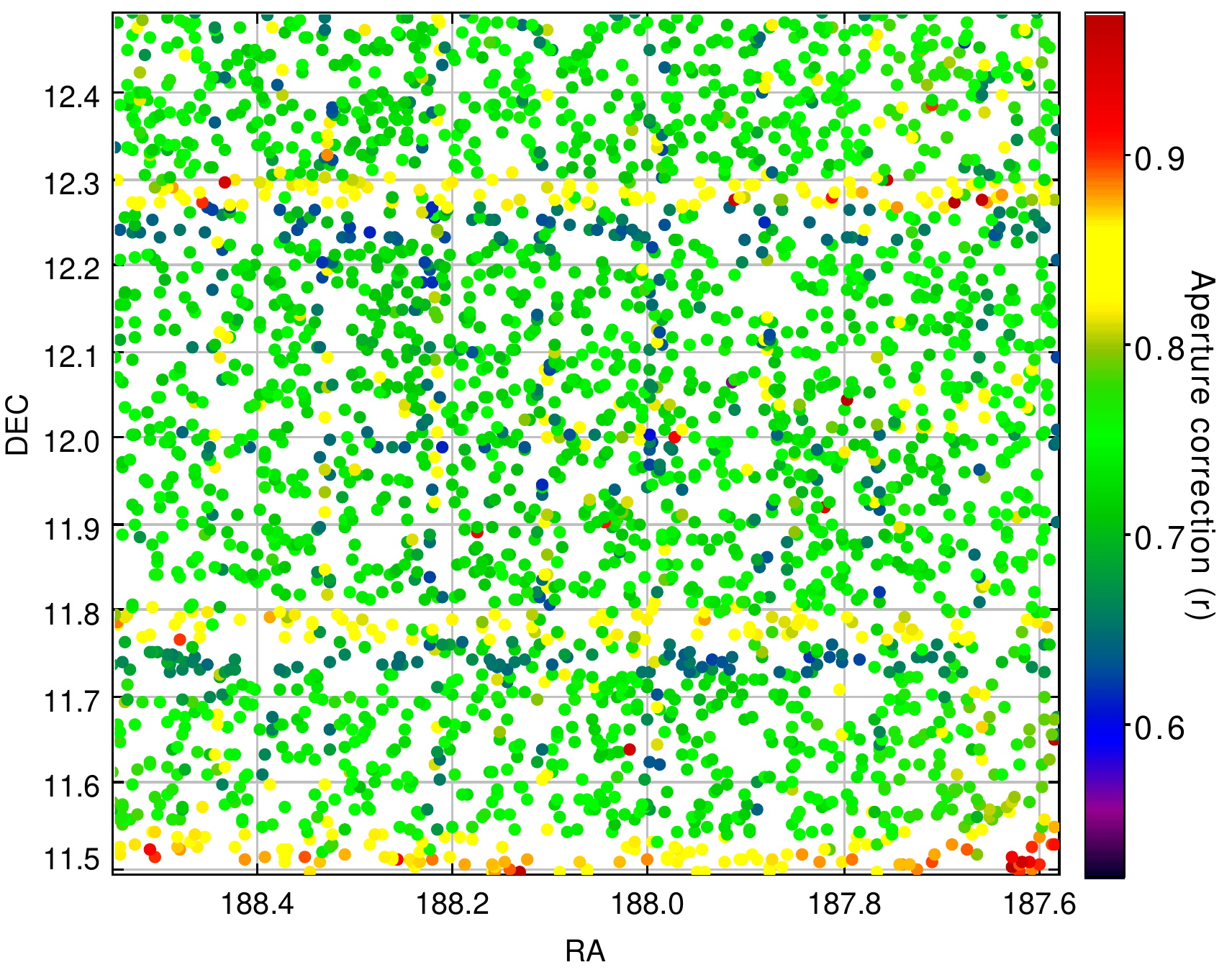}
\caption[]{Illustration of the spatial variations of the point source 
aperture corrections in $r$, for NGVS pointing +0+0 (see \citealt{ferrarese2012} for
pointing numbering). Color maps the difference between magnitudes
measured in apertures of 4 and 8 pixels diameter (0.186\arcsec /pix).}
\label{rapcormap.fig}
\end{center}
\end{minipage}
\begin{minipage}{0.45\linewidth}
\begin{center}
\includegraphics[clip=,width=8cm]{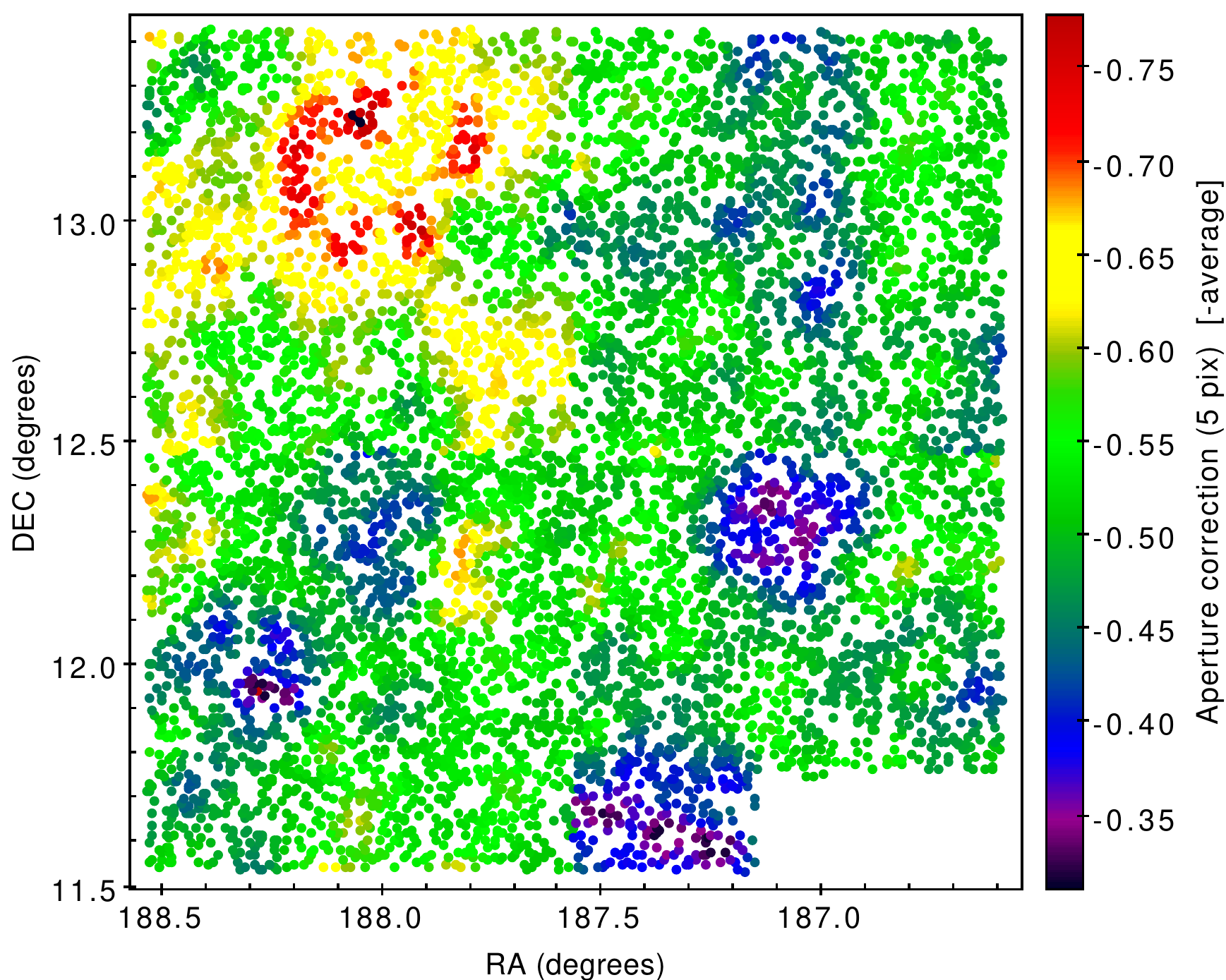}
\caption[]{Aperture corrections for $K_s$ point source photometry 
across the whole Virgo core area (thirty-four WIRCam pointings covering
four MegaCam pointings).}
\label{Kapcormap.fig}
\end{center}
\end{minipage}
\end{figure}

 Figure \ref{rapcormap.fig} shows the aperture correction in NGVS pointing +0+0
\citep{ferrarese2012} in the $r$ band. In this one particular case, unfortunately, the seeing distribution of 
individual stacked images was much broader than usual. As a consequence, aperture
corrections within the gaps between rows of detector chips differ more from
neighbouring values than in typical cases.

Figure \ref{Kapcormap.fig} maps the aperture corrections for $K_s$ photometry
across the 34 WIRCam pointings of the NGVS pilot field ($6\times6 -2$). 
Seeing differences between WIRCam pointings are the dominant cause of structure in this map,
hence the distribution is comparable to the seeing maps shown in Fig.\,8 of
\citet{munoz2014}.

\section{Complementary information on the SLR calibration}
\label{aslr}

\subsection{UV issues}
\label{u_section}

The $u^*$ band raises stronger calibration issues than the others when using SLR
\citep{high2009}.  The $u^*$ transmission extends across
the Balmer jump of stellar spectra. In this region, spectra are
particularly sensitive to stellar parameters such as temperature,
gravity and also metallicity. This is seen clearly 
in the locus of the synthetic stars in color-color planes involving
$u^*$.

\begin{figure*}
\begin{center}
\includegraphics[width=16cm]{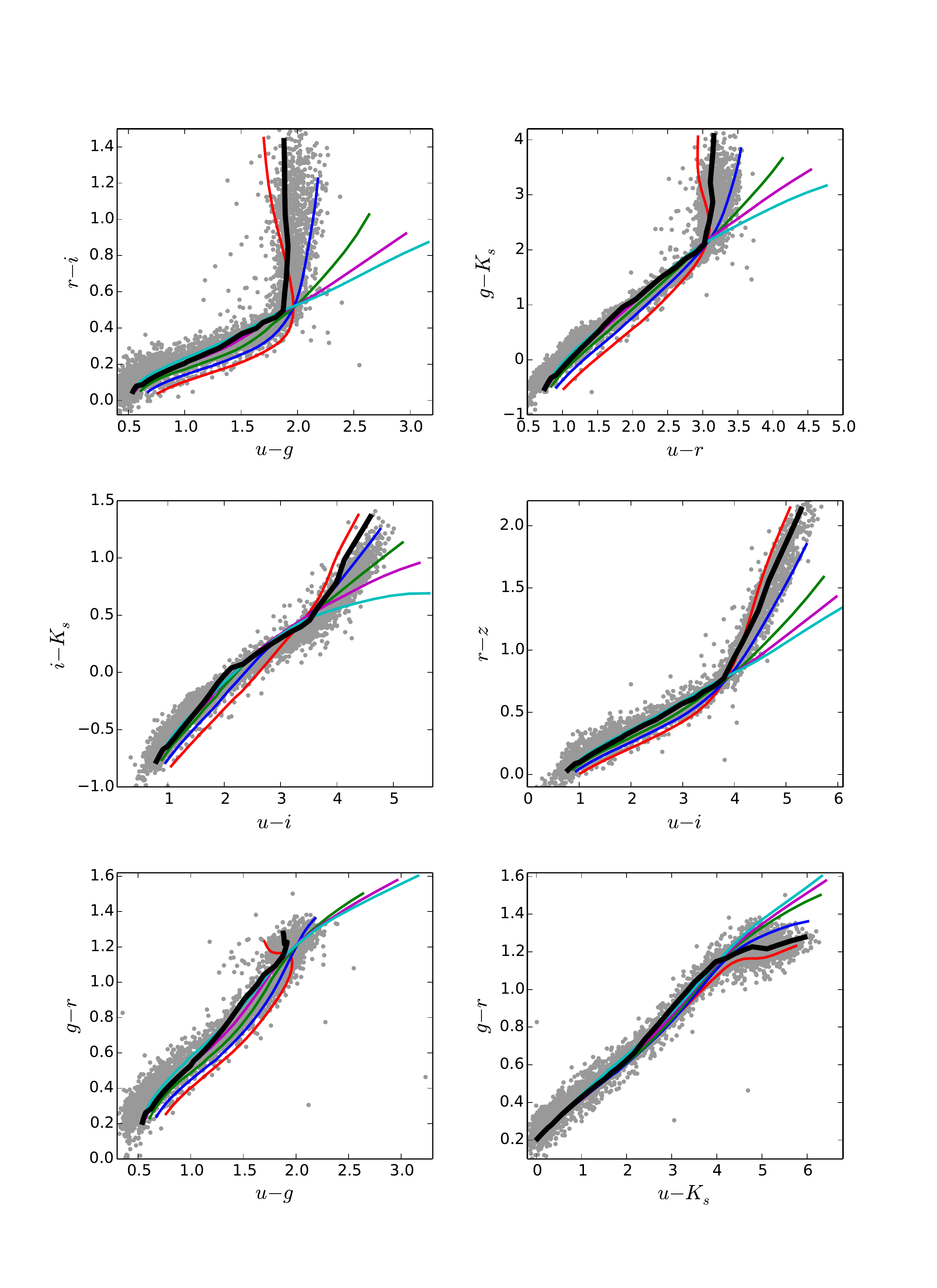}
\caption{\label{u_prob} Color color diagrams which highlight the difficulty of including the $u^*$ filter in the SLR procedure. In gray we show the NGVS stars shifted by the SLR correction vector (best $u^*$ correction included).
The thin lines represent the PHOENIX stars from 3100\,K to 6500\,K, with
log(g)\ =\ 4.5, $[\alpha/Fe]$\ =\ 0.0 and different metallicities
(red\ =\ [Fe/H]\ =\ 0.0, blue\ =\ -0.5, green\ =\ -1, magenta\ =\ -1.5, cyan\ =\ -2). The large black line is the one computed using the PHOENIX library and the Besan\c{c}on model, it is also the one used to compute the applied SLR shifts.
In some diagrams that include $u^*$, the shape of this black locus does not match the observed locus.}
\end{center}
\end{figure*}

Figure \ref{u_prob} shows the effective temperature sequences of
synthetic stars at a given surface gravity, for 
five metallicities (red\ :\ [Fe/H]\ =\ 0.0, blue\ :\ -0.5, 
green\ :\ -1, magenta\ :\ -1.5, cyan\ =\ -2), in six such color-color diagrams.
These sequences illustrate the large sensitivity of the colors to metallicity, 
especially at low effective temperatures (red end of the sequences), thus
creating a strong dependence on the assumed distribution of metallicities
along the observed sequence (i.e. the model for the M dwarf population of the
Milky Way disk and halo). Figure \ref{SLR_difflib} in this paper already showed that the
colors of the lower main sequence are also highly dependent on the adopted 
collection of theoretical spectra; this remains true for the $u^*$ band.

The thick black line in Figure\ \ref{u_prob} shows the locus 
defined by the Besan\c{c}on model of the Milky Way towards Virgo. 
While the match to the NGVS data 
(shifted here by the vector of SLR shifts obtained when including $u^*$ in
the optimization) seems reasonable to the eye, 
an examination of the actual scales of the offsets to the
locus of the Virgo stars is alarming: they amount
several tenths of a magnitude in places.
In the $ugr$ or the $uiK$ diagram,
the preferred model locus does not have a shape compatible with the
observed stellar locus, which indicates that either the filter
transmissions involved are not known well enough or the stellar models,
including the choice of stellar parameters along the sequence,
are not optimal. In view of the estimated uncertainties on the
filter transmissions, the latter reason is believed to be predominant.

As a consequence, we have not used $u^*$ in the calculation
of SLR corrections for GCs.

\subsection{GC color-color diagrams based on the SLR calibration}

\begin{figure}
\begin{center}
\includegraphics[height=10.5cm]{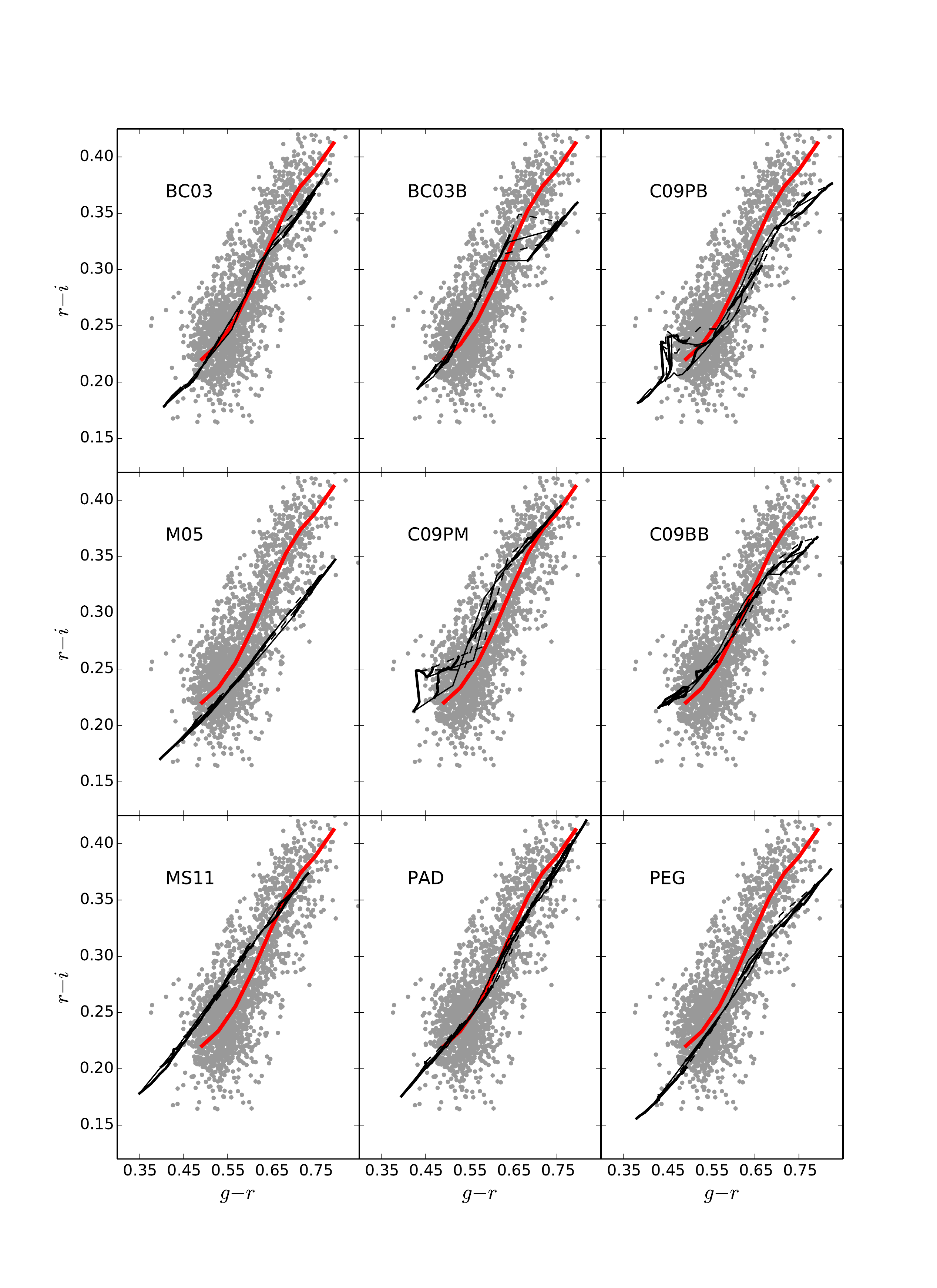}
\includegraphics[height=10.5cm]{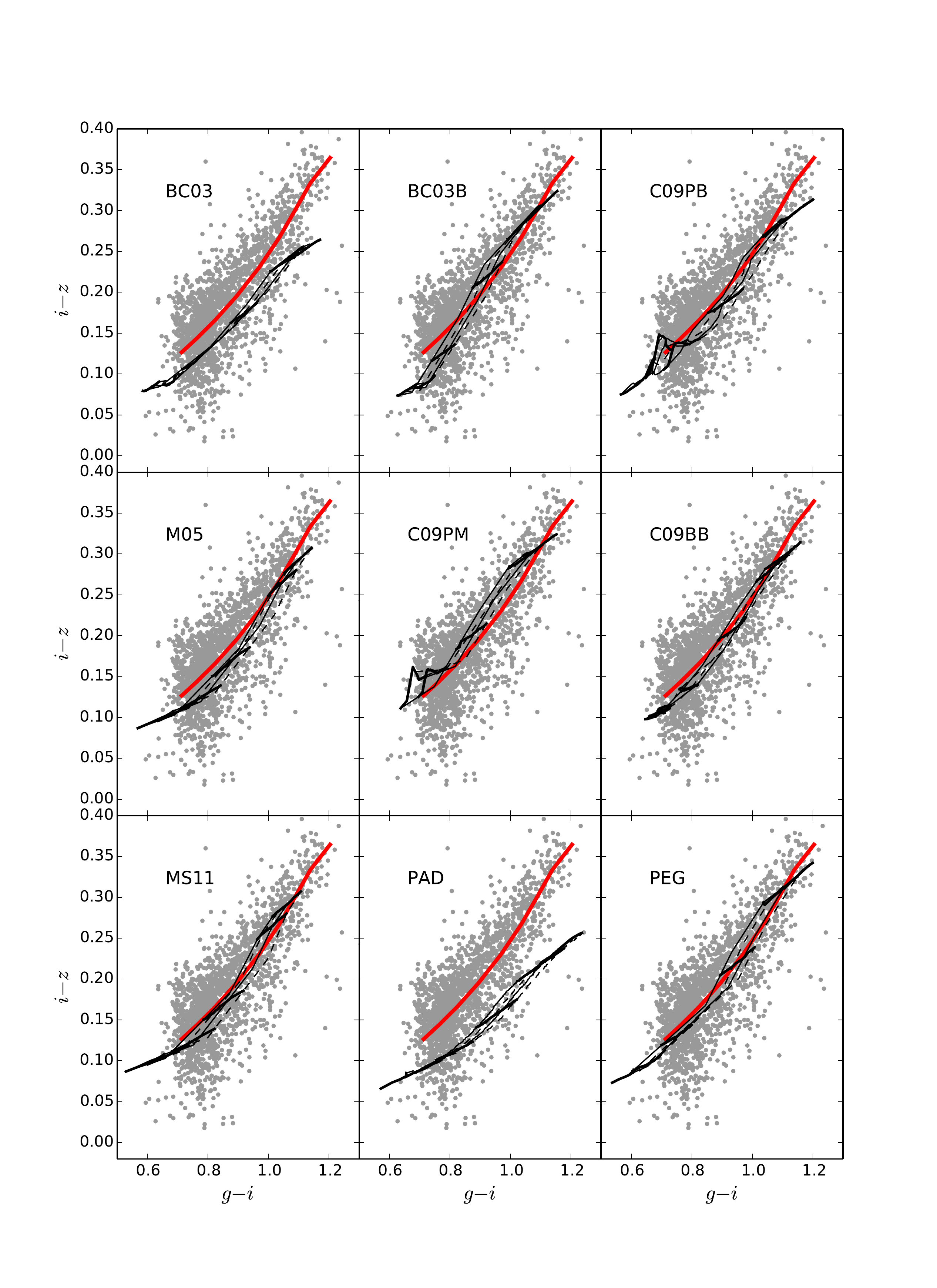}
\includegraphics[height=10.5cm]{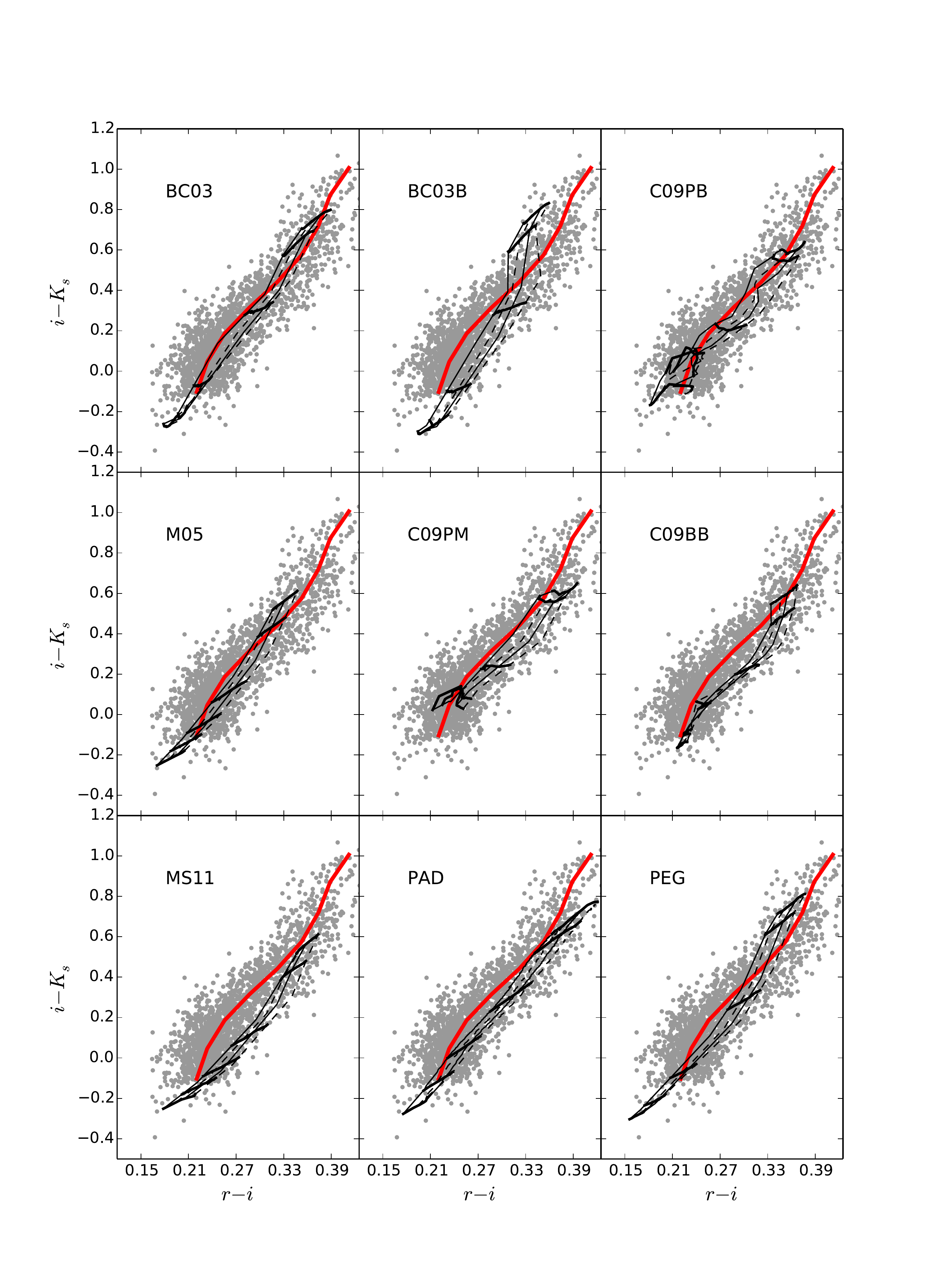}
\includegraphics[height=10.5cm]{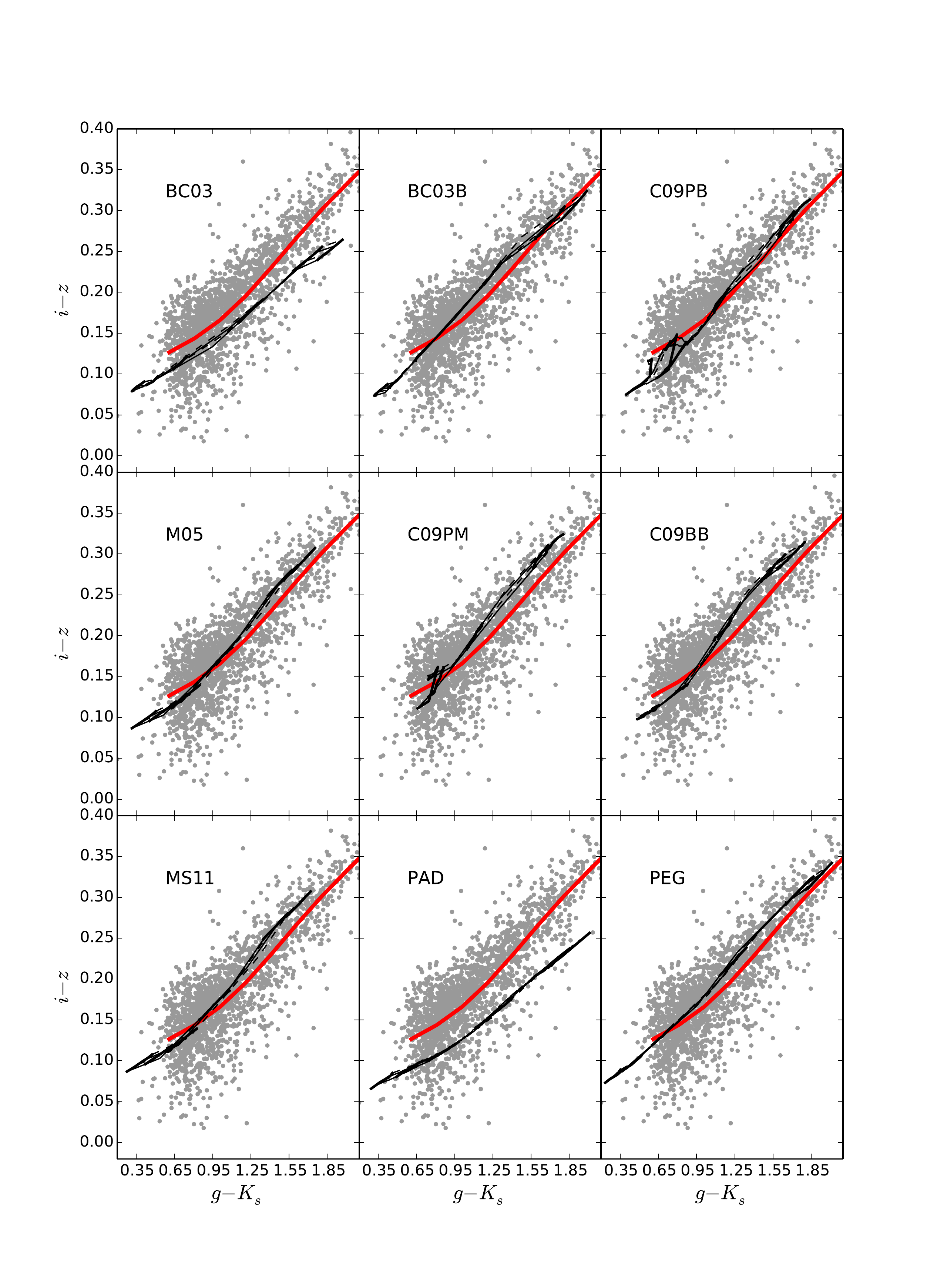}
\caption{\label{ccd4s} Virgo core GC color-color diagrams with SLR-based photometry. 
The symbols and lines are as in Fig.\,\ref{ccd1}.}
\end{center}
\end{figure}

The SLR-based shifts move the NGVS colors of globular colors (and other objects) to redder values. 
Four color-color diagrams obtained with the SLR calibration are shown in Fig.\,\ref{ccd4s}, and briefly 
discussed in Section\ \ref{SLR_res}.

\section{Additional Color-Color diagrams}
\label{app.ccd}

\begin{figure}
\begin{center}
\includegraphics[height=10.5cm]{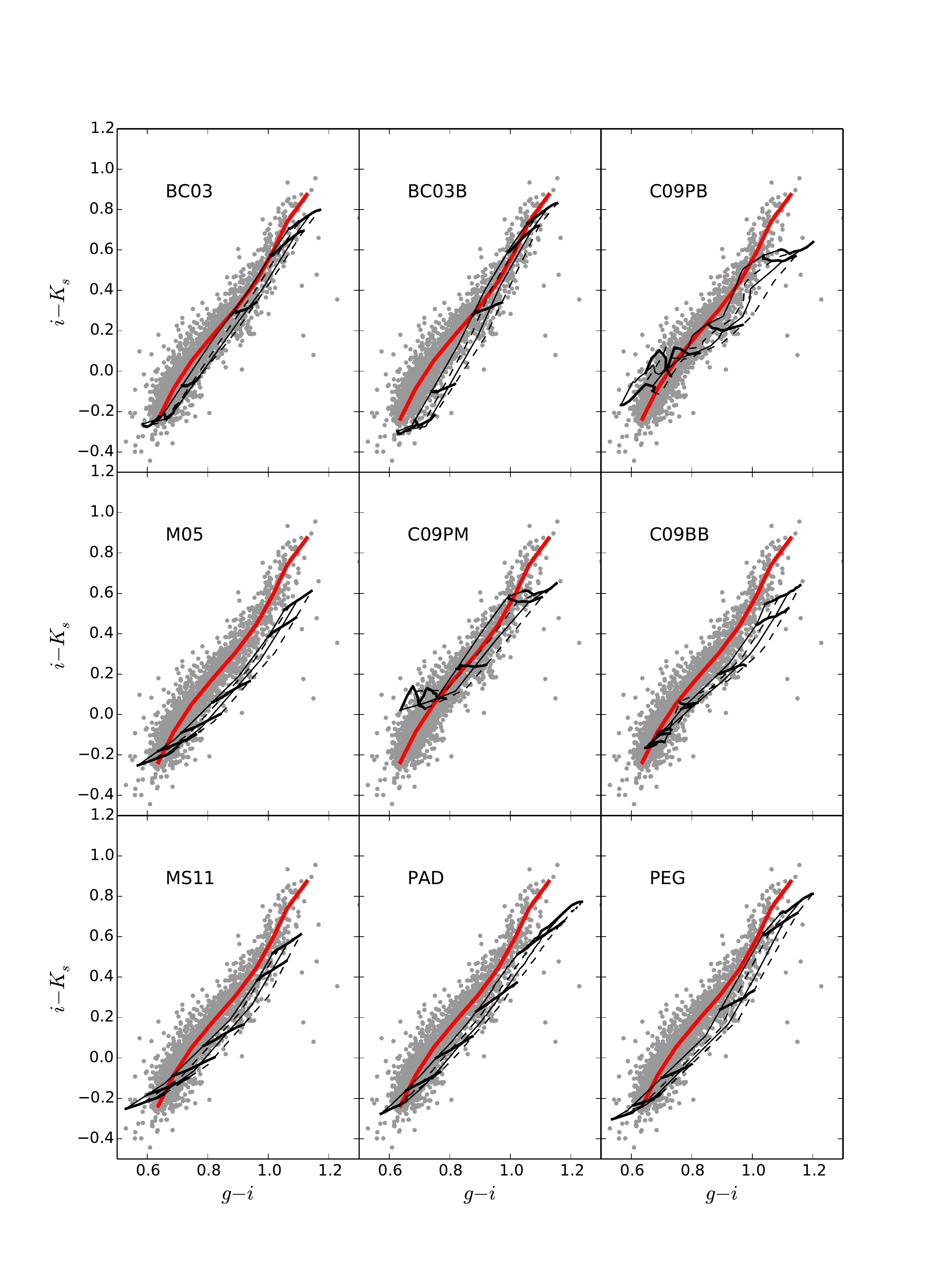} 
\includegraphics[height=10.5cm]{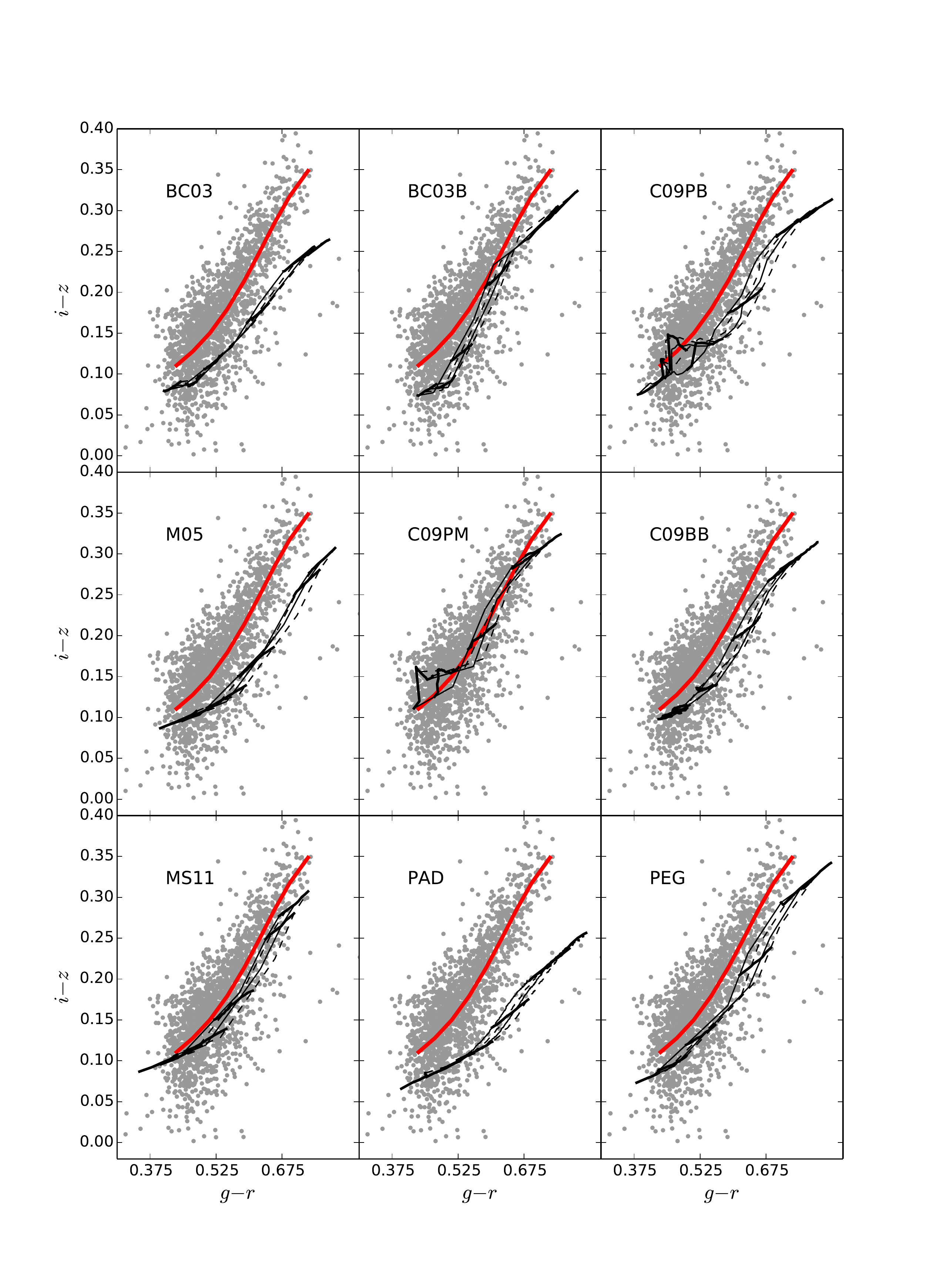}
\caption{\label{ccd_appgik} The $giK$ and $griz$ diagrams of Virgo core GCs, with NGVS photometry 
calibrated against external catalogs. The symbols and lines are as in Fig.\,\ref{ccd1}.}
\end{center}
\end{figure}

 For the convenience of future comparison with other data sets, 
we provide two additional color-color diagrams in Figure\ \ref{ccd_appgik}. Like those discussed
in the main body of the article, they are based on the calibration against external
catalogs, described in Section\ \ref{photo_ext_cat}.

\bibliographystyle{apj.bst}  
\bibliography{biblio} 

\clearpage

\end{document}

%% file: ext_table1.tex
\begin{table}
\caption{Example of all the photometric parameters available for the NGVS GCs catalog.}
\label{catalog_paper}
\begin{center}
\begin{tabular}{c|c|c}
\hline
RA & 187.546 & Right ascension\\
DEC & 13.166 & Declination\\ 
E(B-V) & 0.023     & Exctinction\\
\cline{1-2}
u*mag\_ap8 & 23.713    & \\
gmag\_ap8 & 22.236    & \\
rmag\_ap8 & 21.332    & aperture corrected magnitude \\
imag\_ap8 & 20.900    &  based on 8 pixel aperture\\
zmag\_ap8 & 20.634     & \\
Ksmag\_ap8 & 20.105     & \\
\cline{1-2}
u*mag\_ap8\_0 & 23.608   & \\
gmag\_ap8\_0 & 22.154     & \\
rmag\_ap8\_0 & 21.271    & aperture corrected magnitude, \\
imag\_ap8\_0 & 20.853     & dereddened  \\
zmag\_ap8\_0 &  20.599     & \\
Ksmag\_ap8\_0 & 20.096    & \\
\cline{1-2}
u*err\_ap8 & 0.039    & \\
gerr\_ap8 & 0.011     & \\
rerr\_ap8 & 0.008    & 1$\sigma$ error on magnitude\\
ierr\_ap8 & 0.011    & in 8 pixel aperture \\
zerr\_ap8 & 0.016    &\\
Kserr\_ap8 & 0.051   & \\
\cline{1-2}
(u*-g)\_cal & 1.454    & \\
(g-r)\_cal & 0.883     &  \\
(r-i)\_cal & 0.418    &  color, dereddened \\
(i-z)\_cal & 0.254     &\\
(i-Ks)\_cal &  0.757     & \\
\cline{1-2}
(g-r)\_cal.slr & 0.941     & \\
(r-i)\_cal.slr & 0.437    &  color, dereddened,\\
(i-z)\_cal.slr & 0.270    & after re-calibration with SLR\\
(i-Ks)\_cal.slr & 0.890    & \\

\hline\end{tabular}
\end{center}
\label{catalog_ngvs} {\sc{Notes}}: Magnitudes are aperture corrected, based on measurements in apertures of 8 pixels diameter (1.48\,\arcsec), and calibrated as in
Section\ \ref{photo_ext_cat}. Errors include the correction factors given in Section\ \ref{sec_ptsourcephot}. Only the last four colors are based on SLR color calibration
(Section\ \ref{sec_SLR}).

\end{table}

%% file: ext_table2.tex
\begin{table*}
 \caption{Fiducial SEDs for the GCs in the Virgo core region.}
 \label{sed_ngvs_full}
 \begin{center}
\begin{tabular}{c|cccccccccc}
\hline
  \multicolumn{1}{c|}{ } &
  \multicolumn{1}{c}{Fu*/Fg} &
  \multicolumn{1}{c}{Fg/Fg} &
  \multicolumn{1}{c}{Fr/Fg} &
  \multicolumn{1}{c}{Fi/Fg} &
  \multicolumn{1}{c}{Fz/Fg} &
  \multicolumn{1}{c}{FKs/Fg} &
  \multicolumn{1}{c}{$u-g$} &
  \multicolumn{1}{c}{$g-r$} & 
  \multicolumn{1}{c}{$g-i$} & 
  \multicolumn{1}{c}{$g-z$} \\
   
\hline
  $g-K_s$=0.4 & 0.697 & 1.000 & 0.896 & 0.741 & 0.606 & 0.073 & 0.923 & 0.435 & 0.636 & 0.747 \\
 $g-K_s$=0.6 &  0.658 & 1.000 & 0.928 & 0.777 & 0.645 & 0.088 & 0.985 & 0.472 & 0.687 & 0.814 \\
 $g-K_s$=0.8 &  0.604 & 1.000 & 0.960 & 0.821 & 0.696 & 0.106 & 1.078 & 0.511 & 0.747 & 0.897 \\
 $g-K_s$=1.0 &  0.545 & 1.000 & 0.996 & 0.877 & 0.764 & 0.127 & 1.190 & 0.550 & 0.818 & 0.998 \\
 $g-K_s$=1.2 &  0.489 & 1.000 & 1.032 & 0.938 & 0.844 & 0.153 & 1.309 & 0.589 & 0.892 & 1.106 \\
 $g-K_s$=1.4 & 0.441 & 1.000 & 1.067 & 0.998 & 0.928 & 0.184 & 1.421 & 0.625 & 0.959 & 1.209\\
 $g-K_s$=1.6 & 0.402 & 1.000 & 1.100 & 1.049 & 1.008 & 0.221 & 1.520 & 0.658 & 1.013 & 1.299 \\
 $g-K_s$=1.8 &  0.372 & 1.000 & 1.135 & 1.097 & 1.085 & 0.266 & 1.605 & 0.692 & 1.062 & 1.379 \\
 $g-K_s$=2.0 &  0.344 & 1.000 & 1.179 & 1.165 & 1.186 & 0.320 & 1.691 & 0.734 & 1.127 & 1.475 \\
\hline\end{tabular}
\end{center}
{\sc{Notes}}: The flux ratios are taken in arbitrary units of energy per unit wavelength interval, and the color indices are in
AB magnitudes.
\end{table*}